\documentclass[tighten, times, twocolumn]{aastex62}
\newcommand\apjcls{1}
\newcommand\aastexcls{2}
\newcommand\othercls{3}

\newcommand\papercls{\aastexcls}
\if\papercls \apjcls
\usepackage{apjfonts}
\else\if\papercls \othercls
\usepackage{epsfig}
\usepackage{margin}
\usepackage{times}
\fi\fi
\usepackage{ifthen}
\usepackage{natbib}
\usepackage{amssymb, amsmath}
\usepackage{appendix}
\usepackage{etoolbox}
\usepackage[T1]{fontenc}
\usepackage{paralist}

\if\papercls \apjcls
\newcommand\aas{\ref@jnl{AAS Meeting Abstracts}}% *** added by jh
\newcommand\dps{\ref@jnl{AAS/DPS Meeting Abstracts}}% *** added by jh
\newcommand\maps{\ref@jnl{MAPS}}% *** added by jh
\else\if\papercls \othercls
\usepackage{astjnlabbrev-jh}
\fi\fi

\if\papercls \aastexcls
\hypersetup{citecolor=blue, % color for \cite{...} links
            linkcolor=blue, % color for \ref{...} links
            menucolor=blue, % color for Acrobat menu buttons
            urlcolor=blue}  % color for \url{...} links
\else
\usepackage[%pdftex,      % hyper-references for pdflatex
bookmarks=true,           %%% generate bookmarks ...
bookmarksnumbered=true,   %%% ... with numbers
colorlinks=true,          % links are colored
citecolor=blue,           % color for \cite{...} links
linkcolor=blue,           % color for \ref{...} links
menucolor=blue,           % color for Acrobat menu buttons
urlcolor=blue,            % color for \url{...} links
linkbordercolor={0 0 1},  %%% blue frames around links
pdfborder={0 0 1},
frenchlinks=true]{hyperref}
\fi

\if\papercls \othercls

\else

\fi

\providecommand{\adsurl}[1]{\href{#1}{ADS}}

\makeatletter
\patchcmd{\NAT@citex}
  {\@citea\NAT@hyper@{%
     \NAT@nmfmt{\NAT@nm}%
     \hyper@natlinkbreak{\NAT@aysep\NAT@spacechar}{\@citeb\@extra@b@citeb}%
     \NAT@date}}
  {\@citea\NAT@nmfmt{\NAT@nm}%
   \NAT@aysep\NAT@spacechar\NAT@hyper@{\NAT@date}}{}{}

\patchcmd{\NAT@citex}
  {\@citea\NAT@hyper@{%
     \NAT@nmfmt{\NAT@nm}%
     \hyper@natlinkbreak{\NAT@spacechar\NAT@@open\if*#1*\else#1\NAT@spacechar\fi}%
       {\@citeb\@extra@b@citeb}%
     \NAT@date}}
  {\@citea\NAT@nmfmt{\NAT@nm}%
   \NAT@spacechar\NAT@@open\if*#1*\else#1\NAT@spacechar\fi\NAT@hyper@{\NAT@date}}
  {}{}
\makeatother

\makeatletter
\DeclareRobustCommand{\lowcase}[1]{\@lowcase#1\@nil}
\def\@lowcase#1\@nil{\if\relax#1\relax\else\MakeLowercase{#1}\fi}
\makeatother

\DeclareSymbolFont{UPM}{U}{eur}{m}{n}
\DeclareMathSymbol{\umu}{0}{UPM}{"16}
\let\oldumu=\umu
\renewcommand\umu{\ifmmode\oldumu\else\math{\oldumu}\fi}

\if\papercls \othercls

\else

\fi

\let\oldsim=\sim
\renewcommand\sim{\ifmmode\oldsim\else\math{\oldsim}\fi}
\let\oldpm=\pm
\renewcommand\pm{\ifmmode\oldpm\else\math{\oldpm}\fi}
\newcommand\by{\ifmmode\times\else\math{\times}\fi}
\newbox{\wdbox}
\renewcommand\c{\setbox\wdbox=\hbox{,}\hspace{\wd\wdbox}}
\renewcommand\i{\setbox\wdbox=\hbox{i}\hspace{\wd\wdbox}}

\newcount\timect
\newcount\hourct
\newcount\minct
\newcommand\now{\timect=\time \divide\timect by 60
         \hourct=\timect \multiply\hourct by 60
         \minct=\time \advance\minct by -\hourct
         \number\timect:\ifnum \minct < 10 0\fi\number\minct}

\catcode`@=11

\newcommand\comment[1]{}

\newcommand\commenton{\catcode`\%=14}

\renewcommand\math[1]{$#1$}
\newcommand\mathshifton{\catcode`\$=3}

\let\atab=&
\newcommand\atabon{\catcode`\&=4}

\let\oldmsp=\sp
\let\oldmsb=\sb
\def\sp#1{\ifmmode
           \oldmsp{#1}%
         \else\strut\raise.85ex\hbox{\scriptsize #1}\fi}
\def\sb#1{\ifmmode
           \oldmsb{#1}%
         \else\strut\raise-.54ex\hbox{\scriptsize #1}\fi}
\newbox\@sp
\newbox\@sb
\def\sbp#1#2{\ifmmode%
           \oldmsb{#1}\oldmsp{#2}%
         \else
           \setbox\@sb=\hbox{\sb{#1}}%
           \setbox\@sp=\hbox{\sp{#2}}%
           \rlap{\copy\@sb}\copy\@sp
           \ifdim \wd\@sb >\wd\@sp
             \hskip -\wd\@sp \hskip \wd\@sb
           \fi
        \fi}
\def\msp#1{\ifmmode
           \oldmsp{#1}
         \else \math{\oldmsp{#1}}\fi}
\def\msb#1{\ifmmode
           \oldmsb{#1}
         \else \math{\oldmsb{#1}}\fi}

\def\supon{\catcode`\^=7}

\def\subon{\catcode`\_=8}

\def\supsubon{\supon \subon}

\newcommand\actcharon{\catcode`\~=13}

\newcommand\paramon{\catcode`\#=6}

\comment{And now to turn us totally on and off...}

\newcommand\reservedcharson{ \commenton  \mathshifton  \atabon  \supsubon 
                             \actcharon  \paramon}

\catcode`@=12
\reservedcharson

\if\papercls \apjcls

\else

\fi

\newcommand\chisq{\ifmmode{\chi\sp{2}}\else\math{\chi\sp{2}}\fi}
\newcommand\redchisq{\ifmmode{ \chi\sp{2}\sb{\rm red}}
                    \else\math{\chi\sp{2}\sb{\rm red}}\fi}
\newcommand\Teq{\ifmmode{T\sb{\rm eq}}\else$T$\sb{eq}\fi}
\newcommand\mjup{\ifmmode{M\sb{\rm Jup}}\else$M$\sb{Jup}\fi}
\newcommand\rjup{\ifmmode{R\sb{\rm Jup}}\else$R$\sb{Jup}\fi}
\newcommand\msun{\ifmmode{M\sb{\odot}}\else$M\sb{\odot}$\fi}
\newcommand\rsun{\ifmmode{R\sb{\odot}}\else$R\sb{\odot}$\fi}
\newcommand\mearth{\ifmmode{M\sb{\oplus}}\else$M\sb{\oplus}$\fi}
\newcommand\rearth{\ifmmode{R\sb{\oplus}}\else$R\sb{\oplus}$\fi}
\shorttitle{AGNs}
\shortauthors{Wo\l owska {\em et al.}}
\usepackage{graphicx}
\usepackage{natbib}
\usepackage{lipsum}
\usepackage{longtable}
\usepackage{afterpage}
\usepackage{rotating}
\usepackage{wrapfig}
\usepackage{float}
\floatstyle{boxed} 
\usepackage{amsmath}
\let\oldAA\AA
\renewcommand{\AA}{\text{\normalfont\oldAA}}
\newcommand{\chandra}{{\it Chandra}}
\newcommand{\xmm}{XMM-{\it Newton}}
\DeclareUnicodeCharacter{2212}{-}
\begin{document}

%\title{Caltech-NRAO Stripe 82 Survey (CNSS) Paper V: Switched-on radio AGNs}

\title{Caltech-NRAO Stripe 82 Survey (CNSS) Paper. V. AGNs that transitioned to radio-loud state}

%% AUTHOR/INSTITUTIONS FOR AASTEX6.1:
\author{Aleksandra Wo\l owska}
\affiliation{Institute of Astronomy, Faculty of Physics, Astronomy and Informatics, NCU, Grudzi\k{a}dzka 5/7, 87-100, Toru\'n, Poland}

\author{Magdalena Kunert-Bajraszewska}
\affiliation{Institute of Astronomy, Faculty of Physics, Astronomy and Informatics, NCU, Grudzi\k{a}dzka 5/7, 87-100, Toru\'n, Poland}

\author{Kunal P. Mooley}
\affiliation{National Radio Astronomy Observatory, P.O.\,Box O, Socorro, NM 87801, USA}
\affiliation{Cahill Center for Astronomy, MC 249-17, California Institute of Technology, Pasadena, CA 91125, USA}

\author{Aneta Siemiginowska}
\affiliation{Center for Astrophysics $|$ Harvard \& Smithsonian, Cambridge, MA, USA}

\author{Preeti Kharb}
\affiliation{National Centre for Radio Astrophysics (NCRA) $-$ Tata Institute of Fundamental Research(TIFR), S.P. Pune University Campus, Post Bag 3, Ganeshkhind, Pune 411007, India}

\author{C. H. Ishwara-Chandra}
\affiliation{National Centre for Radio Astrophysics (NCRA) $-$ Tata Institute of Fundamental Research(TIFR), S.P. Pune University Campus, Post Bag 3, Ganeshkhind, Pune 411007, India}

\author{Gregg Hallinan}
\affiliation{Cahill Center for Astronomy, MC 249-17, California Institute of Technology, Pasadena, CA 91125, USA}

\author{Mariusz Gromadzki}
\affiliation{Astronomical Observatory, University of Warsaw, Al. Ujazdowskie 4, 00-478 Warsaw, Poland}

\author{Dorota Kozie\l -Wierzbowska}
\affiliation{Astronomical Observatory of Jagiellonian University, ul. Orla 171, 30-244 Krak\'ow, Poland}

\email{ola@astro.umk.pl}

% %% Extra info for aastex:
% \received{Yesterday}
% \revised{Today}
% \accepted{Tonight}
% \published{Tomorrow}
% \submitjournal{AASJournal}

\begin{abstract}

A recent multiyear Caltech-NRAO Stripe 82 Survey (CNSS) revealed a group of objects that appeared as new radio sources after $>$5--20 yr of absence. They are transient phenomena with respect to the Faint Images of the Radio Sky at Twenty Centimeters (FIRST) survey and constitute the first unbiased sample of renewed radio activity. Here we present a follow-up, radio, optical, and X-ray study of them. The group consists of 12 sources, both quasars and galaxies with wide redshift 
($\rm 0.04 < z < 1.7$) and luminosity ($\rm 22<log_{10}[L_{1.4GHz}/W~Hz^{-1}]>24.5$) distributions. Their radio properties in the first phase of activity, namely the convex spectra and compact morphology, allow them all to be classified as gigahertz-peaked spectrum (GPS) sources. We conclude that the spectral changes are a consequence of the evolution of newly born radio jets. 
Our observations show that over the next few years of activity the GPS galaxies keep the convex shape of the spectrum, while GPS quasars rapidly transform into flat-spectrum sources, which may result in them not being recognized as young sources. The wide range of bolometric luminosities, black hole masses, and jet powers among the transient sources indicates even greater population diversity in the group of young radio objects. 
We also suggest that small changes of the accretion disk luminosity (accretion rate) may be sufficient to ignite low-power radio activity that evolves on the scale of decades.

\end{abstract}

% http://journals.aas.org/authors/keywords2013.html

\section{Introduction}
\label{introduction}

Radio-loud active galactic nuclei (AGNs) constitute a small fraction of the active galaxy population in which accretion onto the central supermassive black hole (SMBH) of a galaxy generates a relativistic jet that transports relativistic plasma and magnetic field out to large distances, creating a characteristic radio morphology \citep{Longair, Kellerman}. As the radio source evolves, it goes through different phases of development.
During the first stages of nuclear activity, called the gigahertz-peaked spectrum (GPS) and compact steep spectrum (CSS) sources, the radio jets have sizes smaller than a few kiloparsecs and reside within the interstellar medium of the host galaxy. Their radio spectra have convex shapes with peak emission frequency inversely proportional to the source size \citep{fanti, Orienti, Odea20}. If the jets are free to expand further, to the intergalactic medium, the source evolves into Fanaroff-Riley type I (FRI) or type II (FRII) morphology \citep{Fanaroff}, reaching sizes of a few hundred kiloparsecs. The large size is thus simply an effect of old age, and this is the essence of the theory of evolution of radio AGNs \citep{fanti, Redhead, Odea2, Snellen2000, MKB10, An}. Moreover the  activity  is  expected  to  be  cyclical  e.g., \citep{Best05, Saikia}.

However, statistical studies have revealed that there is an excess of compact sources with peaked spectra, in comparison to fully developed, large radio galaxies \citep{Odea91, Snellen2000, An}. This indicates that not all GPS and CSS sources evolve into extended structures, and different reasons for this premature cessation of radio activity is debated in the literature. One of the most popular is the presence of a very dense environment in the host galaxy that can significantly hinder or even prevent the development of the source above a certain size, called the frustration scenario \citep{Odea, Wagner, Mukherjee}. This is especially true for weak jets that will not be able to break through the dense clouds of gas.
The recent X-ray observations indicate a possible presence of a dense environment in the youngest and more compact radio galaxies \citep{Siemiginowska,ostorero, Sobolewska}.
In a different scenario the interaction between the jets and the post-merger environment may cause the enhancement of radio emission resulting in their over-abundance in radio selected samples \citep{Morganti, Tadhunter, Dicken}. However, the correlation of radio power with X-ray luminosity (which is considered a measure of AGN bolometric luminosity) does not confirm this hypothesis \citep{MKB14}. And finally the excess of compact sources can be explained by a short time scale episodic activity of radio objects. According to this scenario, the radiation pressure instabilities in an accretion disk make the radio sources transient on timescales $<10^4-10^5$ years \citep{Reynolds, Czerny, MKB10, me, Silpa}. In addition the duration of the radio activity phase may scale with the jet power such that the low-power objects are more short-lived and may stay compact for most of their lifetime \citep{Capetti, Hardcastle, MKB2020}. Perhaps, as it has been recently suggested by \citet{wojtowicz}, the SMBHs behave similarly to galactic X-ray binaries in which the radio jets are produced most efficiently during the high/hard states, and suppressed during the soft states. The dramatic changes in radio luminosity that occurs during the state changes should be visible in radio surveys dedicated to slow transient phenomena. 

To date, there have been rather few radio surveys dedicated to slowly variable and transient sources, and they all have a number of limitations. The majority of these surveys were single, multi-epoch interferometric pointings with a limited field of view, and as a result, the number of variables and transients is low \citep[e.g.,][]{Carilli, Mooley13}.  The Caltech-NRAO Stripe 82 Survey \citep[CNSS;][]{Mooley,Mooley2019} and the Very Large Array (VLA) Sky Survey \citep[VLASS;][]{Lacy} brings a new quality.
The CNSS was a dedicated radio transient survey carried out with the Jansky VLA between 2012  December and 2015 May. Observations of the 270 deg$^2$ of the Sloan Digital Sky Survey (SDSS) Stripe 82 region were carried out over five epochs at the S band (2--4 GHz; center frequency of 3 GHz) and with a uniform thermal rms noise of $\sim80~\mu$Jy per epoch. 
In this survey, many variable/transient sources were discovered, a large fraction of them being of AGN origin. 
Among them, comparison with the existing survey data, namely the Faint Images of the Radio Sky at Twenty Centimeters \citep[FIRST;][]{White} and the VLA survey of the SDSS Southern Equatorial Stripe \citep[VLA-Stripe 82;][]{Hodge}, has revealed  transient sources on timescales $<$20 years that are likely associated with renewed AGN activity.

The first discovered switched-on radio activity was in CNSS quasar VTC233002$-$002736 which increased its flux density about ten times or more at 1.4 GHz over a 15 yr period \citep{Mooley}. Another example of such behavior in the CNSS, found in quasar 013815+00, has been reported recently by \cite{MKB2020}. The transition from radio-quiet to radio-loud phase in this source coincided with changes in its accretion disk luminosity, which we interpret as a result of an enhancement in the SMBH accretion rate. 
Very recently the detection of changing-state AGNs has been also reported by \citet{Nyland} as a result of the VLASS survey. Their quasars show a large flux density increase in the S band and convex radio spectra, as in our CNSS objects. Similar radio behavior has been also reported in a group of narrow-line Seyfert 1 (NLS1) galaxies considered so far to be radio-quiet \citep{Lahteenmaki}.

In this paper we report a total sample of 12 slow transient sources on timescales $<$20 years that were detected in the whole CNSS survey covering the 270 deg$^2$ of the SDSS Stripe 82 region. The sample includes both the first discovered CNSS quasar VTC233002$-$002736 (hereafter 233001$-$00) and quasar 013815+00 already widely discussed by \citet{MKB2020} (also see Table \ref{table1_basic}). We present follow-up comprehensive, multi-epoch, and multiwavelength study of these sources as well as the discussion about their nature and implications of this discovery on the jet formation and triggering mechanism theories.

Throughout this paper we assume a cosmology in which $\rm H_0= 70~km~s^{−1}~Mpc^{−1}$,$\rm \Omega_m= 0.3$, and $\Omega_{\lambda}= 0.7$. The radio spectral index $\alpha$ is defined in the sense $S \propto \nu^{\alpha}$.

\section{Sample Selection}\label{sec:sample}
Using the first three epochs of observations of the CNSS survey \citep[][Mooley et al. 2021, in preparation]{Mooley}, carried out between 2013 December 19 and 2014 February 17, we prepared a deep 3\,GHz image mosaic of the full Stripe 82 region using AIPS {\tt IMAGR} and {\tt FLATN} tasks and a corresponding 5$\sigma$ source catalog using AIPS {\tt SAD}. 
For the $\sim$27,000 sources in the catalog, the median local RMS noise is $\sigma=69\,\mu$Jy.

We then compared this CNSS catalog with the catalogs from the FIRST \citep{White} and NVSS \citep{condon1998} surveys, carried out at 1.4 GHz, in order to find possible new radio sources (i.e. transient candidates) in the CNSS. We retained only those CNSS sources that: 1) are pointlike, i.e. have a ratio of the integrated-to-peak flux densities less than 1.5 \citep[as done by][]{Mooley}, 2) have a signal-to-noise ratio (S/N) larger than 10, 3) have coverage within the FIRST footprint and are absent in the FIRST and NVSS catalogs, 4) have flux density larger than 2.8 mJy (i.e. implied spectral index $\alpha>1.4$, between CNSS and FIRST source detection threshold of around 1 mJy), and 5) are away from bright sources (6\arcmin~region around $>$100 mJy) to mitigate false positives due to imaging artifacts.

\begin{figure*}
\centering
\includegraphics[scale=0.6]{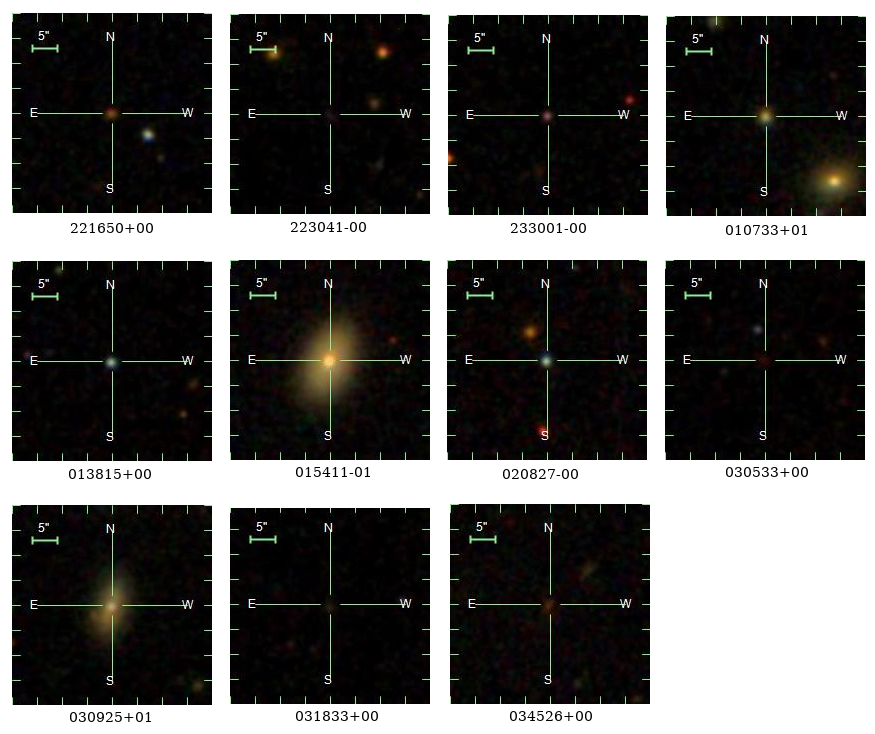}
\caption{Cut-out images from SDSS available for 11 out of 12 sources from our sample.}
\end{figure*}

For the resulting 20 sources, we inspected the FIRST image cutouts and noted that five of them were detected at $\sim$4$\sigma$ significance.
We rejected these sources and prioritized nine out of the remaining 15 (that are relatively bright radio sources and/or have a host galaxy that is detected in the SDSS) for multiwavelength follow-up.
We also included three serendipitous sources (221650$-$00, 010733$+$01 and 013815$+$00; seen to have host galaxies distinctly detected in the SDSS while investigating CNSS sources that were absent in FIRST). The three sources follow the selection criteria mentioned above except 221650$-$00, which has 3 GHz flux density around 2 mJy.
The sample of 12 sources\footnote{The full sample of "switched-on" radio AGN will be reported in Mooley et al. in preparation.} thus selected for multiwavelength follow-up is given in Table~\ref{table1_basic}.

\begin{deluxetable*}{ c c c c c c c c c c c c c c c c}
\tablecaption{Basic properties of the total sample of 12 transient radio sources.}
\tablehead{
Name & R.A.    & Decl. & ID & z & $3\sigma_{FIRST} $ & $\rm S_{1.4}$& $\rm logL_{1.4}$& $\rm S_{5}$&$\rm logL_{5}$& LAS & LLS & $r'$  \\
   & h~m~s  & $\degr$~$\arcmin$~$\arcsec$& &  & (mJy) & (mJy) &  $\rm (W~Hz^{-1})$ & (mJy) & $\rm (W~Hz^{-1})$& (mas) & ($\rm h^{-1}~pc $) & (mag)\\
 (1) & (2) & (3) & (4) & (5) & (6) & (7) & (8) & (9) & (10) & (11) & (12) & (13) \\  
}
\startdata
221650$+$00&22:16:50.44&+00:54:29.14& G &0.55$\ast$&0.39&0.52&23.1&3.20&24.5& 1.1 & 7.1 & 20.98\\
221812$-$01&22:18:12.95&$-$01:03:44.23& $-$&$-$ &0.51&1.60&$-$&14.67&$-$ & 5.4 & $-$ & $-$\\
223041$-$00&22:30:41.44&$-$00:16:44.29&G &0.84$\ast$&0.37&0.98&23.8&5.94&25.2&5.9 &45.0 & 22.54\\
233001$-$00&23:30:01.81&$-$00:27:36.53& Q&1.65&0.32&3.45&24.6&6.89&25.7& 5.7 & 48.3 & 21.32\\
010733$+$01&01:07:33.51&+01:12:22.78& G&0.12&0.47&3.03&22.9 &2.20&22.9& 0.8 & 1.6 & 19.13  \\
013815$+$00&01:38:15.06&+00:29:14.07&Q&0.94&0.32&1.48&24.5&3.68&25.0&8.5&67.0 & 19.77\\
015411$-$01&01:54:11.67&$-$01:11:49.74& G&0.05&0.45&3.79& 22.3&6.38&22.6& 3.0 & 2.9 & 15.41 \\
020827$-$00&02:08:27.06&$-$00:52:08.04& Q&1.34&0.49&1.90&24.2 &8.02&25.7& 0.9 & 7.7 & 19.55 \\
030533$+$00&03:05:33.12&+00:46:09.93& G&0.42$\ast$&0.41&0.88&23.4&5.05&24.5& 1.0 &5.3 & 24.30 \\
030925$+$01&03:09:25.99&+01:14:57.89& G&0.04&0.35&3.69&22.1&9.60&22.6& 0.5 & 0.4 & 16.90 \\
031833$+$00&03:18:33.64&+00:26:35.97& G&0.40$\ast$&0.34&0.79&23.2&4.81&24.5& 1.8 & 9.7 & 21.76\\
034526$+$00&03:45:26.00&+00:41:56.12& G&0.45$\ast$&0.40&1.22&23.5&9.76&24.9& 1.2 & 6.9 & 20.27\\
\enddata

\vspace{0.1 in}
\tablecomments{Columns are listed as follows: (1) source name; (2,3) VLA coordinates in J2000.0; (4) optical ID from SDSS, G - galaxy, Q - quasar, (5) spectroscopic redshift, $\ast$ means photometric redshift; (6) the 3$\sigma$ noise level at the location of the source measured in the 1.4 GHz FIRST images \citep{White}; (7) the latest value of 1.4 GHz flux density measured based on our VLA observations; (8) log of the K-corrected radio luminosity at 1.4 GHz calculated based on the flux density from column 7 and $\alpha_{thick}$ index obtained from spectral modeling; (9) the latest value of 5 GHz flux density measured based on our VLA observations; (10) log of the K-corrected radio luminosity at 5 GHz calculated based on the flux density from column 9 and $\alpha_{thin}$ index obtained from spectral modeling; (11) largest angular size as the deconvolved major axis of the source measured in 4.5 GHz Very Long Baseline Array (VLBA) image; in the case of resolved object 013815+00, the angular size is measured in the 4.5 GHz contour map; (12) largest linear size; and (13) magnitude in SDSS $r'$ filter. Note that the 013815+00 quasar discussed in \citep{MKB2020} is included.}
\label{table1_basic}
\end{deluxetable*}

\section{Observations and Data Reduction}
\label{sec:observations}

We have undertaken a multifrequency follow-up campaign for the 12 transient objects including X-ray (\xmm\ and \chandra\ X-ray Observatory), radio (VLBA, VLA and GMRT), and optical studies (Keck, Southern African Large Telescope, SALT) at different resolutions. We also have multi-epoch VLA spectra for all of our sources. Additionally a set of SDSS photometric and spectroscopic data exists for some some of them. 
The basic parameters of the whole sample of transient sources have been gathered in Table \ref{table1_basic}. 
More details on one of these sources, 013815+00, can be found in a separate publication, CNSS Paper IV \citep{MKB2020}.

\subsection{Radio Data}
\subsubsection{VLA Observations}
\begin{figure}[t]
\centering
\includegraphics[scale=0.4]{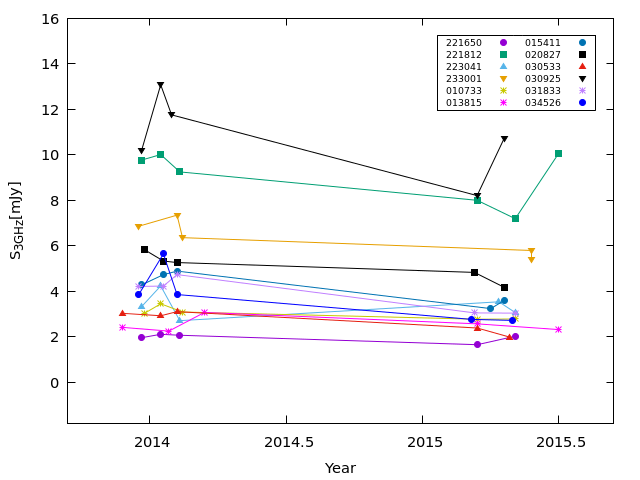}
\caption{3GHz CNSS light curves measured in the 
 five survey epochs.}
\label{light_curves}
\end{figure}

All sources from our sample detected by the CNSS project are transient with respect to the FIRST survey. Their 3 GHz VLA flux density measurements in five survey epochs are presented in Figure \ref{light_curves} and show that after the burst of radio activity, the flux density stabilizes at an average level of a few to a dozen millijanskys.
In addition, multifrequency observations of some of the transient sources were carried out with the VLA in the B configuration as part of the CNSS survey (2012-2015), using five receivers covering the spectrum from 1000 to 16,884 MHz (L, S, C, X, and Ku). The observing setup was the correlator with 16 spectral windows and 64 2-MHz-wide channels. 
Observations were continued in 2016 November under the project name VLA/16B-047 (6 hr), in the A configuration, using the same receivers and observing setup. The last epoch of observations was obtained in 2019 November (project  VLA/19B-209, 6 hr) in the D configuration using four receivers (S, C, X, and Ku). 

The target sources divided into groups, based on their coordinates, were observed with integration time from 2 to 10 minutes depending on the configuration and band. 3C48 was the primary flux density calibrator, and several phase calibrators chosen from the VLA calibrator manual were also observed during the run.
Then the detailed calibration and imaging of VLA data was carried out using CASA\footnote{http://casa.nrao.edu} software.
The data from single spectral windows (128 MHz each) for each source were extracted to be imaged and measured separately. In order to obtain good spectral coverage while maintaining a high S/N, four adjacent spectral windows were averaged in each band to measure average flux, resulting in four measurements of a flux per four bands, and two measurements for the L band, which consists of only eight spectral windows. The final measurements with error estimations are presented for 11 sources in Figure \ref{vla_spectra} and are gathered in Table \ref{VLA_measurement_points}.

\subsubsection{VLBA Observations}
VLBA C band observations were conducted from 2016 February 26 to May 2 under the project name VLBA/16A-007. Each of 12 sources was observed on separate days along with the phase calibrator and fringe-finder with the integration time varied from 2 to 4 hr. The total time assigned for a project was 45 hours. Additionally we divided the available bandwidth at C band receiver into two sub-bands centered at 4.5 and 7.5 GHz. This strategy allowed us to obtain images of our sources at two frequencies in one scan. Data reduction (including editing, amplitude calibration, instrumental phase corrections, and fringe-fitting) was performed with the standard procedure using the NRAO AIPS\footnote{http://www.aips.nrao.edu} software.
After this stage, the AIPS task IMAGR was used to produce the final total intensity images. Most of the sources appeared to be very compact, unresolved or slightly resolved with VLBA even at the higher frequency. The slightly extended sources were fitted by two Gaussian models using task JMFIT, and the solution was accepted only if the separation between the two components was larger than about one beam size; otherwise, the object was classified as one single extended source. We considered only source components with peak brightness $\geq 3\sigma$. Finally, for four of our sources, the components fulfilling the above criteria were found. These are: 010733+01, 034526+00, 030925+01 (see section \ref{notes} and Figure \ref{vlba_images}) and 013815+00 \citep{MKB2020}. The estimated angular and linear sizes of the sources measured in the 4.5 GHz image are listed in Table \ref{table1_basic}.

\subsubsection{GMRT Observations}
To complement the spectra at low frequencies, sub-gigahertz observations with upgraded GMRT in Band-3 (250-500 MHz) and Band-4 (550-850 MHz) were carried out on 2018 March 16 and 19, as a project 33\_016. The total allocated time was 26 hr.
Unfortunately, the Band-3 data were corrupted, and we could not obtain meaningful results.
Good-quality data were obtained only for Band-4. Eight out of 12 observed sources were detected and analyzed using a CASA-based pipeline (\href{http://www.ncra.tifr.res.in/~ishwar/pipeline.html}{http://www.ncra.tifr.res.in/$\sim$ishwar/pipeline.html}). The details of the pipeline are available in  \citet{Ishwara2020}.   In order to obtain as many measurement points as possible, while maintaining a sufficient S/N ratio, data were divided into a few spectral windows, each one processed separately, and added to the radio spectrum plot (Figure \ref{vla_spectra}).
The exception is the source 030533+00, for which we were unable to perform the correct flux calibration. The GMRT flux density measurements are presented in Table \ref{gmrt_points}.

%tu byla tablica X-rays

\subsection{Spectral Modeling}
\label{Spectral_modeling}
In order to characterize the significant changes of the radio spectra of our sources, we have fitted each spectrum with the analytical function.
Since the standard nonthermal power-law model is not enough to represent peaked sources showing significant curvature in their spectra, we followed the approach by \citet{Snellen} and used the modified power-law model:
\begin{equation}
S(\nu)=\frac{S_{p}}{(1-e^{-1})} \times (\frac{\nu}{\nu_{p}})^{\alpha_\mathrm{thick}} \times (1-e^{-(\frac{\nu}{\nu_{p}})^{\alpha_\mathrm{thin}-\alpha_\mathrm{thick}}})
\end{equation}
\label{powerlaw}
where $\alpha_{thick}$ is the optically thick spectral index, $\alpha_{thin}$ is the optically thin spectral index, and $S_{p}$ and $\nu_{p}$ are the peak flux density and peak frequency, respectively.
In some cases, the model does not fit the lowest- and highest-frequency sides of the spectrum perfectly. For those sources, the model was fitted using the approximate values for some of the free parameters. 
All fitted models are accurate to within the limits of the error. The obtained values for each epoch of observation are given in Table \ref{table2_spectra}, and the fitted spectra are presented in Figure \ref{vla_spectra}. 

Since the GMRT observations were carried out at a different time compared to the VLA observations, only VLA data were fitted in each epoch. However, the low-frequency GMRT data are also indicated in the plot. The fits made characterize the significant changes that occur mainly in the optically thin part of the spectra of our objects. At lower frequencies, as might be expected, the spectrum is more stable. It is also very likely that there are no changes at the GMRT frequencies. Therefore, the location of these points in relation to the modeled VLA radio spectrum curve influenced the choice of the best model. 
\subsection{X-ray observations}

We obtained X-ray observations of a subsample of the transient radio sources with \xmm\ and \chandra\
(see Table~\ref{chandra}).
Two objects were observed in 2016 using XMM-{\it Newton}/EPIC (Program 078345) with an exposure time of $\sim$24 ks on each source. The third one was available in the archive (ObsID 0673002341) and was observed as part of the Stripe 82 X-ray (82X) survey \citep{LaMassa13}.

Five objects were observed by {\it Chandra} with ACIS-S3 using the standard aim point location and the one-eight subarray readout mode, with the $\sim$ 15 ks exposure time on each source. The observations were carried out in a few months at the turn of 2017 and 2018. The {\it Chandra} data were reduced using CIAO 4.11 \citep{Fruscione} with the calibration files from CALDB  4.8.5. 
We assumed a circular extraction region for each source, with the radius $1\arcsec.5$, which also contains the entire radio emission. The background regions consisted of an annulus with the radius between $4\arcsec-8\arcsec$ 
centered on the source. Two sources have been detected in these observations. In the case of no detection, upper limits on the flux densities have been estimated using the {\tt aprates} tool in CIAO (see Table \ref{chandra}).

We calculated the hardness ratios for the detected sources following the method of \citep{park2006} and assuming $HR = {{H-S} \over {H+S}}$, where S and H are the source counts for the soft (0.5-2keV) and hard (2-7keV) bands, respectively (Table~\ref{chandra}).

\subsection{SALT Spectroscopy}

Three transient objects (010733$+$01, 015411$-$01, and 030925$+$01) 
have been observed with \citep[SALT,][]{Buckley06} under the project 2019-1-SCI-023, using the Robert Stobie Spectrograph \citep[RSS,][]{Burgh03}. 
The observations were carried out on 2019 July 10 (010733$+$01, 015411$-$01) and August 9.
The on-source exposure times were 1800/600/300 s, respectively, and were done in the long slit mode with a slit width of 1".5. An additional 4 s was used on an exposure of the Xenon calibration lamp.
We used the $\rm 4x2$ binning readout option, which allowed us to reach an S/N=100. 
In order to cover the main lines of interest, we used PG0900 VPH grating with a 15.8725 degree tilt angle, giving us wavelength coverage from $\rm 4463\AA$ to $\rm 7523\AA$. For the central wavelength of $\rm 5900\AA$, the spectral resolution was R=1065.
Order blocking was done with the UV PC03850 blocking filter. The sources were observed on a clear night with $\sim 2"$ seeing.

The preliminary data reduction (gain and amplifier cross-talk corrections, bias subtraction, amplifier mosaicking, and cosmetic corrections) was done with a semiautomated pipeline from the SALT PyRAFpackage\footnote{http://pysalt.salt.ac.za} by the SALT observatory staff \citep{Crawford10}. We performed  further reduction including wavelength calibration, background subtraction, extraction of 1D spectra, and flux calibration using the IRAF package. 
We could only obtain relative flux calibrations from observing spectrophotometric standards during twilight on other nights, due to the SALT design, which has a time-varying, asymmetric, and underfilled entrance pupil. As a result there is a difference in the continuum level between the SALT and SDSS observations. To remove this effect, we took only a part of the SALT spectra corresponding to the SDSS aperture into consideration. Further, we rescaled the SALT spectrum using a scaling function found by comparing the SALT and SDSS spectra.
However, the spectrum of the source 015411$-$01 turned out to be of poor quality for reliable measurements due to a low signal to noise ratio (S/N$<3$) and strong sky lines distorting H$\alpha$ and [N II] emissions.
Finally, we prepared the SALT spectra of two objects, 010733$+$01 and 030925$+$01, for the dispersion and line measurements described in the next section. The SALT spectra are presented in Figure \ref{image_optical_spectra}.

\subsection{Emission Line Measurements}
\label{emission_lines}
To track the brightness changes in the optical and UV range of the presented sources, we have collected all available photometric data points since 1992 until 2008 from the SDSS and data gathered since 2005 to 2014 from the Catalina Sky Survey (CRTS; \citep{Drake}. Both CRTS and SDSS data for all investigated sources showed no significant changes over the whole period of observations. Slight fluctuations visible in SDSS photometric measurements are within the limit of error for all sources.

The good-quality SDSS spectroscopic observations are available for five out of 12 of our objects. These are 010733$+$01, 015411$-$01, 020827$-$00 and 030925$+$01 which we analyze in this paper, and the quasar 013815$+$00 presented in \citet{MKB2020}. Additionally the Keck (DEIMOS) spectroscopic observation of source 233001$-$00 \citep{Mooley} and the SALT (RSS) observation of sources 010733$+$01 and  030925+00 are also processed and analyzed in this work. All spectra were corrected for Galactic extinction with the reddening map of \citet{Schlafly}, and shifted to the rest-frame wavelength by using the SDSS redshift (Figure \ref{image_optical_spectra}).

In the case where the continuum of the source was dominated by AGN emission, the
decomposition of the spectra has been done using the IRAF package and assuming the following components: power law (representing the emission from an accretion disk), FeII pseudo-continuum, and suite of Lorentzian and Gaussian components to model the broad and narrow emission lines. In order to remove the contribution from FeII emission, an iron template from \citep{Bruhweiler} was fitted to both spectra. The template was convolved with Gaussian profile with different dispersion values for kinematic broadening of FeII lines, in order to find the most accurate one. 

Spectra strongly contaminated by the host galaxy starlight were processed with the code STARLIGHT \citep{Starlight} in order to fit their continua and measure the stellar velocity dispersion. The code allows the user to adjust the model to the observed spectrum using the Markov Chain Monte Carlo method, and a database containing 150 spectra of star populations of different ages and metallicity. Obtained results provide abundant information about the source, such as percentage share of individual star populations in the observed spectrum, their masses, and the line broadening parameter, which allows us to calculate the star dispersion coefficient using a relation:

\begin{equation}
\sigma^2_* = vd^2 - \sigma^2_{inst} + \sigma^2_{base}
\end{equation}
\label{dispeq}
where $vd$ is the line broadening parameter, and $\sigma_{base}$ and $\sigma_{inst}$ are stellar model and instrumental velocity dispersion, respectively.

The emission lines were fitted with a Lorentzian/Gaussian function (with single or multiple components, depending on a line) to determine fluxes and FWHMs. 
All of the widths of the narrow lines were corrected by the instrumental resolution.
The properties of the emission lines resulting from the fit to each spectrum are listed in Table \ref{table_emission_lines}.

The error estimation of the continuum flux density was made using the rms method. 
The uncertainty of the line flux measurements has been estimated using the standard formula for noise averaging
$\rm \sigma_f = \sigma_{c} L / \sqrt{N}$, where $\rm \sigma_c$ is the rms of the continuum flux density, L is the integration interval, and N is a number of spectrum samples.
The error of the line width has been calculated by finding the minimum and maximum width of the Lorentz/Gaussian line fit at which the integral of the fit changes by $\rm \sigma_f$ keeping the amplitude of the fit fixed.

\section{Calculations of Physical Parameters}
\label{physical_parameters}

The k-corrected luminosities of our sources are calculated with:
\begin{equation}
\mathrm{L=4\pi D^2_L S_{\nu}(1+z)^{-(1+\alpha)}~~W~Hz^{-1}}
\end{equation}
where $D_L$ is the luminosity distance, $S_{\nu}$ is the flux density at a given frequency, $z$ is the redshift and $\alpha$ is a thin or thick spectral index, depending on the spectrum shape at a given frequency.

In order to calculate the jet kinetic power, we used the relation established for evolved radio sources in the form discussed by \citep{Rusinek}; namely,
\begin{equation}
\mathrm{P_{j}=5\times 10^{22} \left(\frac{L_{1.4GHz}}{W~Hz^{-1}} \right)^{6/7}~~erg~s^{-1}}
\label{rusinek_equation}
\end{equation}
and the corresponding 1.4 GHz luminosity is calculated based on our VLA measurements.

The black hole mass of quasars was estimated from the MgII$\lambda$2799 line and luminosity at 3000$\rm \AA$, using the following relation \citep{Trakhtenbrot}:
\begin{equation}
{\frac{\mathrm{M_{BH}}}{\mathrm{M_{\odot}}}=5.6\times10^6 \left( \frac{\lambda L_{3000}}{10^{44} ~\mathrm{erg~s}^{-1}} \right)^{0.62}
\left[ \frac{\mathrm{FWHM(MgII)}}{10^3~\mathrm{km~s}^{-1}} \right]^2} \\
\end{equation}

Then the bolometric AGN luminosity was calculated (Table \ref{table_physical_parameters}) using $\rm \lambda L_{3000}$ and a conversion factor of 5.3 to convert from monochromatic to bolometric luminosity \citep{Runnoe}.

In the case of sources with a significant host contribution in the spectra, we used the established scaling relation $M_{BH}-\sigma_{*}$ \citep{Kormendy} to get their black hole masses:

\begin{equation}
\rm
log \left(\frac{M_{BH}}{M_{\odot}} \right)=8.49+4.38\times log \left(\frac{\sigma_{*}}{200kms^{-1}} \right)
\end{equation}

However, for two sources, the measurement of stellar velocity dispersion turned out to be problematic. In this case, to estimate the mass of the black hole, we used the velocity dispersion of the [O III] line core ($\rm \sigma_{ [OIII]} = FWHM_{[OIII]} /2.35$) which can serve as a surrogate for the $\sigma _ {*}$ \citep{Liao}.

Then, in order to calculate their bolometric luminosities, we used either $H\alpha$ or $H\beta$ line measurements corrected for the reddening in host galaxies by requiring $H\alpha/H\beta = 3$ and the bolometric correction factor defined by \citep{Netzer} for a narrow $H_{\beta}$ line:

\begin{equation}
\rm
k_{bol}=4580\times \left(\frac{L_{H\beta}}{10^{42}ergs^{-1}}\right)^{0.18}
\end{equation}

The methods  used to estimate the black hole mass and bolometric luminosity for each source are listed in Table \ref{table_physical_parameters}.

%For comparative purposes, in Figure \ref{jet_power} additional samples were placed and that require explanation. 
For the discussion in section \ref{jet_and_accretion}, we used additional source samples that require a short description.
Sources classified as narrow line radio galaxies (NLRGs), broad line radio galaxies (BLRGs) and radio-loud quasars (RLQs) are from \citet{Rusinek}. All of these objects, with a few exceptions that we have removed from the samples, have FRII radio morphologies. A sample of FRI-type radio sources is from \citet{Fricat}. 
Their bolometric luminosity was calculated based on the [OIII] luminosity with the correction factor $\rm L_{bol} = 3500~L_{[OIII]}$ \citep{Best}.
The group of objects marked as young radio sources (YRSs) is a collection of CSS and GPS sources from \citet{Liao} and \citet{wojtowicz}. We cross matched the objects from the sample of \citet{Liao} with available radio databases to get their 1.4 flux density. This resulted in a 21\% reduction in the sample size. The spectroscopic data have been taken from the articles directly. 

The jet kinetic power of all sources discussed in this article was computed using Equation \ref{rusinek_equation}. 
It is a calibrated \citet{Willott} formula based on calorimetry of radio lobes. But some studies suggest that in young radio sources, the enhancement of the radiative efficiency of compact radio-emitting jets and lobes may occur due to their direct interaction with the interstellar medium of host galaxies \citep{Tadhunter,Dicken}. This in turn can cause the overestimation of their jet power when using the calorymetric formula \citep{wojtowicz}. However, this hypothesis is not confirmed by the correlation of radio power with, considered as a measure of AGN bolometric luminosity, the X-ray luminosity \citep{MKB14}. Hence we calculated the jet power for all of our sources in the same way.

\begin{figure*}[htp]
\centering
\includegraphics[scale=0.37]{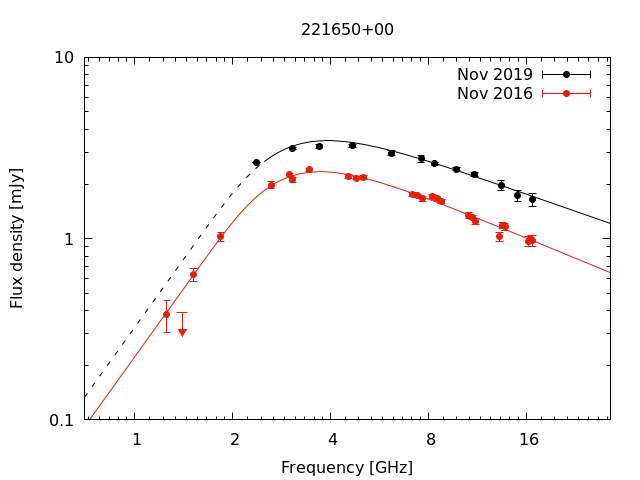}
\includegraphics[scale=0.37]{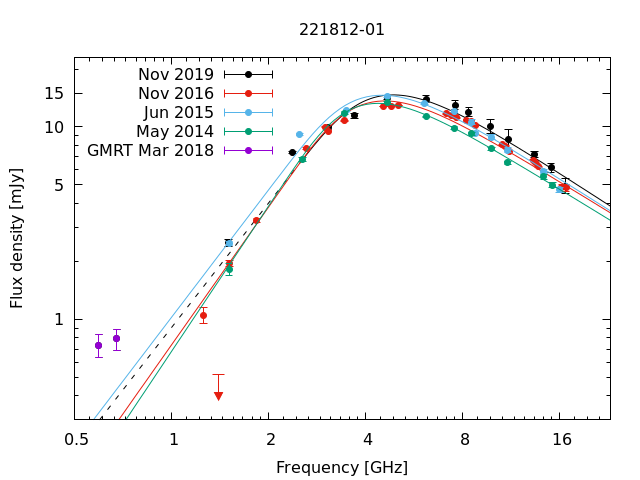}
\includegraphics[scale=0.37]{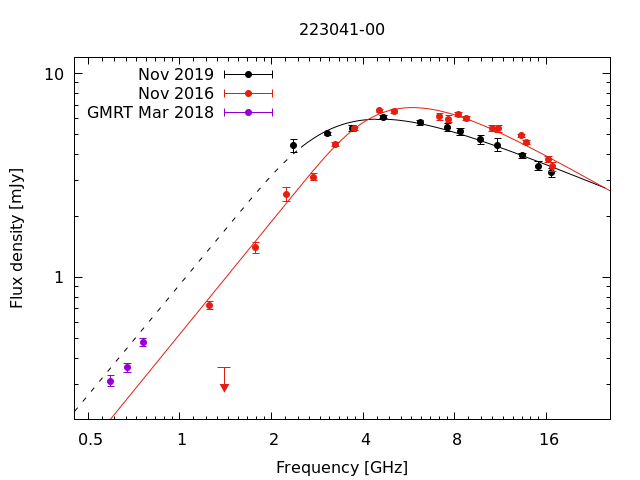}
\includegraphics[scale=0.37]{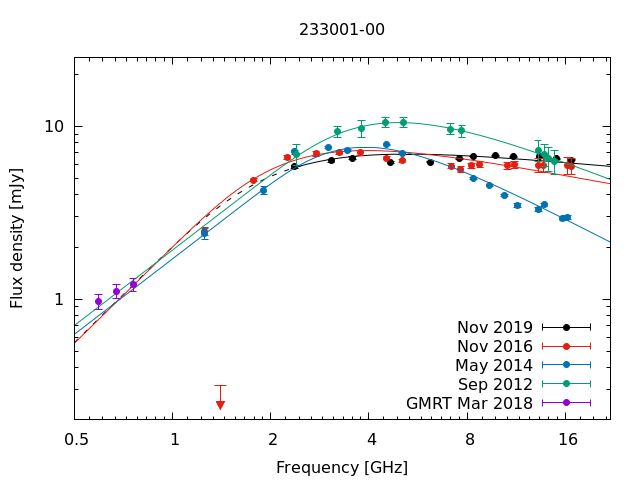}
\includegraphics[scale=0.37]{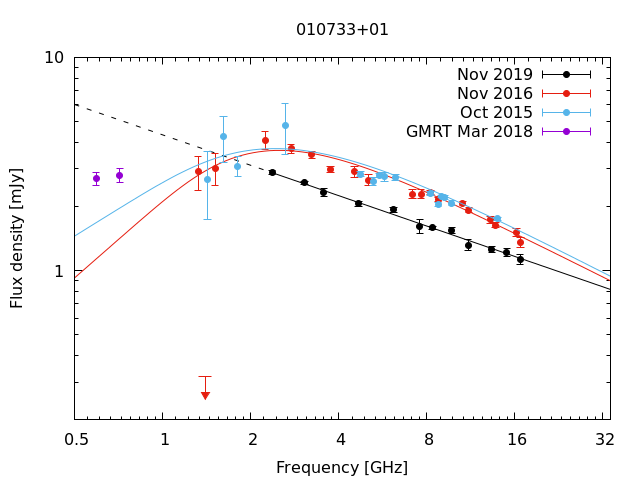}
\includegraphics[scale=0.37]{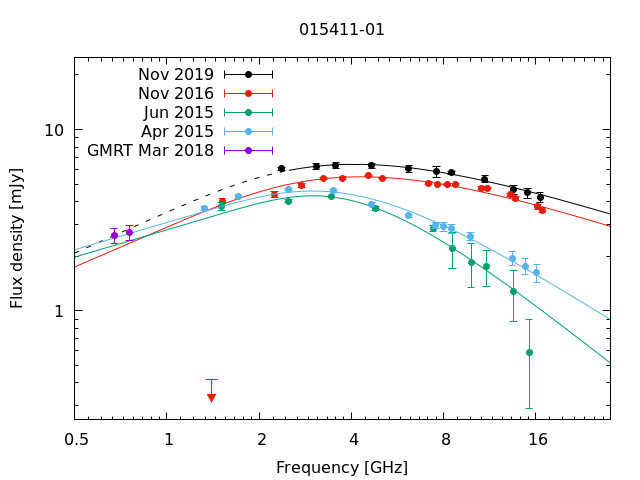}
\caption{Multi-epoch radio spectra with fitted models, including VLA measurements in the range 1-20 GHz and GMRT points in the range 550-850 MHz. The red arrow indicates the 3$\sigma$ upper limit at 1.4 GHz from the FIRST survey.}
\label{vla_spectra}
\end{figure*}

\setcounter{figure}{2}
\begin{figure*}[!htp]
\centering
\includegraphics[scale=0.37]{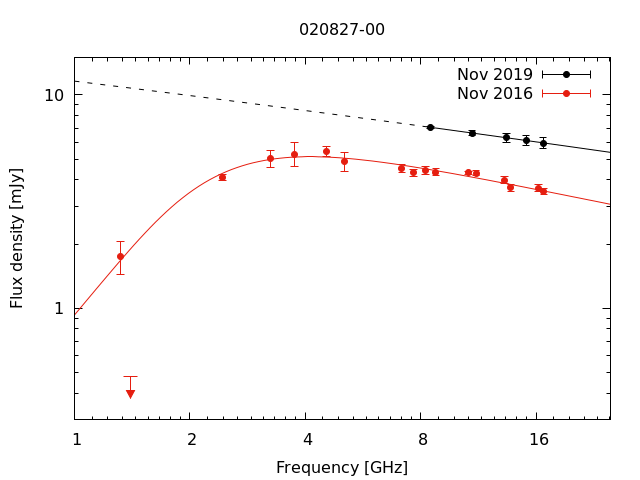}
\includegraphics[scale=0.37]{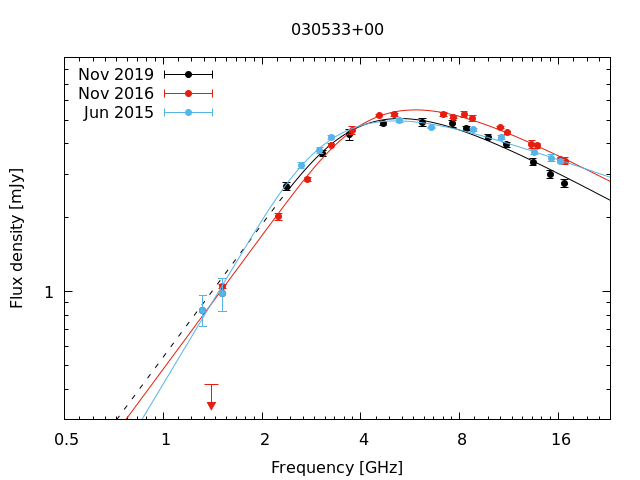}
\includegraphics[scale=0.37]{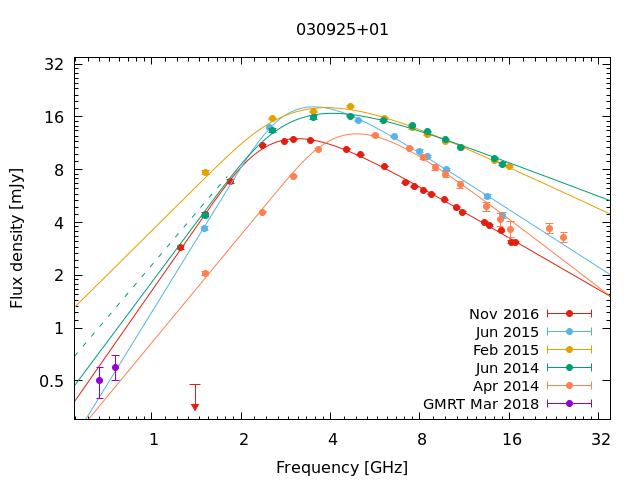}
\includegraphics[scale=0.37]{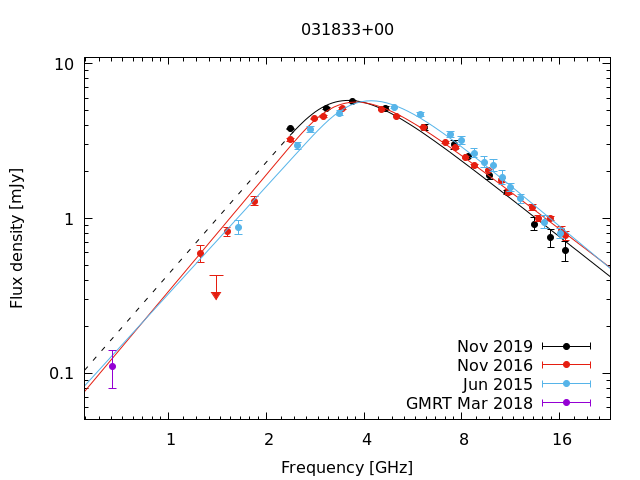}
\includegraphics[scale=0.37]{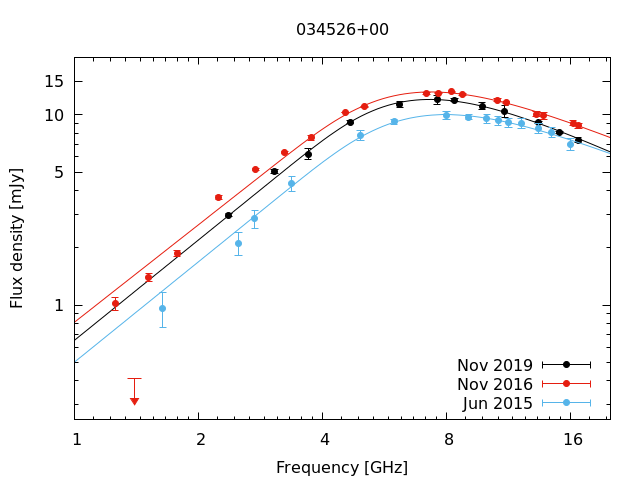}
\caption{Continued}
\end{figure*}

\section{Notes on Individual Sources}
\label{notes}
All sources are transients with respect to the FIRST survey. They  were  undetected  in  the  FIRST (1995-2011) at a sensitivity level of $<0.5$ mJy at 1.4 GHz but discovered in the CNSS observations (2012-2015) to have brightened significantly (an average of a factor of 6.5 for quasars and a factor of 4.5 for galaxies) and become radio-loud sources.
%and they were discovered by comparing the CNSS source catalog with the FIRST source catalog. FIRST observations of most of the sources were done in 1995 except one, 034526$+$00, which was observed using EVLA in 2011. 
Three out of 12 sources, namely 221812$-$01, 223041$-$00 and 010733$+$01, had already been detected in the earlier deep VLA-Stripe 82 survey performed in 2007-2009 \citep{Hodge}. The others were found based on the CNSS survey between 2012 and 2013 \citep{Mooley2019}. 
They were detected at the millijansky level, ranging from 1 to 13 mJy.

{\bf 221650$+$00.}
This is a very compact source, unresolved in high-resolution VLBA observations (Figure \ref{vlba_images}). The two VLA spectra taken over a period of 3 yr show a further increase in source brightness with a slight change in spectral slope (Figure \ref{vla_spectra}).
Careful inspection of the VLA-Stripe 82 image from 2009 may indicate a marginal detection of this source at the level of 0.43$\pm$0.12 mJy. It is below the catalog detection limit at this position, which is 0.54 mJy/beam \citep{Hodge}.
Only SDSS photometry is available for this object.

{\bf 221812$-$01.}
This source was detected in 2009 in the VLA-Stripe 82 survey \citep{Hodge} with integrated flux density of 1.39$\pm$0.11 mJy at 1.4 GHz, which is in agreement with our observations. There is also radio emission at these coordinates in FIRST observations from 1995 at the 3$\sigma$ noise level (much below the catalog detection limit at this position), which excludes reliable measurement. This however, indicates about a threefold increase in the source flux density noticed several years later.

The source is unresolved in  the high-resolution VLBA observations (Figure \ref{vlba_images}) and has a typical GPS spectrum that does not show large changes over time (Figure \ref{vla_spectra}).

There is no optical identification for this object, and it has not been detected in X-rays (Table \ref{chandra}).

{\bf 223041$-$00.}
A detection of this source is already present in the VLA-Stripe 82 survey. The observation was made in 2008, and the integrated flux of the source amounts to 0.86$\pm$0.06 mJy which is in agreement with our observations. The VLA measurements show the time-varying GPS spectrum (Figure \ref{vla_spectra}). The source is unresolved in VLBA observations (Figure \ref{vlba_images}).
Only SDSS photometry is available in case of this object.

\begin{deluxetable}{ c c c c c c c }
\tabletypesize{\scriptsize}
\tablecaption{Results of spectral modeling of VLA radio data.}
\tablehead{
Name  & Epoch &$S_{p}$&$\nu_{p}$& $\nu_{p}$(1+z) &$\alpha_{thin}$&$\alpha_{thick}$\\
    &       &  (mJy)  &   (GHz)&   (GHz) & & \\
(1) &  (2)  &   (3)    &   (4)            &    (5)                &   (6) & (7)         \\
}
\startdata
221650$+$00 & Nov 2016 & 2.20\pm 0.04 & 3.00\pm 0.10 & 4.65 & -0.74\pm 0.03 & 2.50\pm 0.15\\
          & Nov 2019 & 3.20\pm 0.06 & 3.00\pm 0.09 & 4.65 & -0.63\pm 0.03 & 2.50*\\
221812$-$01 & May 2014 & 12.93\pm 0.22 & 3.88\pm 0.13 &- &-1.03\pm 0.05 & 2.61\pm 0.08\\
          & Jun 2015 & 14.43\pm 0.45 &4.10\pm 0.44 & - &-1.06\pm 0.14 & 2.20*\\
          & Nov 2016 & 13.37\pm 0.25 & 4.10\pm 0.21 & - & -1.03\pm 0.06 & 2.38\pm 0.09 \\
          & Nov 2019 & 14.53\pm 0.53 & 4.50\pm 0.17 & - & -1.09\pm 0.09 & 2.15* \\
223041$-$00 & Nov 2016 & 6.70\pm 0.18 & 5.13\pm 0.18 & 9.44 & -0.85\pm 0.07 & 1.83\pm 0.15\\
            & Nov 2019 & 5.76\pm 0.17 & 3.69\pm 0.15 & 6.79 & -0.63\pm 0.05  & 1.80*\\
233001$-$00 & Sep 2012& 10.39\pm 0.07 & 4.39\pm 0.05 & 11.63 & -0.74\pm 0.07 & 1.45 \pm 0.25\\
          & May 2014& 7.55\pm 0.21 & 3.82\pm 0.15 & 10.12 & -0.98\pm 0.06 & 1.45*\\
          & Nov 2016& 6.60\pm 0.15 & 2.40\pm 0.19 & 6.36 & -0.35\pm 0.05 & 1.85 \pm 0.22\\
          & Nov 2019& 5.60\pm 0.22 & 2.20\pm 0.16 & 5.83 & -0.17\pm 0.03 & 1.85*\\
010733$+$01 & Oct 2015 & 3.73\pm 0.21 & 2.60\pm 0.37 & 2.91 & -0.71\pm 0.05 &0.85*\\
          & Nov 2016 &3.65\pm 0.17 & 2.30\pm 0.21 & 2.58 & -0.69\pm 0.05 & 1.20 \pm 0.15\\
          & Nov 2019 & - & - & - & -0.68\pm 0.04 & - \\
015411$-$01 & Apr 2015 & 4.16\pm 0.05 & 4.60\pm 0.12 & 4.83 &-1.09\pm 0.04 & 0.47 \pm 0.03\\
          & Jun 2015 & 3.79\pm 0.13 & 4.60\pm 0.37 & 4.83 & -1.35\pm 0.15 & 0.47 \pm 0.05\\
          & Nov 2016 & 5.48\pm 0.07 & 4.50\pm 0.26 & 4.73 &-0.57\pm 0.04 & 0.73 \pm 0.03\\
          & Nov 2019 & 6.43\pm 0.06 & 4.15\pm 0.26 & 4.36 &-0.55\pm 0.03 & 0.75*\\
020827$-$00 & Nov 2016 & 4.48\pm 0.22 & 2.58\pm 0.26 & 6.03 & -0.37\pm 0.05 & 2.15 \pm 0.33\\
            & Nov 2019 & - & - & - & -0.31\pm 0.01 & -\\
030533$+$00 & Jun 2015 & 4.46\pm 0.13 & 3.54\pm 0.16 & 5.03 & -0.47\pm 0.04 & 2.23 \pm 0.16 \\ 
          & Nov 2016 & 5.33\pm 0.08 & 4.85\pm 0.20 & 6.89 & -0.70\pm 0.04 & 1.82 \pm 0.11\\
          & Nov 2019 & 4.93\pm 0.09 & 4.40\pm 0.12 & 6.25 & -0.72\pm 0.08 & 1.80*\\
030925$+$01 & Apr 2014 & 12.45\pm 0.33 & 4.46\pm 0.10 & 4.64 & -1.15\pm 0.06 & 2.15 \pm 0.20\\
          & Jun 2014 & 15.71\pm 0.34 & 3.16\pm 0.09 & 3.29 & -0.64\pm 0.03 & 2.27 \pm 0.25\\
          & Feb 2015& 17.70\pm 0.41 & 3.13\pm 0.10 & 3.25 & -0.75\pm 0.04 & 2.10*\\
          & Jun 2015& 17.77\pm 0.47 & 3.12\pm 0.10 & 3.24 & -1.09\pm 0.05 & 2.75 \pm 0.22\\
          & Nov 2016& 11.72\pm 0.11 & 2.72\pm 0.06 & 2.83 & -0.98\pm 0.02 & 2.43 \pm 0.10\\
031833$+$00 & Jun 2015 & 5.77\pm 0.23 & 4.25\pm 0.10 & 5.95 & -1.75\pm 0.12 & 2.30 \pm 0.09\\
          & Nov 2016 & 5.65\pm 0.09 & 3.70\pm 0.05 & 5.18 & -1.60\pm 0.04 & 2.50 \pm 0.07\\
          & Nov 2019 & 5.80\pm 0.13 & 3.55\pm 0.08 & 4.97 & -1.65\pm 0.07 & 2.40*\\
034526$+$00 & Jun 2015 & 10.23\pm 0.10 & 7.33\pm 0.13 & 10.27 & -0.99\pm 0.08 & 1.78 \pm 0.03\\
          & Nov 2016 & 13.06\pm 0.15 & 6.80\pm 0.09 & 9.72 & -0.93\pm 0.03 & 1.70*\\
          & Nov 2019 & 11.70\pm 0.12 & 6.77\pm 0.09 & 9.72 & -0.95\pm 0.03 & 1.75 \pm 0.04\\
\enddata
\vspace{0.1 in}
\tablecomments{Columns are listed as follows: (1) source name; (2) epoch of observation; (3) the peak flux density; (4) observed frequency of the peak (spectral turnover); (5) rest-frame peak frequency; (6) spectral index of the optically thin part of the radio spectrum; (7) spectral index of the optically thick part of the radio spectrum, * indicates index values that have been fixed to obtain a proper fit.
The values quoted in Column 3-7 were obtained via modeling described in Section \ref{Spectral_modeling}.}
\label{table2_spectra}
\end{deluxetable}

{\bf 233001$-$00.}
This is a compact quasar unresolved in high-resolution VLBA observations (Figure \ref{vlba_images}). Its radio spectrum is characterized by large changes that are traced from 2012 when this source was discovered \citep{Mooley}. 
During the first few years of observations, the peak of the spectrum moved toward lower frequencies, and the radio spectrum evolved from a steep to flat spectrum (Figure \ref{vla_spectra}).
The good-quality optical-UV spectroscopic observation of this quasar has been made with DEIMOS at Keck II in 2012 \citep{Mooley}, i.e. in its radio-loud phase (Figure \ref{image_optical_spectra}). The spectrum shows a strong broad MgII line typical of unobscured AGNs (Tables \ref{table_emission_lines} and \ref{table_physical_parameters}).
There is also a detection of 233001-00 in X-rays (Table \ref{chandra}). 
The value of the hardness ratio indicates the soft X-ray spectrum, but detailed spectral studies require deeper X-ray observations.

{\bf 010733$+$01.}
Radio emission of 010733+01 is not present in the FIRST image, but the source was detected in 2007 in the VLA-Stripe 82 survey with an integrated flux density of 2.80$\pm$0.06 mJy which is in agreement with our observations. The multi-epoch VLA measurements show the time-varying GPS spectrum (Figure \ref{vla_spectra}). The 7.5 GHz VLBA observations revealed the presence of two components in the source: the central one, which is probably a radio core with flux density 1.63$\pm$0.01 mJy and the NE jet-like feature with flux density 0.42$\pm$0.02 mJy (Figure \ref{vlba_images}). This source has not been detected in X-rays 
(Table \ref{chandra}).

\begin{figure}[t]
\centering
\includegraphics[scale=0.48]{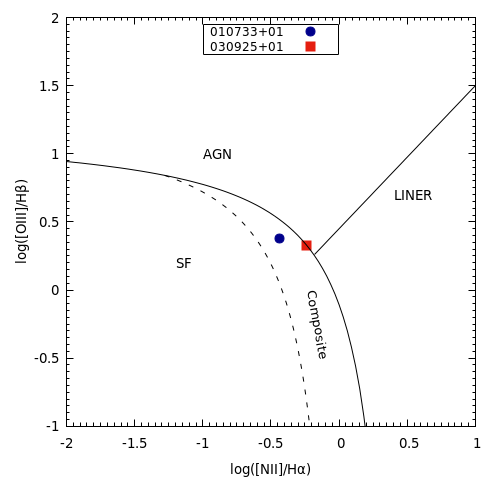}
\caption{Emission line ratio diagram \citep{Baldwin}. The different lines,
taken from \citet{Kewley01}, \citet{Kauffmann}, and \citet{Schawinski}, illustrate boundaries between sources classified as SF regions/galaxies, AGNs, low ionization
nuclear emission line regions (LINERs), and 'composite' sources. }
\label{bpt_diagram}
\end{figure}

The optical spectroscopic observations of this galaxy were made in 2015 (SDSS) and 2019 (SALT), both in its radio-loud phase (Tables \ref{table_emission_lines} and \ref{table_physical_parameters}). 
The comparison of the two spectra show that there are no significant differences between the measurements. 
The emission line ratios we have measured indicate that, in this galaxy, the ionizing radiation comes from an AGN and star-forming (SF) regions (Figure \ref{bpt_diagram}).

{\bf 015411$-$01.}
The SE extension is detected at both 4.5 and 7.5 GHz VLBA images of this object and may suggest the presence of a small jet (Figure \ref{vlba_images}). The VLA measurements show 
a significant flattening of the spectrum with time in its optically thin part (Figure \ref{vla_spectra}). 
There is also weak X-ray detection of this source with an intermediate value of the hardness ratio (Table \ref{chandra}). 
The spectroscopic observations of this object were made by SDSS in 2000, i.e. in its radio-quiet phase and after the burst of its radio activity in 2019 with SALT (Tables \ref{table_emission_lines} and \ref{table_physical_parameters}). 
However, the poor quality of the SALT spectrum made reliable measurements impossible.

{\bf 020827$-$00.}
This is a compact quasar unresolved on VLBA resolution (Figure \ref{vlba_images}). Its radio spectrum is flat with a peak at about 5 GHz (Figure \ref{vla_spectra}). 
The SDSS spectroscopic observation of it has been made in its radio-quiet phase (Tables \ref{table_emission_lines} and \ref{table_physical_parameters}). It exhibits strong broad MgII emission line typical for unobscured AGNs (Figure \ref{image_optical_spectra}).

{\bf 030533$+$00.}
Compact source with a small SW extension visible at 7.5 GHz in the VLBA observations (Figure \ref{vlba_images}). The VLA observations show a typical GPS spectrum (Figure \ref{vla_spectra}).
Only SDSS photometry is available in case of this object.

{\bf 030925$+$01.}
There are two epochs of VLBA observations made for this source (Fig. \ref{vlba_images}). The 4.5 and 7.5 GHz VLBA images from June 2015, when the source was in its high state, show the presence of two compact features: the central one with flux densities 11.52$\pm$0.04 and 8.68$\pm$0.03 mJy, and the NW component with flux densities 3.84$\pm$0.04 and 2.00$\pm$0.04 mJy at 4.5 and 7.5 GHz, respectively. The radio spectral indices amount to 0.56 and 1.24 for the central and NW features, respectively. Thus the initial radio structure of 030925+01 is core-jet-type with the central component being a radio core.
However, the same observations performed almost one year later (2016 March) show only a single compact object (Fig. \ref{vlba_images}). 
The multi-epoch VLA observations performed in 2014-2016 revealed significant changes in flux density and peak frequency of the spectrum (Figure \ref{vla_spectra}).

Spectroscopic observations of this object were made by SDSS several times between 2000 and 2001., i.e. in its radio-quiet phase, and after the burst of its radio activity in 2019 with SALT (Tables \ref{table_emission_lines} and \ref{table_physical_parameters}). Out of all SDSS spectra, we chose the spectrum from MJD 52258 (plate 412, fiber 402), indicated by SDSS as that with the best quality.
The spectra from both (2001 and 2019) epochs are very similar and display many narrow lines typical for obscured AGNs. 
According to the available line ratios in both phases, 030925$+$01 is probably a 
"composite" emission source, which includes radiation from an accreting SMBH in addition to young stars (Figure \ref{bpt_diagram}).

The SDSS optical image shows that 030925$+$01 has a disturbed
structure, which may indicate a recent merger event and thus greater presence of dust and gas in this galaxy.

Additionally, a weak X-ray detection of this source was made in 2017 November (Table \ref{chandra}). The X-ray spectrum appears to be relatively hard, although more counts are needed to confirm this result.

{\bf 031833$+$00.}
This is a slightly extended source without any clear features in VLBA observations (Fig. \ref{vlba_images}). 
Its radio spectrum has a typical GPS shape (Figure \ref{vla_spectra}). Only SDSS photometry of this source is available.

{\bf 034526$+$00.}
This object has been resolved into two components in 7.5 GHz VLBA observations (Fig. \ref{vlba_images}). The central component, which is probably the radio core, has a flux density of 6.81$\pm$0.03 mJy. The SW component is a jet-like feature with flux density 2.36$\pm$0.03 mJy. 

The VLA measurements show changes in source flux density that have occurred over several years. However, the spectrum retains the typical GPS shape with peak at very high rest-frame frequency of $\sim$10 GHz (Figure \ref{vla_spectra}). This source has not been detected in X-rays (Table \ref{chandra}).

\begin{figure*}[htp]
\centering
\includegraphics[scale=0.42]{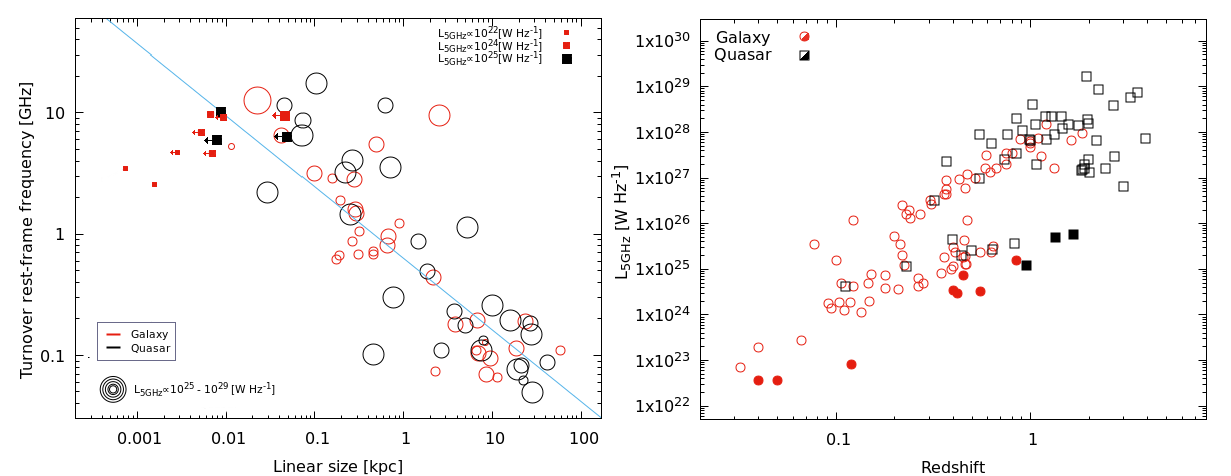}
\caption {{\bf Left}: the intrinsic turnover frequency vs. linear size for the CSS/GPS bulk sample from \citet{Odea2}, \citet{Snellen}, \citet{devries}, \citet{Stranghellini}, and \citet{fanti}, 
circles, with GPS sources presented in this work (squares). The sizes of the circles/squares correspond with the k-corrected radio luminosity at 5GHz, and arrows indicate maximum linear sizes for unresolved sources. The blue solid line indicates the linear relationship log~$\nu_p = -0.21 - 0.59\times \mathrm{log~LLS}$, found by \citet{Orienti}. 
{\bf Right}: redshift vs. luminosity at 5GHz for the bulk sample combined with the low-luminosity CSS sources from \citet{MKB10}, empty points, and transient objects presented in this work (filled points) with a distinction between quasars and galaxies. }
\label{odea_plot}
\end{figure*}

\section{Discussion}
The radio sources we studied in this project were discovered by CNSS \citep{Mooley} and are transient with respect to the FIRST survey \citep{White}. Together with the quasar 013815$+$00 recently published by \citet{MKB2020} they form a group of 12 objects. As the CNSS survey covers the area of $\rm \sim 270~deg^2$ of Stripe 82, this means a detection rate on the level of one such source at about $\rm 20~deg^2$. The current radio luminosity of these objects of $\rm log_{10}[L_{1.4GHz}/W~Hz^{-1}]>22.5$ indicates that they are now in a radio-loud state \citep{Kellerman}. The exceptions are the two sources (015411$-$01 and 030925$+$01) that are on the luminosity boundary between the objects being powered by the SMBH and those whose radio emission is powered by star formation in their host galaxies (Table \ref{table1_basic}). However, this limit is not sharp \citep{Mauch,Malefahlo}. 
We found that the radio characteristic of these objects and the X-ray detection favour an AGN origin of their emission.

The whole sample of 12 transient radio sources consists of three quasars with a redshift of $z>0.9$ and eight galaxies. One object has no optical identification. The redshift distribution of our sample is shown in Figure \ref{odea_plot} and compared to other samples of compact objects. It is clear that the composition of the samples is similar, although all of our objects are less radio luminous. Regardless of the luminosity, however, locally the population of compact sources is dominated by low-power galaxies. Quasars are brighter than galaxies, and thus it is easier to detect them at higher redshift. This is in agreement with the  previous studies of the redshift distribution of radio sources and unification by an orientation scheme  \citep{Jackson, Berton17}. 

Both groups of objects, galaxies and quasars, differ in radio, optical, and X-ray properties, as well as the character and rate of changes observed since the beginning of the radio activity.

\subsection{Analysis of Radio Properties}
The follow-up multifrequency VLA observations of our sources revealed a convex spectra peaking in the range 2$-$12 GHz at rest frame (Figure \ref{vla_spectra} and Table \ref{table2_spectra}). During the few years of monitoring, which are the first years of the radio-loud phase of the sources, most of them show variability of the radio flux density, mainly in the optically thin part of the spectrum. The most significant changes occurred in the case of two nearby galaxies 015411$-$01 and 030925$+$01 and two quasars 233001$-$00 and 020827$-$00. In the case of the latter, a progressive flattening of the spectrum is visible. In addition, sources 233001$-$00, 223041$-$00 and 030925$+$01 also have a very strong peak shift toward lower frequencies. Ultimately, after the first few years of life, the new radio quasars can be classified as flat-spectrum objects while galaxies keep the convex shape of the spectrum. We interpret the GPS spectra of our sources, for both quasars and galaxies, as a burst of new radio jet activity. The observed changes and flattening of the spectra probably reflect the changes taking place in the jet itself, and thus its expansion and dissipation of energy. Moreover, the formation of a new component has a large impact on the radio spectrum mostly in the case of sources oriented at a small angle to the line of sight. This causes the source to be observed as variable at high frequencies and may explain the changes observed in some of our sources.

The changes in the optically thick part of the synchrotron spectrum of our transient sources are definitely less significant. In most cases the estimated value of their optically thick spectral index differs from the theoretical limit of $\alpha_{thick}=2.5$ for a uniform source of synchrotron radiation (Table \ref {table2_spectra}). However, about half of our sources have very inverted spectra with $\alpha_{thick}\geq2$ and for the three sources, $\alpha_{thick}$ even reaches the canonical value of 2.5 in one of the epochs of observations, which is in agreement with their compact, unresolved morphology.

The high-resolution VLBA observations of our objects do not show any well-resolved
components in most cases. The presence of a small jet is visible in four sources (Figure \ref{vlba_images}) including quasar 013815+00 \citep{MKB2020}. However, since the majority of these observations were made about 3 yr after the burst of radio activity, it could have been enough time for complete jet energy dissipation for most of the objects, and the jet is no longer visible in the radio image. The two epoch VLBA observations of galaxy 030925$+$01 seem to confirm such a scenario. The small jet is visible on the 2015 images of 030925$+$01 while in 2016 only a very compact single component is present (Figure \ref{vlba_images}). 
In conclusion, our radio sources can be classified as either core-jet or pointlike objects.
\subsection{The Peak Frequency - Linear Size  Relationship}
It has been shown based on different samples of radio sources that there is a continuous distribution of young AGNs along the frequency of the peak (turnover frequency) $\nu_p - l$ linear size plane \citep{Odea,Orienti,Sotnikova}. This relationship suggests that the physical properties of the CSS and GPS sources are similar, and the only variable that depends directly on their size is their turnover frequency. The radio properties of our transient sources in their very initial phase of activity, namely the convex spectra and compact morphology, place them in the upper-left part of the $\nu_p - l$ plane and cause most of them to follow the established relationship quite accurately (Figure \ref{odea_plot}). The exceptions are three sources: 010733$+$01, 015411$-$00 and 030925$+$01,with the lowest radio luminosity $\rm log_{10}[L_{1.4 GHz}/W~Hz^{-1}]<23$.

There are two main processes causing turnover in the spectra of CSS and GPS sources. These are synchrotron self-absorption (SSA) and free-free absorption (FFA) caused by the external or internal environment of the synchrotron emitting volume \citep{deKool, Odea2, Bicknell}. There are examples of individual objects whose spectra are better described by FFA rather than by SSA \citep{Callingham17,Collier,Keim}, or in which both mechanisms could be at work \citep{Kameno}. Nevertheless, the relation between the frequency of the peak $\nu_p$ (turnover frequency) and the linear size $l$ is better reproduced by SSA over the large range of linear sizes \citep{Jeyakumar}. 

For our sample the values of the optically thick
spectral indices do not exceed the SSA limit of 2.5, but are close to the limit in a few objects.
This indicates that probably a nonhomogeneous synchrotron component rather than the presence of FFA is responsible for the low-frequency absorption of their radio spectra. However, we cannot exclude the presence of thermal plasma, either external or internal to the synchrotron emitting volume, which can cause the FFA effects at some stage (epoch) of the source evolution.

We conclude that many radio spectral properties of our low-luminosity objects in the early stages of evolution are very similar to those of high-power GPS and CSS sources. 
However, most of them are more compact with  a relatively high-frequency turnover (Figure~\ref{odea_plot}).
Their position on the size vs. turnover frequency plane may imply a different, more steep or parallel path of a radio galaxy development, 
e.g. remain compact for a longer period, or even for most of its life.
In turn, in the case of quasars, a much faster and a more significant change of their radio spectral properties means that they appear on $ \nu_p - l $ plane only for a short period of time. This is in agreement with previous studies of GPS sources saying that over time, even most quasar-type GPS sources no longer meet the criteria for belonging to this class of objects \citep{Dallacasa, Orienti12, Torniainen,Sotnikova}. This behavior prevents many young GPS quasars from being recognized as such, and therefore they may remain hidden in the flat-spectrum quasar population.

\begin{figure}[t]
\centering
\includegraphics[scale=0.4]{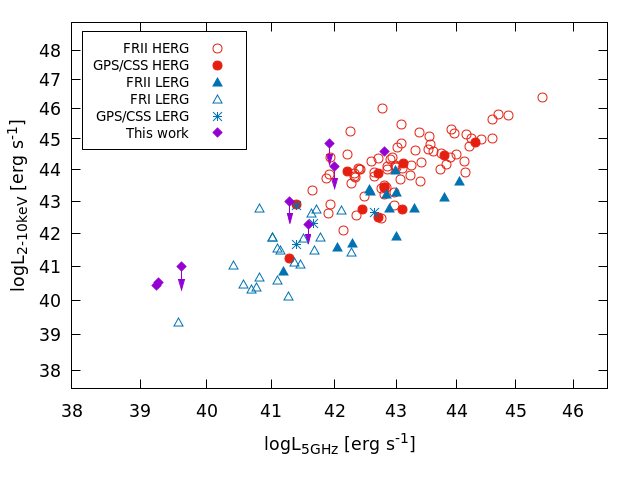}
\caption {X-ray 2-10 keV vs. radio 5GHz luminosity diagram for the sample of AGNs \citep{MKB14} classified as high-excitation radio galaxies (HERGs, red) and low excitation radio galaxies (LERGs, blue), combined with sources presented in this work (purple diamonds). FRII HERG and FRII LERG sources are indicated with open red circles and blue triangles, respectively. GPS/CSS HERG and GPS/CSS LERG sources are indicated with red circles and blue stars, respectively. FRI LERG sources are indicated with empty blue triangles. Luminosities for sources presented in this work estimated from X-ray upper limit (Table \ref{chandra}) are indicated with arrows.}
\label{herg_lerg}
\end{figure}

\subsection{The Origin of X-ray Emission}
\label{section_X}
We detected three of the nine sources observed in X-rays 
(Table \ref{chandra}, counting an undetected quasar 013815$+$00 \citep{MKB2020}. These are two nearby galaxies, 015411$-$01 and 030925$+$01, with a small number of counts and a much brighter quasar 233001$-$00 with the X-ray luminosity of $4\times 10^{44} \rm erg~s^{-1}$ in the 2-10\,keV energy range. These modest data prevent us from conducting a more detailed analysis of individual objects; however, they do allow us to trace certain emission relationships.

\citet{MKB14} compared the X-ray properties of GPS and CSS sources with large-scale FRI and FRIIs galaxies on the radio$-$X-ray luminosity plane. The results suggested that there is a continuity in the properties of the small and large-scale sources. 
Additionally, 
the separation into low- and high-excitation radio galaxies (HERGs and LERGs, respectively) suggests that the two different X-ray emission modes, viz. X-rays originating from the base of the relativistic jet (LERG) and X-rays being dominated by the emission from an accretion disk (HERG), are 
present among the younger compact AGNs.
The optical spectroscopic data of our sources are not sufficient for the sources' allocation into LERG or HERG types. However, the location of the sources on the radio$-$X-ray luminosity plane provides a preliminary classification, which, moreover, is arranged along the partition between strong and weak radio luminosity sources
(Figure \ref{herg_lerg}). Thus, the two galaxies detected in X-rays, 
015411$-$01 and 030925$+$01, can probably be classified as LERGs and a quasar 233001$-$00 can be classified as an HERG.
In addition, the low [OIII]$\lambda5007$ line luminosity of 030925$+$01 agrees with the interpretation of this object as an LERG.

We note that the location of X-ray emitting radio sources on the radio$-$X-ray plane may strongly depend on the epoch of the observation. There are many examples of objects, especially flat-spectrum radio quasars, in which the flaring event is detected simultaneously in the radio, optical-UV, X-ray, and even $\gamma$-ray range \citep{Berton}. According to the known model of such events, the new jet component is ejected during the high state, and in its initial phase, it carries enough energy to produce high-energy photons. It loses this ability when it becomes optically thin for synchrotron emission. In this phase, we observe the peak of the radio luminosity. It is reasonable to suspect that in the case of our transient sources, the burst of radio activity followed this pattern. However, since the observations of most of our objects took place several years after the ignition of the radio activity, we were unable to register the associated burst of the X-ray emission. 

The exception is the quasar 233001$-$00, which was observed by XMM-Newton \citep{LaMassa13} about 2 months before the radio detection, and two nearby galaxies, 015411$-$01 and 030925$+$01. We  
infer that in the case of the galaxies, the X-ray emission recorded in our Chandra observations, 4-5 yr after the new radio jet ejection, 
originates from the base of jet. However, in the case of the quasar 233001$-$00, it is directly related to the new ejection event. 
What is more, the X-ray power is higher than the radio power in our objects with a larger difference detected in the quasar 233001$-$00. Such a trend is also visible in other high-excitation objects (Figure \ref{herg_lerg}). This may indicate the presence of an additional X-ray source in HERG sources, which could be related to the accreting gas.

In general, the results of our X-ray observations indicate that the X-ray emission is rather weak in these transient radio sources.
The X-ray luminosity observed in the galaxies is much lower than $\rm 10^{42}~erg~s^{-1}$ which is consistent with the presence of a radiatively inefficient accretion onto a black hole with a mass within $\sim 10^6-10^8 M_{\odot}$ \citep{Paggi}. It is also possible that the total X-ray emission that we observe is a combination of the emission produced by the AGN and the stellar and hot interstellar medium (ISM) emission (see the discussion in next section).

\begin{figure}[t]
\centering
\includegraphics[scale=0.5]{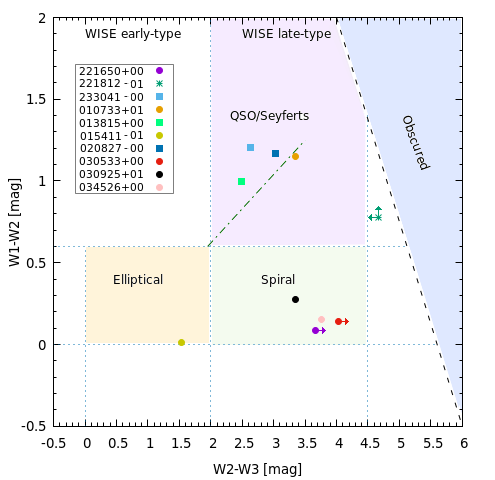}
\caption{WISE color-color plot for sources presented in this work with early-type and late-type galaxies distinction. 013815+00 quasar from \citet{MKB2020} is included. Each source is marked individually, and color filters are W1 - 3.35$\mu$m, W2 - 4.6$\mu$m and W3 - 12$\mu$m. Regions for particular galaxy populations are bounded by dashed lines and indicated with colors. The dashed dark green line indicates the WISE blazar sequence from \citet{Massaro}. The border limiting the region for obscured sources is determined by the linear relation $(W1-W2) + 1.25(W2-W3) > 7$ from \citet{Londsale}. Upper limits for particular bands are indicated with arrows pointing in adequate directions.}
\label{WISE}
\end{figure}

\begin{figure*}[t]
\centering
\includegraphics[scale=0.65]{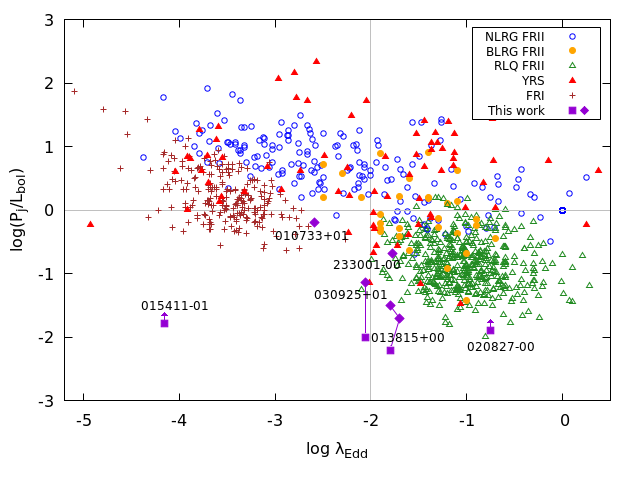}
\caption{Distribution of AGNs on a $\lambda_{Edd}$ vs $P_j/L_{bol}$ plane.
The NRLGs are indicated with blue circles, BLRGs) with yellow circles, and RLQs with green triangles 
\citep{Rusinek}. The FRI sample is shown by 
brown crosses \citep{Fricat}. 
The group of YRSs, which consists of CSS and GPS objects from samples of \citet{wojtowicz} and \citet{Liao}, is indicated with red triangles. 
More information on samples can be found in section \ref{physical_parameters}.
Sources presented in this work are marked with purple points, with activity stage indicated by squares for radio-quiet and diamonds for radio-loud phases. In the radio-quiet phase, the upper limits of radio emission at 1.4 GHz were used as a starting point for source changes presented in the plot.
The horizontal line corresponds to the level where $P_j$ equals $L_{bol}$ and the vertical line marks an approximate value of the Eddington ratio at which the accretion mode is changing from the radiatively inefficient (left side) to the radiatively efficient (right side; \citet{Best, Mingo}). 
}
\label{jet_power}
\end{figure*}

\subsection{The WISE Color-Color Diagram}
Near- and mid-infrared photometry is available for most of our sources (Table \ref{table_wise}) from the Wide-field Infrared Survey Explorer (WISE) mission \citep{Cutri}. Figure \ref{WISE} shows  a  WISE  color-color plot for them, and the dotted horizontal and vertical lines divide for different group of objects  \citep{Wright, Sadler}.  
We have also adopted the host galaxy naming convention introduced by \citet{Sadler} for elliptical and spiral galaxies as the 'WISE early-type' and 'WISE late-type' objects, respectively. 

As can be seen in the Figure \ref{WISE}, the vast majority of our sources are 'WISE late-type' objects. However, as pointed out by \citet{Sadler}, this class is not homogeneous but is a mixture of low- and high-excitation galaxies, which is probably also the case with our sample of objects. Four of our objects are classified as AGNs based on the infrared photometry \citep{Assef} in agreement with SDSS classification. 

Nevertheless, regardless of the HERG/LERG division, all objects with $\rm W2 - W3 > 2$ are thought to have gas- and dust-rich environments, although probably distributed in different ways, either forming a dusty torus (HERGs) or being a set of individual absorbing clouds (LERGs). This has been recently confirmed by the high, and similar, HI detection rate found in both
'WISE late-type' LERGs and HERGs with a compact radio structure \citep{Chandola}. Since LERGs and HERGs are thought to differ in accretion mode, the presence of large amounts of gas alone does not imply a high accretion rate. This difference is probably due to the feeding mechanism \citep{Chandola}.

\subsection{Jet Power and Accretion Process}
\label{jet_and_accretion}

We have analyzed the available optical 
spectra of our sources, and estimated the bolometric luminosities $\rm L_{bol}$ of their accretion disks, their black hole masses, $\rm M_{BH}$ and Eddington ratio $\rm \lambda_{Edd}=L_{bol}/L_{Edd}$ ( Table \ref{table_physical_parameters}). 
In addition, using the relation between the jet power
and radio luminosity at 1.4 GHz discussed
by \citet{Rusinek}, we estimated the
power of the newborn jets in these transient objects.
Whenever possible, estimates were made for both the radio-loud and radio-quiet phases of each source.
In the case of the radio-quiet phase, depending on the source, we either assumed that the 1.4 GHz marginal radio emission was of AGN origin or we adopted an upper limit of radio emission as a starting point for the discussion of source changes. And so we were able to determine the physical values for both phases for two sources: 030925$+$01 and 013815$+$00. For objects 015411$-$01 and 020827$-$00, only the radio-quiet phase is calculated, and for objects 010733$+$01 and 233001$-$00, we only have data for the radio-loud phase (Table \ref {table_physical_parameters}).
The most complete set of observations gathered so far relates to the quasar 013815$+$00 \citep{MKB2020} and
allow us to trace the changes in its radio and optical emission with greater accuracy.

In  Figure \ref{jet_power}  we  plot  the ratio $\rm P_j/L_{bol}$ against  the  Eddington  scaled  accretion  luminosity, or  Eddington ratio $\rm \lambda_{Edd}$ for the samples of different groups of radio-loud AGNs taken from the literature, and for our sources. 
The value of the Eddington ratio $\rm log\lambda_{Edd}=-2$ is thought to be the approximate boundary between objects operating in different accretion modes, i.e. radiatively inefficient ($\rm log\lambda_{Edd}<-2$) and radiatively efficient ($\rm log\lambda_{Edd}>-$2). In the radiatively inefficient regime, most of the gravitational energy is probably channeled into the jets, rather than radiative output \citep{Mingo}.
In the case of our sources, we can see large differentiation in the values of the Eddington ratio. However, the quasars clearly group to the right of the dividing line. The calculated values of the jet kinetic power also show a wide range in agreement with those previously obtained for other young radio sources \citep{FanWu, Liao, wojtowicz}.

It is also worth noting here that
the location of a given source on the plot probably depends on its phase of activity. As has been recently discussed by \citet{MKB2020} on the example of a quasar 013815$+$00, the bolometric luminosity is a measure of accretion rate, which can change on the much shorter timescales than the lifetime of the radio source. 
The initial phase of high radio activity in the sources coincides with an increase of its accretion disk luminosity, which we interpret as an enhancement of the accretion process leading to the launch of a radio jet. In the following next few years, the disk brightness returns to its pre-event level while the radio luminosity still remains high.
As a consequence we observe a movement of the sources along the distribution of objects on the $P_j/L_{bol} - \lambda_{Edd}$ plane as discussed by \citet{Rusinek} and shown in the example of 013815$+$00 (Figure \ref{jet_power}).

\section{Summary}
The transient sources presented
in this article constitute the first unbiased sample of newly born radio sources discovered by the Caltech-NRAO Stripe 82 Survey. The sample consists of three quasars, eight galaxies, and one object without optical identification. One of the sources, quasar 013815+00, has already been discussed in more detail in a separate paper \citep{MKB2020}. We have performed a comprehensive, multi-epoch, and multifrequency study of this sample, which we summarize as follows:

\begin{itemize}
\item 
%The radio activity ignition has been observed in these objects between 2007 and 2013, and for majority of them, their radio luminosity exceeds the limit $\rm log_{10}[L_{1.4GHz}/W~Hz^{-1}]\sim22.5$ above which the source is considered to be radio-loud.

Radio activity was discovered in these objects between 2007 and 2013, and 
for the majority of them,
their radio luminosity exceeds the $\rm log_{10}[L_{1.4GHz}/W~Hz^{-1}]\sim22.5$ limit
above which the source is considered to be radio-loud.
They might have transitioned from a radio-quiet to radio-loud state either as a result of the increase in radio power or its ignition.

\item
All of the sources, in their initial phase of activity, show convex radio spectra peaking at a few gigahertz, and compact morphology typical for young AGNs, and thus can be classified as GPS sources. The spectra change with time as a result of jet expansion, which has been confirmed for some sources in the VLBA images. This transforms GPS quasars, which are likely to be seen at smaller angles, into flat-spectrum objects, while galaxies keep the convex shape of the spectrum. Thus, we conclude that many of the young quasars can be hidden in the flat-spectrum quasar population.

\item The transient sources are less radio luminous than the GPS objects studied so far. However, their placement on the $\nu_p - l$ plane, in comparison with more powerful GPS and CSS objects, shows that their behavior is similar, with slight discrepancy from the established relation in the case of the weakest sources, suggesting their slower growth in size.

\item
Modeling of the radio spectra shows that these transient sources do not reach the SSA limit of the optically thick spectral index, although several objects are close to this limit.
We conclude that their radio spectra consist of more than one radiative component, and the turnover of the spectrum is due to the synchrotron self-absorption process.

\item 
The X-ray luminosities of the studied sources show a wide range of values $\rm 40<log_{10}[L_X/erg~s^{-1}]<45$ which indicates the diversity of young objects in our sample. The lower values are consistent  with an inefficient accretion mode. 

\item WISE infrared colors and optical observations imply that the majority of the host galaxies of our sources are spirals or other dusty, late-type galaxies with some ongoing star-formation.  

\item 
The transient sources show a wide range of bolometric luminosities, black hole masses, and jet powers, suggesting that young AGNs belong to several different subclasses of objects.
The ignition of radio activity coincides with relatively small changes of bolometric luminosity and hence Eddington ratio. The changes in the accretion disk happen on the much shorter timescales than the lifetime of the newborn radio source, in what is visible as a movement along the $\rm P_j/L_{bol} - \lambda_{Edd}$ plane. 
\end{itemize}

\acknowledgments
We are grateful to Anna W\'ojtowicz and \L ukasz Stawarz for the fruitful discussion.
The National Radio Astronomy Observatory is a facility of the National Science Foundation operated under cooperative agreement by Associated Universities, Inc. We thank the staff of the VLBA and VLA for carrying out these observations in their usual efficient manner.
This work made use of the Swinburne University of Technology software correlator, developed as part of the Australian Major National Research Facilities Programme and operated under licence.
We thank the staff of the GMRT who have made these observations possible. The GMRT is run by the National Centre for Radio Astrophysics of the Tata Institute of Fundamental Research. P.K. and I.C.H. acknowledge the support of the Department of Atomic Energy, Government of India, under the project 12-R\&D-TFR-5.02-0700.
Some of the observations reported in this paper were obtained with SALT under program 2019-1-SCI-023 (PI: A. Wo\l owska). Polish participation in SALT is funded by grant No. MNiSW DIR/WK/2016/07.
This project was supported in part by NASA grant GO8-19081X (Chandra) and by NASA contract NAS8-03060 to the Chandra X-ray Center (A.S.). 
M.K.B. and A.W. acknowledge support from the "National Science Centre, Poland" under grant No. 2017/26/E/ST9/00216.
K.P.M. is a Jansky Fellow of the National Radio Astronomy Observatory.
K.P.M. and G.H. acknowledge support from the National Science Foundation grant AST-1911199. M.G. is supported by the Polish NCN MAESTRO grant 2014/14/A/ST9/00121.

\software{CASA \citep{McMullin}, AIPS \citep{vanMoorsel}, IRAF \citep{Tody86,Tody93}, STARLIGHT \citep{Starlight}, CIAO \citep{Fruscione}, Sherpa \citep{Freeman}. }

\bibliographystyle{aasjournal}
\bibliography{ms}

\begin{thebibliography}{}
\expandafter\ifx\csname natexlab\endcsname\relax\def\natexlab#1{#1}\fi

\bibitem[{{An and Baan}(2012)}]{An}
{An and Baan}. 2012, \apj, 760, 77

\bibitem[{{Assef} {et~al.}(2018){Assef}, {Stern}, {Noirot}, {Jun}, {Cutri}, \&
  {Eisenhardt}}]{Assef}
{Assef}, R.~J., {Stern}, D., {Noirot}, G., {et~al.} 2018, VizieR Online Data
  Catalog, J/ApJS/234/23

\bibitem[{{Baldwin} {et~al.}(1981){Baldwin}, {Phillips}, \&
  {Terlevich}}]{Baldwin}
{Baldwin}, J.~A., {Phillips}, M.~M., \& {Terlevich}, R. 1981, \pasp, 93, 5

\bibitem[{{Berton} {et~al.}(2017){Berton}, {Foschini}, {Caccianiga}, {Ciroi},
  {Congiu}, {Cracco}, {Frezzato}, {La Mura}, \& {Rafanelli}}]{Berton17}
{Berton}, M., {Foschini}, L., {Caccianiga}, A., {et~al.} 2017, Frontiers in
  Astronomy and Space Sciences, 4, 8

\bibitem[{{Berton} {et~al.}(2018){Berton}, {Liao}, {La Mura},
  {J{\"a}rvel{\"a}}, {Congiu}, {Foschini}, {Frezzato}, {Ramakrishnan}, {Fan},
  {L{\"a}hteenm{\"a}ki}, {Pursimo}, {Abate}, {Bai}, {Calcidese}, {Ciroi},
  {Chen}, {Cracco}, {Li}, {Tornikoski}, \& {Rafanelli}}]{Berton}
{Berton}, M., {Liao}, N.~H., {La Mura}, G., {et~al.} 2018, \aap, 614, A148

\bibitem[{{Best} {et~al.}(2005){Best}, {Kauffmann}, {Heckman}, {Brinchmann},
  {Charlot}, {Ivezi{\'c}}, \& {White}}]{Best05}
{Best}, P.~N., {Kauffmann}, G., {Heckman}, T.~M., {et~al.} 2005, \mnras, 362,
  25

\bibitem[{{Best and Heckman}(2012)}]{Best}
{Best and Heckman}. 2012, \mnras, 421, 1569

\bibitem[{{Bicknell} {et~al.}(1997){Bicknell}, {Dopita}, \& {O'Dea}}]{Bicknell}
{Bicknell}, G.~V., {Dopita}, M.~A., \& {O'Dea}, C. P.~O. 1997, \apj, 485, 112

\bibitem[{{Bruhweiler} \& {Verner}(2008)}]{Bruhweiler}
{Bruhweiler}, F., \& {Verner}, E. 2008, \apj, 675, 83

\bibitem[{{Buckley} {et~al.}(2006){Buckley}, {Swart}, \& {Meiring}}]{Buckley06}
{Buckley}, D. A.~H., {Swart}, G.~P., \& {Meiring}, J.~G. 2006, in Society of
  Photo-Optical Instrumentation Engineers (SPIE) Conference Series, Vol. 6267,
  Society of Photo-Optical Instrumentation Engineers (SPIE) Conference Series,
  ed. L.~M. {Stepp}, 62670Z

\bibitem[{{Burgh} {et~al.}(2003){Burgh}, {Nordsieck}, {Kobulnicky}, {Williams},
  {O'Donoghue}, {Smith}, \& {Percival}}]{Burgh03}
{Burgh}, E.~B., {Nordsieck}, K.~H., {Kobulnicky}, H.~A., {et~al.} 2003, in
  Society of Photo-Optical Instrumentation Engineers (SPIE) Conference Series,
  Vol. 4841, Instrument Design and Performance for Optical/Infrared
  Ground-based Telescopes, ed. M.~{Iye} \& A.~F.~M. {Moorwood}, 1463--1471

\bibitem[{{Callingham} {et~al.}(2017){Callingham}, {Ekers}, {Gaensler}, {Line},
  {Hurley-Walker}, {Sadler}, {Tingay}, {Hancock}, {Bell}, {Dwarakanath}, {For},
  {Franzen}, {Hindson}, {Johnston-Hollitt}, {Kapi{\'n}ska}, {Lenc}, {McKinley},
  {Morgan}, {Offringa}, {Procopio}, {Staveley-Smith}, {Wayth}, {Wu}, \&
  {Zheng}}]{Callingham17}
{Callingham}, J.~R., {Ekers}, R.~D., {Gaensler}, B.~M., {et~al.} 2017, \apj,
  836, 174

\bibitem[{{Capetti} {et~al.}(2019){Capetti}, {Baldi}, {Brienza}, {Morganti}, \&
  {Giovannini}}]{Capetti}
{Capetti}, A., {Baldi}, R.~D., {Brienza}, M., {Morganti}, R., \& {Giovannini},
  G. 2019, \aap, 631, A176

\bibitem[{{Capetti} {et~al.}(2017){Capetti}, {Massaro}, \& {Baldi}}]{Fricat}
{Capetti}, A., {Massaro}, F., \& {Baldi}, R.~D. 2017, \aap, 598, A49

\bibitem[{{Carilli} {et~al.}(2003){Carilli}, {Ivison}, \& {Frail}}]{Carilli}
{Carilli}, C.~L., {Ivison}, R.~J., \& {Frail}, D.~A. 2003, \apj, 590, 192

\bibitem[{{Chandola} {et~al.}(2020){Chandola}, {Saikia}, \& {Li}}]{Chandola}
{Chandola}, Y., {Saikia}, D.~J., \& {Li}, D. 2020, \mnras, 494, 5161

\bibitem[{{Cid Fernandes} {et~al.}(2011){Cid Fernandes}, {Mateus}, {Sodr{\'e}},
  {Stasinska}, \& {Gomes}}]{Starlight}
{Cid Fernandes}, R., {Mateus}, A., {Sodr{\'e}}, L., {Stasinska}, G., \&
  {Gomes}, J.~M. 2011, {STARLIGHT: Spectral Synthesis Code}, , , ascl:1108.006

\bibitem[{{Collier} {et~al.}(2018){Collier}, {Tingay}, {Callingham}, {Norris},
  {Filipovi{\'c}}, {Galvin}, {Huynh}, {Intema}, {Marvil}, {O'Brien}, {Roper},
  {Sirothia}, {Tothill}, {Bell}, {For}, {Gaensler}, {Hancock}, {Hindson},
  {Hurley-Walker}, {Johnston-Hollitt}, {Kapi{\'n}ska}, {Lenc}, {Morgan},
  {Procopio}, {Staveley-Smith}, {Wayth}, {Wu}, {Zheng}, {Heywood}, \&
  {Popping}}]{Collier}
{Collier}, J.~D., {Tingay}, S.~J., {Callingham}, J.~R., {et~al.} 2018, \mnras,
  477, 578

\bibitem[{{Condon} {et~al.}(1998){Condon}, {Cotton}, {Greisen}, {Yin},
  {Perley}, {Taylor}, \& {Broderick}}]{condon1998}
{Condon}, J.~J., {Cotton}, W.~D., {Greisen}, E.~W., {et~al.} 1998, \aj, 115,
  1693

\bibitem[{{Crawford} {et~al.}(2010){Crawford}, {Still}, {Schellart}, {Balona},
  {Buckley}, {Dugmore}, {Gulbis}, {Kniazev}, {Kotze}, {Loaring}, {Nordsieck},
  {Pickering}, {Potter}, {Romero Colmenero}, {Vaisanen}, {Williams}, \&
  {Zietsman}}]{Crawford10}
{Crawford}, S.~M., {Still}, M., {Schellart}, P., {et~al.} 2010, in Society of
  Photo-Optical Instrumentation Engineers (SPIE) Conference Series, Vol. 7737,
  Observatory Operations: Strategies, Processes, and Systems III, ed. D.~R.
  {Silva}, A.~B. {Peck}, \& B.~T. {Soifer}, 773725

\bibitem[{{Cutri} \& {et al.}(2013)}]{Cutri}
{Cutri}, R.~M., \& {et al.} 2013, VizieR Online Data Catalog, II/328

\bibitem[{{Czerny} {et~al.}(2009){Czerny}, {Siemiginowska}, {Janiuk},
  {Nikiel-Wroczy{\'n}ski}, \& {Stawarz}}]{Czerny}
{Czerny}, B., {Siemiginowska}, A., {Janiuk}, A., {Nikiel-Wroczy{\'n}ski}, B.,
  \& {Stawarz}, {\L}. 2009, \apj, 698, 840

\bibitem[{{Dallacasa} {et~al.}(2000){Dallacasa}, {Stanghellini}, {Centonza}, \&
  {Fanti}}]{Dallacasa}
{Dallacasa}, D., {Stanghellini}, C., {Centonza}, M., \& {Fanti}, R. 2000, \aap,
  363, 887

\bibitem[{{de Kool} \& {Begelman}(1989)}]{deKool}
{de Kool}, M., \& {Begelman}, M.~C. 1989, \nat, 338, 484

\bibitem[{{de Vries} {et~al.}(1997){de Vries}, {Barthel}, \& {O'Dea}}]{devries}
{de Vries}, W.~H., {Barthel}, P.~D., \& {O'Dea}, C.~P. 1997, \aap, 321, 105

\bibitem[{{Dicken} {et~al.}(2012){Dicken}, {Tadhunter}, {Axon}, {Morganti},
  {Robinson}, {Kouwenhoven}, {Spoon}, {Kharb}, {Inskip}, {Holt}, {Ramos
  Almeida}, \& {Nesvadba}}]{Dicken}
{Dicken}, D., {Tadhunter}, C., {Axon}, D., {et~al.} 2012, \apj, 745, 172

\bibitem[{{Drake} {et~al.}(2009){Drake}, {Djorgovski}, {Mahabal}, {Beshore},
  {Larson}, {Graham}, {Williams}, {Christensen}, {Catelan}, {Boattini},
  {Gibbs}, {Hill}, \& {Kowalski}}]{Drake}
{Drake}, A.~J., {Djorgovski}, S.~G., {Mahabal}, A., {et~al.} 2009, \apj, 696,
  870

\bibitem[{{Fan} \& {Wu}(2019)}]{FanWu}
{Fan}, X.-L., \& {Wu}, Q. 2019, \apj, 879, 107

\bibitem[{{Fanaroff} \& {Riley}(1974)}]{Fanaroff}
{Fanaroff}, B.~L., \& {Riley}, J.~M. 1974, \mnras, 167, 31P

\bibitem[{{Fanti} {et~al.}(1990){Fanti}, {Fanti}, {Schilizzi}, {Spencer}, {Nan
  Rendong}, {Parma}, {van Breugel}, \& {Venturi}}]{fanti}
{Fanti}, R., {Fanti}, C., {Schilizzi}, R.~T., {et~al.} 1990, \aap, 231, 333

\bibitem[{{Freeman} {et~al.}(2001){Freeman}, {Doe}, \&
  {Siemiginowska}}]{Freeman}
{Freeman}, P., {Doe}, S., \& {Siemiginowska}, A. 2001, in Society of
  Photo-Optical Instrumentation Engineers (SPIE) Conference Series, Vol. 4477,
  Astronomical Data Analysis, ed. J.-L. {Starck} \& F.~D. {Murtagh}, 76--87

\bibitem[{{Fruscione} {et~al.}(2006){Fruscione}, {McDowell}, {Allen},
  {Brickhouse}, {Burke}, {Davis}, {Durham}, {Elvis}, {Galle}, {Harris},
  {Huenemoerder}, {Houck}, {Ishibashi}, {Karovska}, {Nicastro}, {Noble},
  {Nowak}, {Primini}, {Siemiginowska}, {Smith}, \& {Wise}}]{Fruscione}
{Fruscione}, A., {McDowell}, J.~C., {Allen}, G.~E., {et~al.} 2006, in Society
  of Photo-Optical Instrumentation Engineers (SPIE) Conference Series, Vol.
  6270, \procspie, 62701V

\bibitem[{{Hardcastle} {et~al.}(2019){Hardcastle}, {Williams}, {Best},
  {Croston}, {Duncan}, {R{\"o}ttgering}, {Sabater}, {Shimwell}, {Tasse},
  {Callingham}, {Cochrane}, {de Gasperin}, {G{\"u}rkan}, {Jarvis}, {Mahatma},
  {Miley}, {Mingo}, {Mooney}, {Morabito}, {O'Sullivan}, {Prandoni},
  {Shulevski}, \& {Smith}}]{Hardcastle}
{Hardcastle}, M.~J., {Williams}, W.~L., {Best}, P.~N., {et~al.} 2019, \aap,
  622, A12

\bibitem[{{Hodge} {et~al.}(2011){Hodge}, {Becker}, {White}, {Richards}, \&
  {Zeimann}}]{Hodge}
{Hodge}, J.~A., {Becker}, R.~H., {White}, R.~L., {Richards}, G.~T., \&
  {Zeimann}, G.~R. 2011, \aj, 142, 3

\bibitem[{{Ishwara-Chandra} {et~al.}(2020){Ishwara-Chandra}, {Taylor}, {Green},
  {Stil}, {Vaccari}, \& {Ocran}}]{Ishwara2020}
{Ishwara-Chandra}, C.~H., {Taylor}, A.~R., {Green}, D.~A., {et~al.} 2020,
  \mnras, 497, 5383

\bibitem[{{Jackson} \& {Wall}(1999)}]{Jackson}
{Jackson}, C.~A., \& {Wall}, J.~V. 1999, \mnras, 304, 160

\bibitem[{{Jeyakumar}(2016)}]{Jeyakumar}
{Jeyakumar}, S. 2016, \mnras, 458, 3786

\bibitem[{{Kameno} {et~al.}(2003){Kameno}, {Inoue}, {Wajima}, {Sawada-Satoh},
  \& {Shen}}]{Kameno}
{Kameno}, S., {Inoue}, M., {Wajima}, K., {Sawada-Satoh}, S., \& {Shen}, Z.-Q.
  2003, \pasa, 20, 213

\bibitem[{{Kauffmann} {et~al.}(2003){Kauffmann}, {Heckman}, {Tremonti},
  {Brinchmann}, {Charlot}, {White}, {Ridgway}, {Brinkmann}, {Fukugita}, {Hall},
  {Ivezi{\'c}}, {Richards}, \& {Schneider}}]{Kauffmann}
{Kauffmann}, G., {Heckman}, T.~M., {Tremonti}, C., {et~al.} 2003, \mnras, 346,
  1055

\bibitem[{{Keim} {et~al.}(2019){Keim}, {Callingham}, \&
  {R{\"o}ttgering}}]{Keim}
{Keim}, M.~A., {Callingham}, J.~R., \& {R{\"o}ttgering}, H.~J.~A. 2019, \aap,
  628, A56

\bibitem[{{Kellermann} {et~al.}(2016){Kellermann}, {Condon}, {Kimball},
  {Perley}, \& {Ivezi{\'c}}}]{Kellerman}
{Kellermann}, K.~I., {Condon}, J.~J., {Kimball}, A.~E., {Perley}, R.~A., \&
  {Ivezi{\'c}}, {\v{Z}}. 2016, \apj, 831, 168

\bibitem[{{Kewley} {et~al.}(2001){Kewley}, {Heisler}, {Dopita}, \&
  {Lumsden}}]{Kewley01}
{Kewley}, L.~J., {Heisler}, C.~A., {Dopita}, M.~A., \& {Lumsden}, S. 2001,
  \apjs, 132, 37

\bibitem[{{Kormendy} \& {Ho}(2013)}]{Kormendy}
{Kormendy}, J., \& {Ho}, L.~C. 2013, \araa, 51, 511

\bibitem[{{Kunert-Bajraszewska} {et~al.}(2010){Kunert-Bajraszewska},
  {Gawro{\'n}ski}, {Labiano}, \& {Siemiginowska}}]{MKB10}
{Kunert-Bajraszewska}, M., {Gawro{\'n}ski}, M.~P., {Labiano}, A., \&
  {Siemiginowska}, A. 2010, \mnras, 408, 2261

\bibitem[{{Kunert-Bajraszewska} {et~al.}(2014){Kunert-Bajraszewska}, {Labiano},
  {Siemiginowska}, \& {Guainazzi}}]{MKB14}
{Kunert-Bajraszewska}, M., {Labiano}, A., {Siemiginowska}, A., \& {Guainazzi},
  M. 2014, \mnras, 437, 3063

\bibitem[{{Kunert-Bajraszewska} {et~al.}(2020){Kunert-Bajraszewska},
  {Wo{\l}owska}, {Mooley}, {Kharb}, \& {Hallinan}}]{MKB2020}
{Kunert-Bajraszewska}, M., {Wo{\l}owska}, A., {Mooley}, K., {Kharb}, P., \&
  {Hallinan}, G. 2020, \apj, 897, 128

\bibitem[{{Lacy} {et~al.}(2020){Lacy}, {Gates}, {Brandt}, {Clarke}, {Gaensler},
  {Kimball}, {Law}, {Lazio}, {O'Dea}, {Schinzel}, {Vaccari}, {White}, {Yoon},
  {Zhu}, \& {Vlass Team}}]{Lacy}
{Lacy}, M., {Gates}, E., {Brandt}, W., {et~al.} 2020, in American Astronomical
  Society Meeting Abstracts, American Astronomical Society Meeting Abstracts,
  306.16

\bibitem[{{L{\"a}hteenm{\"a}ki} {et~al.}(2018){L{\"a}hteenm{\"a}ki},
  {J{\"a}rvel{\"a}}, {Ramakrishnan}, {Tornikoski}, {Tammi}, {Vera}, \&
  {Chamani}}]{Lahteenmaki}
{L{\"a}hteenm{\"a}ki}, A., {J{\"a}rvel{\"a}}, E., {Ramakrishnan}, V., {et~al.}
  2018, \aap, 614, L1

\bibitem[{{LaMassa} {et~al.}(2013){LaMassa}, {Urry}, {Cappelluti}, {Civano},
  {Ranalli}, {Glikman}, {Treister}, {Richards}, {Ballantyne}, {Stern},
  {Comastri}, {Cardamone}, {Schawinski}, {B{\"o}hringer}, {Chon}, {Murray},
  {Green}, \& {Nandra}}]{LaMassa13}
{LaMassa}, S.~M., {Urry}, C.~M., {Cappelluti}, N., {et~al.} 2013, \mnras, 436,
  3581

\bibitem[{{Liao} \& {Gu}(2020)}]{Liao}
{Liao}, M., \& {Gu}, M. 2020, \mnras, 491, 92

\bibitem[{{Longair} {et~al.}(1973){Longair}, {Ryle}, \& {Scheuer}}]{Longair}
{Longair}, M.~S., {Ryle}, M., \& {Scheuer}, P.~A.~G. 1973, \mnras, 164, 243

\bibitem[{{Lonsdale} {et~al.}(2015){Lonsdale}, {Lacy}, {Kimball}, {Blain},
  {Whittle}, {Wilkes}, {Stern}, {Condon}, {Kim}, {Assef}, {Tsai}, {Efstathiou},
  {Jones}, {Eisenhardt}, {Bridge}, {Wu}, {Lonsdale}, {Jones}, {Jarrett}, \&
  {Smith}}]{Londsale}
{Lonsdale}, C.~J., {Lacy}, M., {Kimball}, A.~E., {et~al.} 2015, \apj, 813, 45

\bibitem[{{Malefahlo} {et~al.}(2020){Malefahlo}, {Santos}, {Jarvis}, {White},
  \& {Zwart}}]{Malefahlo}
{Malefahlo}, E., {Santos}, M.~G., {Jarvis}, M.~J., {White}, S.~V., \& {Zwart},
  J. T.~L. 2020, \mnras, 492, 5297

\bibitem[{{Massaro} {et~al.}(2012){Massaro}, {D'Abrusco}, {Tosti}, {Ajello},
  {Gasparrini}, {Grindlay}, \& {Smith}}]{Massaro}
{Massaro}, F., {D'Abrusco}, R., {Tosti}, G., {et~al.} 2012, \apj, 750, 138

\bibitem[{{Mauch} \& {Sadler}(2007)}]{Mauch}
{Mauch}, T., \& {Sadler}, E.~M. 2007, \mnras, 375, 931

\bibitem[{{McMullin} {et~al.}(2007){McMullin}, {Waters}, {Schiebel}, {Young},
  \& {Golap}}]{McMullin}
{McMullin}, J.~P., {Waters}, B., {Schiebel}, D., {Young}, W., \& {Golap}, K.
  2007, in Astronomical Society of the Pacific Conference Series, Vol. 376,
  Astronomical Data Analysis Software and Systems XVI, ed. R.~A. {Shaw},
  F.~{Hill}, \& D.~J. {Bell}, 127

\bibitem[{{Mingo} {et~al.}(2014){Mingo}, {Hardcastle}, {Croston}, {Dicken},
  {Evans}, {Morganti}, \& {Tadhunter}}]{Mingo}
{Mingo}, B., {Hardcastle}, M.~J., {Croston}, J.~H., {et~al.} 2014, \mnras, 440,
  269

\bibitem[{{Mooley} {et~al.}(2013){Mooley}, {Frail}, {Ofek}, {Miller},
  {Kulkarni}, \& {Horesh}}]{Mooley13}
{Mooley}, K.~P., {Frail}, D.~A., {Ofek}, E.~O., {et~al.} 2013, \apj, 768, 165

\bibitem[{{Mooley} {et~al.}(2019){Mooley}, {Myers}, {Frail}, {Hallinan},
  {Butler}, {Kimball}, \& {Golap}}]{Mooley2019}
{Mooley}, K.~P., {Myers}, S.~T., {Frail}, D.~A., {et~al.} 2019, \apj, 870, 25

\bibitem[{{Mooley} {et~al.}(2016){Mooley}, {Hallinan}, {Bourke}, {Horesh},
  {Myers}, {Frail}, {Kulkarni}, {Levitan}, {Kasliwal}, {Cenko}, {Cao}, {Bellm},
  \& {Laher}}]{Mooley}
{Mooley}, K.~P., {Hallinan}, G., {Bourke}, S., {et~al.} 2016, \apj, 818, 105

\bibitem[{{Morganti} {et~al.}(2011){Morganti}, {Holt}, {Tadhunter}, {Ramos
  Almeida}, {Dicken}, {Inskip}, {Oosterloo}, \& {Tzioumis}}]{Morganti}
{Morganti}, R., {Holt}, J., {Tadhunter}, C., {et~al.} 2011, \aap, 535, A97

\bibitem[{{Mukherjee} {et~al.}(2016){Mukherjee}, {Bicknell}, {Sutherland }, \&
  {Wagner}}]{Mukherjee}
{Mukherjee}, D., {Bicknell}, G.~V., {Sutherland }, R., \& {Wagner}, A. 2016,
  \mnras, 461, 967

\bibitem[{{Netzer}(2019)}]{Netzer}
{Netzer}, H. 2019, \mnras, 488, 5185

\bibitem[{{Nyland} {et~al.}(2020){Nyland}, {Dong}, {Patil}, {Lacy}, {van
  Velzen}, {Kimball}, {Sarbadhicary}, {Hallinan}, {Baldassare}, {Clarke},
  {Goulding}, {Greene}, {Hughes}, {Kassim}, {Kunert-Bajraszewska}, {Maccarone},
  {Mooley}, {Mukherjee}, {Peters}, {Petrov}, {Polisensky}, {Rujopakarn},
  {Whittle}, \& {Vaccari}}]{Nyland}
{Nyland}, K., {Dong}, D.~Z., {Patil}, P., {et~al.} 2020, \apj, 905, 74

\bibitem[{{O'Dea}(1998)}]{Odea2}
{O'Dea}, C.~P. 1998, \pasp, 110, 493

\bibitem[{{O'Dea} \& {Baum}(1997)}]{Odea}
{O'Dea}, C.~P., \& {Baum}, S.~A. 1997, \aj, 113, 148

\bibitem[{{O'Dea} {et~al.}(1991){O'Dea}, {Baum}, \& {Stanghellini}}]{Odea91}
{O'Dea}, C.~P., {Baum}, S.~A., \& {Stanghellini}, C. 1991, \apj, 380, 66

\bibitem[{{O'Dea} \& {Saikia}(2020)}]{Odea20}
{O'Dea}, C.~P., \& {Saikia}, D.~J. 2020, arXiv e-prints, arXiv:2009.02750

\bibitem[{{Orienti} \& {Dallacasa}(2012)}]{Orienti12}
{Orienti}, M., \& {Dallacasa}, D. 2012, \mnras, 424, 532

\bibitem[{{Orienti} \& {Dallacasa}(2014)}]{Orienti}
---. 2014, \mnras, 438, 463

\bibitem[{{Paggi} {et~al.}(2016){Paggi}, {Fabbiano}, {Civano}, {Pellegrini},
  {Elvis}, \& {Kim}}]{Paggi}
{Paggi}, A., {Fabbiano}, G., {Civano}, F., {et~al.} 2016, \apj, 823, 112

\bibitem[{{Park} {et~al.}(2006){Park}, {Kashyap}, {Siemiginowska}, {van Dyk},
  {Zezas}, {Heinke}, \& {Wargelin}}]{park2006}
{Park}, T., {Kashyap}, V.~L., {Siemiginowska}, A., {et~al.} 2006, \apj, 652,
  610

\bibitem[{{Readhead} {et~al.}(1996){Readhead}, {Taylor}, {Pearson}, \&
  {Wilkinson}}]{Redhead}
{Readhead}, A.~C.~S., {Taylor}, G.~B., {Pearson}, T.~J., \& {Wilkinson}, P.~N.
  1996, \apj, 460, 634

\bibitem[{{Reynolds} \& {Begelman}(1997)}]{Reynolds}
{Reynolds}, C.~S., \& {Begelman}, M.~C. 1997, \apj, 487, L135

\bibitem[{{Runnoe} {et~al.}(2012){Runnoe}, {Brotherton}, \& {Shang}}]{Runnoe}
{Runnoe}, J.~C., {Brotherton}, M.~S., \& {Shang}, Z. 2012, \mnras, 422, 478

\bibitem[{{Rusinek} {et~al.}(2017){Rusinek}, {Sikora}, {Kozie{\l}-Wierzbowska},
  \& {Godfrey}}]{Rusinek}
{Rusinek}, K., {Sikora}, M., {Kozie{\l}-Wierzbowska}, D., \& {Godfrey}, L.
  2017, \mnras, 466, 2294

\bibitem[{{Sadler} {et~al.}(2014){Sadler}, {Ekers}, {Mahony}, {Mauch}, \&
  {Murphy}}]{Sadler}
{Sadler}, E.~M., {Ekers}, R.~D., {Mahony}, E.~K., {Mauch}, T., \& {Murphy}, T.
  2014, \mnras, 438, 796

\bibitem[{{Saikia} \& {Jamrozy}(2009)}]{Saikia}
{Saikia}, D.~J., \& {Jamrozy}, M. 2009, Bulletin of the Astronomical Society of
  India, 37, 63

\bibitem[{{Schawinski} {et~al.}(2007){Schawinski}, {Thomas}, {Sarzi},
  {Maraston}, {Kaviraj}, {Joo}, {Yi}, \& {Silk}}]{Schawinski}
{Schawinski}, K., {Thomas}, D., {Sarzi}, M., {et~al.} 2007, \mnras, 382, 1415

\bibitem[{{Schlafly} \& {Finkbeiner}(2011)}]{Schlafly}
{Schlafly}, E.~F., \& {Finkbeiner}, D.~P. 2011, \apj, 737, 103

\bibitem[{{Silpa} {et~al.}(2020){Silpa}, {Kharb}, {Ho}, {Ishwara-Chandra},
  {Jarvis}, \& {Harrison}}]{Silpa}
{Silpa}, S., {Kharb}, P., {Ho}, L.~C., {et~al.} 2020, \mnras, 499, 5826

\bibitem[{{Snellen} {et~al.}(1998){Snellen}, {Schilizzi}, {de Bruyn}, {Miley},
  {Rengelink}, {Roettgering}, \& {Bremer}}]{Snellen}
{Snellen}, I.~A.~G., {Schilizzi}, R.~T., {de Bruyn}, A.~G., {et~al.} 1998,
  \aaps, 131, 435

\bibitem[{{Snellen} {et~al.}(2000){Snellen}, {Schilizzi}, {Miley}, {de Bruyn},
  {Bremer}, \& {R{\"o}ttgering}}]{Snellen2000}
{Snellen}, I.~A.~G., {Schilizzi}, R.~T., {Miley}, G.~K., {et~al.} 2000, \mnras,
  319, 445

\bibitem[{{Sotnikova} {et~al.}(2019){Sotnikova}, {Mufakharov}, {Majorova},
  {Mingaliev}, {Udovitskii}, {Bursov}, \& {Semenova}}]{Sotnikova}
{Sotnikova}, Y.~V., {Mufakharov}, T.~V., {Majorova}, E.~K., {et~al.} 2019,
  Astrophysical Bulletin, 74, 348

\bibitem[{{Stanghellini} {et~al.}(1998){Stanghellini}, {O'Dea}, {Dallacasa},
  {Baum}, {Fanti}, \& {Fanti}}]{Stranghellini}
{Stanghellini}, C., {O'Dea}, C.~P., {Dallacasa}, D., {et~al.} 1998, \aaps, 131,
  303

\bibitem[{{Tadhunter} {et~al.}(2011){Tadhunter}, {Holt}, {Gonz{\'a}lez
  Delgado}, {Rodr{\'\i}guez Zaur{\'\i}n}, {Villar-Mart{\'\i}n}, {Morganti},
  {Emonts}, {Ramos Almeida}, \& {Inskip}}]{Tadhunter}
{Tadhunter}, C., {Holt}, J., {Gonz{\'a}lez Delgado}, R., {et~al.} 2011, \mnras,
  412, 960

\bibitem[{{Tody}(1986)}]{Tody86}
{Tody}, D. 1986, in Society of Photo-Optical Instrumentation Engineers (SPIE)
  Conference Series, Vol. 627, Instrumentation in astronomy VI, ed. D.~L.
  {Crawford}, 733

\bibitem[{{Tody}(1993)}]{Tody93}
{Tody}, D. 1993, in Astronomical Society of the Pacific Conference Series,
  Vol.~52, Astronomical Data Analysis Software and Systems II, ed. R.~J.
  {Hanisch}, R.~J.~V. {Brissenden}, \& J.~{Barnes}, 173

\bibitem[{{Torniainen} {et~al.}(2005){Torniainen}, {Tornikoski},
  {Ter{\"a}sranta}, {Aller}, \& {Aller}}]{Torniainen}
{Torniainen}, I., {Tornikoski}, M., {Ter{\"a}sranta}, H., {Aller}, M.~F., \&
  {Aller}, H.~D. 2005, \aap, 435, 839

\bibitem[{{Trakhtenbrot} \& {Netzer}(2012)}]{Trakhtenbrot}
{Trakhtenbrot}, B., \& {Netzer}, H. 2012, \mnras, 427, 3081

\bibitem[{{van Moorsel} {et~al.}(1996){van Moorsel}, {Kemball}, \&
  {Greisen}}]{vanMoorsel}
{van Moorsel}, G., {Kemball}, A., \& {Greisen}, E. 1996, in Astronomical
  Society of the Pacific Conference Series, Vol. 101, Astronomical Data
  Analysis Software and Systems V, ed. G.~H. {Jacoby} \& J.~{Barnes}, 37

\bibitem[{{Wagner} \& {Bicknell}(2011)}]{Wagner}
{Wagner}, A.~Y., \& {Bicknell}, G.~V. 2011, \apj, 728, 29

\bibitem[{{White} {et~al.}(1997){White}, {Becker}, {Helfand}, \&
  {Gregg}}]{White}
{White}, R.~L., {Becker}, R.~H., {Helfand}, D.~J., \& {Gregg}, M.~D. 1997,
  VizieR Online Data Catalog, VIII/48

\bibitem[{{Willott} {et~al.}(1999){Willott}, {Rawlings}, {Blundell}, \&
  {Lacy}}]{Willott}
{Willott}, C.~J., {Rawlings}, S., {Blundell}, K.~M., \& {Lacy}, M. 1999,
  \mnras, 309, 1017

\bibitem[{{W{\'o}jtowicz} {et~al.}(2020){W{\'o}jtowicz}, {Stawarz}, {Cheung},
  {Ostorero}, {Kosmaczewski}, \& {Siemiginowska}}]{wojtowicz}
{W{\'o}jtowicz}, A., {Stawarz}, {\l}., {Cheung}, C.~C., {et~al.} 2020, \apj,
  892, 116

\bibitem[{{Wo{\l}owska} {et~al.}(2017){Wo{\l}owska}, {Kunert-Bajraszewska},
  {Mooley}, \& {Hallinan}}]{me}
{Wo{\l}owska}, A., {Kunert-Bajraszewska}, M., {Mooley}, K., \& {Hallinan}, G.
  2017, Frontiers in Astronomy and Space Sciences, 4, 38

\bibitem[{{Wright} {et~al.}(2010){Wright}, {Eisenhardt}, {Mainzer}, {Ressler},
  {Cutri}, {Jarrett}, {Kirkpatrick}, {Padgett}, {McMillan}, {Skrutskie},
  {Stanford}, {Cohen}, {Walker}, {Mather}, {Leisawitz}, {Gautier}, {McLean},
  {Benford}, {Lonsdale}, {Blain}, {Mendez}, {Irace}, {Duval}, {Liu}, {Royer},
  {Heinrichsen}, {Howard}, {Shannon}, {Kendall}, {Walsh}, {Larsen}, {Cardon},
  {Schick}, {Schwalm}, {Abid}, {Fabinsky}, {Naes}, \& {Tsai}}]{Wright}
{Wright}, E.~L., {Eisenhardt}, P. R.~M., {Mainzer}, A.~K., {et~al.} 2010, \aj,
  140, 1868

\end{thebibliography}
\newpage
%\begin{appendices}
%\restartappendixnumbering
\appendix
\numberwithin{table}{section}
\section{Tables}
Here we report the multiband VLA (Table \ref{VLA_measurement_points}), GMRT (Table \ref{gmrt_points}) and WISE (Table \ref{table_wise}) measurements, the emission line measurements (Table \ref{table_emission_lines}), the results of X-ray study (Table \ref{chandra}) and astrophysical properties (Table \ref{table_physical_parameters}) of our sources with
detailed calculations described in Section \ref{physical_parameters}.
\begin{deluxetable}{ c c c c c c c c c c c c c }[hb!]
\tablecolumns{13}
\tablewidth{0pt}
\tablecaption{VLA measurement points}
\label{VLA_measurement_points}
\tablehead{
& & & & & & {\bf2019} & & & & & &  \\
\hline
&$\nu_{S1}$&$S_{S1}$&$\nu_{S2}$&$S_{S2}$&$\nu_{S3}$&$S_{S3}$&$\nu_{C1}$&$S_{C1}$&$\nu_{C2}$&$S_{C2}$&$\nu_{C3}$&$S_{C3}$\\
Name&(GHz)&(mJy)&(GHz)&(mJy)&(GHz)&(mJy)&(GHz)&(mJy)&(GHz)&(mJy)&(GHz)&(mJy)\\
  &(1)&(2)&(3)&(4)&(5)&(6)&(7)&(8)&(9)&(10)&(11)&(12)}
  \startdata
221650$+$00&2.37&2.63$\pm$0.02&3.06&3.14$\pm$0.04&3.69&3.22$\pm$0.07&4.67&3.25$\pm$0.07&6.16&2.96$\pm$0.10&7.59&2.76$\pm$0.12\\
221812$-$01&2.37&7.37$\pm$0.07&3.06&9.89$\pm$0.16&3.69&11.41$\pm$0.37&4.67&14.20$\pm$0.36&6.16&13.97$\pm$0.57&7.59&13.00$\pm$0.73\\
223041$-$00&2.37&4.42$\pm$0.31&3.06&5.06$\pm$0.09&3.69&5.38$\pm$0.16&4.67&6.11$\pm$0.09&6.16&5.75$\pm$0.16&7.59&5.42$\pm$0.25\\
233001$-$00&2.37&5.82$\pm$0.12&3.06&6.29$\pm$0.10&3.56&6.52$\pm$0.13&4.67&6.17$\pm$0.07&6.16&6.2$\pm$0.05&7.59&6.49$\pm$0.05\\
010733$+$01&2.37&2.89$\pm$0.05&3.06&2.58$\pm$0.05&3.56&2.32$\pm$0.10&4.67&2.06$\pm$0.06&6.16&1.93$\pm$0.05&7.59&1.62$\pm$0.12\\
013815$+$00&2.37&2.62$\pm$0.06&3.06&2.95$\pm$0.07&3.56&3.24$\pm$0.09&4.67&3.62$\pm$0.02&6.16&3.78$\pm$0.10&7.59&3.83$\pm$0.08\\
015411$-$01&2.37&6.07$\pm$0.13&3.06&6.25$\pm$0.21&3.56&6.36$\pm$0.25&4.67&6.30$\pm$0.19&6.16&6.07$\pm$0.25&7.59&5.87$\pm$0.41\\
030533$+$00&2.37&2.68$\pm$0.04&3.06&3.66$\pm$0.10&3.69&4.35$\pm$0.22&4.67&4.81$\pm$0.10&6.16&4.92$\pm$0.18&7.59&4.82$\pm$0.17\\
031833$+$00&2.37&3.80$\pm$0.03&3.06&5.12$\pm$0.07&3.69&5.73$\pm$0.08&4.67&5.18$\pm$0.14&6.16&3.86$\pm$0.17&7.59&3.00$\pm$0.20\\
034526$+$00&2.37&2.97$\pm$0.05&3.06&5.05$\pm$0.13&3.69&6.22$\pm$0.39&4.67&9.14$\pm$0.18&6.16&11.35$\pm$0.45&7.59&11.99$\pm$0.70\\
\hline
&$\nu_{X1}$&$S_{X1}$&$\nu_{X2}$&$S_{X2}$&$\nu_{X3}$&$S_{X3}$&$\nu_{Ku1}$&$S_{Ku1}$&$\nu_{Ku2}$&$S_{Ku2}$&$\nu_{Ku3}$&$S_{Ku3}$\\
\hline
 221650$+$00&8.37&2.59$\pm$0.04&9.76&2.42$\pm$0.04&11.09&2.26$\pm$0.07&13.37&1.96$\pm$0.12&15.06&1.72$\pm$0.12&16.69&1.64$\pm$0.14\\
 221812$-$01&8.37&11.96$\pm$0.68&9.76&10.10$\pm$0.84&11.09&8.63$\pm$1.10&13.37&7.21$\pm$0.27&15.06&6.14$\pm$0.31&16.69&4.93$\pm$0.44\\
 223041$-$00&8.37&5.19$\pm$0.20&9.76&4.73$\pm$0.22&11.09&4.46$\pm$0.32&13.37&3.95$\pm$0.11&15.06&3.52$\pm$0.18&16.69&3.26$\pm$0.17\\
 233001$-$00&8.37&6.69$\pm$0.03&9.76&6.75$\pm$0.05&11.09&6.65$\pm$0.05&13.37&6.64$\pm$0.03&15.06&6.47$\pm$0.04&16.69&6.37$\pm$0.05\\
 010733$+$01&8.37&1.59$\pm$0.02&9.76&1.55$\pm$0.05&11.09&1.32$\pm$0.08&13.37&1.26$\pm$0.04&15.06&1.22$\pm$0.05&16.69&1.13$\pm$0.06\\
 013815$+$00&8.44&3.88$\pm$0.05&9.76&3.95$\pm$0.06&11.03&3.83$\pm$0.04&13.31&3.75$\pm$0.12&14.94&3.62$\pm$0.16&16.63&3.50$\pm$0.17\\
 015411$-$01&8.49&5.76$\pm$0.09&10.90&5.32$\pm$0.23&-&-&13.50&4.65$\pm$0.25&15.00&4.47$\pm$0.28&16.50&4.24$\pm$0.28\\
 020827$-$00&8.49&7.05$\pm$0.07&10.90&6.62$\pm$0.18&-&-&13.37&6.3$\pm$0.28&15.06 &6.12$\pm$0.33&16.69&5.95$\pm$0.36\\
 030533$+$00&8.37&4.62$\pm$0.08&9.76&4.26$\pm$0.11&11.09&3.97$\pm$0.05&13.37&3.37$\pm$0.10&15.06&3.01$\pm$0.11&16.69&2.76$\pm$0.10\\
 031833$+$00&8.37&2.53$\pm$0.07&9.76&1.91$\pm$0.11&11.09&1.48$\pm$0.05&13.37&0.92$\pm$0.09&15.06&0.75$\pm$0.10&16.69&0.62$\pm$0.09\\
 034526$+$00&8.37&11.94$\pm$0.27&9.76&11.11$\pm$0.42&11.09&10.41$\pm$0.74&13.37&9.11$\pm$0.13&15.06&8.08$\pm$0.13&16.69&7.37$\pm$0.22\\
\hline
& & & & & & {\bf2016} & & & & & & \\
\hline
&$\nu_{L1}$&$S_{L1}$&$\nu_{L2}$&$S_{L2}$&$\nu_{L3}$&$S_{L3}$&$\nu_{S1}$&$S_{S1}$&$\nu_{S2}$&$S_{S2}$&$\nu_{S3}$&$S_{S3}$\\
\hline
221650$+$00&1.26&0.55$\pm$0.03&1.78&1.20$\pm$0.04&-&-&2.24&2.29$\pm$0.05&2.76&2.46$\pm$0.09&3.24&2.44$\pm$0.03\\
221812$-$01&1.52&1.68$\pm$0.17&1.71&3.07$\pm$0.12&-&-&2.24&7.00$\pm$0.23&2.76&8.79$\pm$0.15&3.24&10.83$\pm$0.19\\
223041$-$00&1.26&0.73$\pm$0.03&1.78&1.40$\pm$0.09&-&-&2.24&2.55$\pm$0.20&2.76&3.10$\pm$0.12&3.24&4.49$\pm$0.10\\
233001$-$00&1.26&2.46$\pm$0.13&1.78&4.87$\pm$0.10&-&-&2.24&6.62$\pm$0.18&2.76&6.93$\pm$0.21&3.24&7.00$\pm$0.14\\
010733$+$01&1.32&2.92$\pm$0.53&1.52&3.02$\pm$0.51&-&-&2.24&4.10$\pm$0.40&2.76&3.73$\pm$0.17&3.24&3.49$\pm$0.12\\
013815$+$00&1.52&1.72$\pm$0.11&-&-&-&-&2.24&3.29$\pm$0.16&-&-&3.24&4.36$\pm$0.17\\
015411$-$01&1.52&4.00$\pm$0.17&-&-&-&-&2.24&4.40$\pm$0.17&2.76&4.92$\pm$0.13&3.24&5.38$\pm$0.07\\
020827$-$00&1.32&1.75$\pm$0.31&-&-&-&-&2.44&4.11$\pm$0.15&-&-&3.24&5.02$\pm$0.46\\
030533$+$00&1.52&1.05$\pm$0.02&-&-&-&-&2.24&2.02$\pm$0.07&2.76&2.87$\pm$0.06&3.24&3.95$\pm$0.05\\
030925$+$01&1.26&2.90$\pm$0.08&1.52&4.45$\pm$0.11&1.84&6.86$\pm$0.17&2.37&11.02$\pm$0.09&2.81&11.62$\pm$0.04&2.99&11.90$\pm$0.06\\
031833$+$00&1.26&0.70$\pm$0.04&1.78&1.46$\pm$0.09&-&-&2.24&3.01$\pm$0.10&2.76&4.41$\pm$0.06&3.24&5.18$\pm$0.09\\
034526$+$00&1.26&1.02$\pm$0.08&1.52&1.40$\pm$0.07&1.78&1.87$\pm$0.08&2.24&3.68$\pm$0.09&2.76&5.14$\pm$0.06&3.24&6.35$\pm$0.07\\
\hline
&$\nu_{S4}$&$S_{S4}$&$\nu_{C1}$&$S_{C1}$&$\nu_{C2}$&$S_{C2}$&$\nu_{C3}$&$S_{C3}$&$\nu_{C4}$&$S_{C4}$&$\nu_{X1}$&$S_{X1}$\\
\hline
221650$+$00&3.76&2.54$\pm$0.05&4.54&2.50$\pm$0.04&5.06&2.34$\pm$0.03&7.14&1.98$\pm$0.04&7.66&1.90$\pm$0.05&8.24&1.89$\pm$0.05\\
221812$-$01&3.76&12.16$\pm$0.21&4.54&13.82$\pm$0.11&5.06&13.96$\pm$0.10&7.14&12.64$\pm$0.23&7.66&12.26$\pm$0.26&8.24&12.07$\pm$0.19\\
223041$-$00&3.76&5.35$\pm$0.08&4.54&6.58$\pm$0.11&5.06&6.53$\pm$0.15&7.14&6.14$\pm$0.23&7.66&5.98$\pm$0.22&8.24&6.27$\pm$0.15\\
233001$-$00&3.76&7.00$\pm$0.19&4.54&6.53$\pm$0.12&5.06&6.32$\pm$0.12&7.14&5.88$\pm$0.18&7.66&5.62$\pm$0.21&8.24&5.89$\pm$0.21\\
010733$+$01&3.76&2.99$\pm$0.09&4.54&2.91$\pm$0.16&5.06&2.66$\pm$0.16&7.14&2.29$\pm$0.12&7.66&2.29$\pm$0.11&8.24&2.323$\pm$0.05\\
013815$+$00&3.76&4.96$\pm$0.16&4.54&5.29$\pm$0.18&5.06&5.35$\pm$0.20&7.14&4.58$\pm$0.15&7.66&4.54$\pm$0.17&8.24&4.53$\pm$0.09\\
015411$-$01&3.76&5.39$\pm$0.09&4.54&5.55$\pm$0.03&5.06&5.39$\pm$0.03&7.14&5.07$\pm$0.09&7.66&4.98$\pm$0.04&8.24&4.96$\pm$0.05\\
020827$-$00&3.76&5.29$\pm$0.68&4.54&5.44$\pm$0.30&5.06&4.89$\pm$0.49&7.14&4.52$\pm$0.19&7.66&4.31$\pm$0.17&8.24&4.43$\pm$0.18\\
030533$+$00&3.76&4.53$\pm$0.16&4.54&5.24$\pm$0.05&5.06&5.25$\pm$0.06&7.14&5.26$\pm$0.09&7.66&5.14$\pm$0.09&8.24&5.27$\pm$0.14\\
\hline
\enddata
\end{deluxetable}

\setcounter{table}{0}
\begin{deluxetable*}{ c c c c c c c c c c c c c }[hbt!]
\tablecolumns{13}
\tablewidth{0pt}
\tablecaption{VLA measurement points - continued.}
\tablehead{&$\nu_{S4}$&$S_{S4}$&$\nu_{C1}$&$S_{C1}$&$\nu_{C2}$&$S_{C2}$&$\nu_{C3}$&$S_{C3}$&$\nu_{C4}$&$S_{C4}$&$\nu_{X1}$&$S_{X1}$}
\startdata
\hline
030925$+$01&3.44&11.74$\pm$0.06&4.54&10.51$\pm$0.05&6.10&8.33$\pm$0.03&7.14&6.76$\pm$0.04&7.66&6.41$\pm$0.05&8.24&6.07$\pm$0.05\\
031833$+$00&3.76&5.31$\pm$0.24&4.54&5.49$\pm$0.08&5.06&4.94$\pm$0.09&7.14&3.48$\pm$0.08&7.66&3.06$\pm$0.09&8.24&2.55$\pm$0.17\\
034526$+$00&3.76&7.56$\pm$0.20&4.54&10.28$\pm$0.05&5.06&11.02$\pm$0.06&7.14&12.89$\pm$0.13&7.66&12.95$\pm$0.07&8.24&13.18$\pm$0.13\\
\hline
&$\nu_{X2}$&$S_{X2}$&$\nu_{X3}$&$S_{X3}$&$\nu_{X4}$&$S_{X4}$&$\nu_{Ku1}$&$S_{Ku1}$&$\nu_{Ku2}$&$S_{Ku2}$&$\nu_{Ku3}$&$S_{Ku3}$\\
\hline
221650$+$00&8.76&1.66$\pm$0.03&10.643&1.54$\pm$0.03&11.16&1.38$\pm$0.05&13.24&1.33$\pm$0.05&13.76&1.40$\pm$0.06&16.24&1.27$\pm$0.09\\
221812$-$01&8.76&11.49$\pm$0.21&10.64&9.27$\pm$0.27&11.16&8.72$\pm$0.30&13.24&7.55$\pm$0.08&13.76&7.09$\pm$0.12&16.24&4.83$\pm$0.22\\
223041$-$00&8.76&6.01$\pm$0.13&10.64&5.37$\pm$0.18&11.16&5.35$\pm$0.21&13.24&4.95$\pm$0.09&13.76&4.58$\pm$0.11&16.24&3.79$\pm$0.15\\
233001$-$00&8.76&5.99$\pm$0.22&10.64&5.90$\pm$0.29&11.16&5.97$\pm$0.30&13.24&5.93$\pm$0.54&13.76&5.94$\pm$0.55&16.24&5.90$\pm$0.65\\
010733$+$01&8.76&2.15$\pm$0.05&10.64&2.06$\pm$0.04&11.16&1.92$\pm$0.06&13.24&1.73$\pm$0.07&13.76&1.64$\pm$0.05&16.24&1.51$\pm$0.06\\
013815$+$00&8.76&4.14$\pm$0.10&10.64&3.98$\pm$0.13&11.16&3.99$\pm$0.10&13.24&3.57$\pm$0.08&13.76&3.50$\pm$0.07&16.24&3.39$\pm$0.12\\
015411$-$01&8.76&4.97$\pm$0.05&10.64&4.76$\pm$0.07&11.16&4.76$\pm$0.05&13.24&4.36$\pm$0.06&13.76&4.17$\pm$0.10&16.24&3.76$\pm$0.12\\
020827$-$00&8.76&4.34$\pm$0.16&10.64&4.31$\pm$0.06&11.16&4.30$\pm$0.11&13.24&3.98$\pm$0.15&13.76&3.67$\pm$0.14&16.24&3.66$\pm$0.14\\
030533$+$00&8.76&5.07$\pm$0.13&10.64&4.66$\pm$0.03&11.16&4.44$\pm$0.06&13.24&3.97$\pm$0.14&13.76&3.93$\pm$0.09&16.24&3.46$\pm$0.07\\
030925$+$01&9.70&5.40$\pm$0.03&10.64&4.91$\pm$0.05&11.16&4.57$\pm$0.05&13.24&4.00$\pm$0.05&13.76&3.86$\pm$0.05&15.00&3.59$\pm$0.03\\
031833$+$00&8.76&2.40$\pm$0.15&10.64&1.88$\pm$0.06&11.16&1.60$\pm$0.05&13.24&1.29$\pm$0.04&13.76&1.13$\pm$0.04&16.24&0.92$\pm$0.06\\
034526$+$00&8.76&12.78$\pm$0.13&10.64&11.94$\pm$0.17&11.16&11.60$\pm$0.13&13.24&10.08$\pm$0.35&13.76&9.86$\pm$0.40&16.24&9.04$\pm$0.24\\
\hline
&$\nu_{Ku4}$&$S_{Ku4}$\\
\hline
221650$+$00&16.76&1.19$\pm$0.06\\
221812$-$01&16.76&4.82$\pm$0.21\\
223041$-$00&16.76&3.48$\pm$0.17\\
233001$-$00&16.76&5.86$\pm$0.63\\
010733$+$01&16.76&1.36$\pm$0.08\\
013815$+$00&16.76&3.32$\pm$0.16\\
015411$-$01&16.76&3.59$\pm$0.11\\
020827$-$00&16.76&3.52$\pm$0.11\\
030533$+$00&16.76&3.40$\pm$0.12\\
030925$+$01&16.76&3.08$\pm$0.05\\
031833$+$00&16.76&0.90$\pm$0.05\\
034526$+$00&16.76&8.77$\pm$0.27\\
\hline
& & & & & & {\bf2015} & & & & & &\\
\hline
&$\nu_{L1}$&$S_{L1}$&$\nu_{L2}$&$S_{L2}$&$\nu_{L3}$&$S_{L3}$&$\nu_{S1}$&$S_{S1}$&$\nu_{S2}$&$S_{S2}$&$\nu_{S3}$&$S_{S3}$\\
\hline
221812$-$01&1.64&0.88$\pm$0.09&-&-&-&-&2.50&9.15$\pm$0.11&-&-&3.50&12.17$\pm$0.18\\
010733$+$01&1.42&2.67$\pm$0.94&1.62&4.25$\pm$1.02&1.81&3.10$\pm$0.34&2.63&4.78$\pm$1.29&-&-&-&-\\
015411$-$01A&1.32&3.67$\pm$0.15&1.71&4.29$\pm$0.13&-&-&2.50&4.66$\pm$0.10&3.50&4.62$\pm$0.07&-&-\\
015411$-$01J&1.51&3.77$\pm$0.20&-&-&2.50&4.03$\pm$0.06&3.44&4.25$\pm$0.13&-&-&-&-\\
030533$+$00&1.32&0.84$\pm$0.12&1.52&0.98$\pm$0.15&-&-&2.63&3.27$\pm$0.11&3.00&3.76$\pm$0.10&3.24&4.24$\pm$0.08\\
030925$+$01F&1.52&7.72$\pm$1.50&-&-&-&-&2.55&15.63$\pm$0.60&3.50&17.11$\pm$0.11&-&-\\
030925$+$01J&1.51&4.18$\pm$0.11&-&-&-&-&2.50&11.53$\pm$0.15&-&-&-&-\\
031833$+$00&1.58&0.76$\pm$0.17&-&-&-&-&2.50&2.95$\pm$0.16&2.74&3.76$\pm$0.18&3.37&4.80$\pm$0.19\\
034526$+$00&1.64&0.96$\pm$0.20&-&-&-&-&2.50&2.12$\pm$0.31&2.74&2.84$\pm$0.28&3.37&4.37$\pm$0.41\\
\hline
&$\nu_{C1}$&$S_{C1}$&$\nu_{C2}$&$S_{C2}$&$\nu_{C3}$&$S_{C3}$&$\nu_{C4}$&$S_{C4}$&$\nu_{C5}$&$S_{C5}$&$\nu_{X1}$&$S_{X1}$\\
\hline
221812$-$01&4.67&14.40$\pm$0.15&-&-&6.10&13.29$\pm$0.14&7.53&12.03$\pm$0.17&-&-&8.55&10.60$\pm$0.28\\
010733$+$01&4.74&2.84$\pm$0.07&5.26&2.62$\pm$0.11&5.50&2.80$\pm$0.07&5.74&2.76$\pm$0.14&6.26&2.75$\pm$0.08&8.24&2.31$\pm$0.05\\
015411$-$01A&4.67&3.87$\pm$0.08&6.16&3.34$\pm$0.08&7.53&2.95$\pm$0.10&7.99&2.90$\pm$0.15&-&-&8.49&2.83$\pm$0.16\\
015411$-$01J&4.80&3.66$\pm$0.09&7.40&2.85$\pm$0.10&-&-&-&-&-&-&8.55&2.21$\pm$0.08\\
030533+00&5.25&4.98$\pm$0.11&6.55&4.65$\pm$0.03&-&-&-&-&-&-&8.80&4.57$\pm$0.06\\
030925$+$01F&4.67&18.17$\pm$1.00&6.10&15.67$\pm$0.18&7.53&13.91$\pm$0.25&-&-&-&-&8.49&12.65$\pm$0.45\\
030925$+$01J&4.95&12.76$\pm$0.11&6.55&11.62$\pm$0.06&7.99&10.22$\pm$0.17&-&-&-&-&8.49&10.42$\pm$0.16\\
031833$+$00&4.95&5.20$\pm$0.09&5.97&4.70$\pm$0.15&7.40&3.48$\pm$0.15&7.99&3.20$\pm$0.17&-&-&8.74&2.65$\pm$0.17\\
034526$+$00&4.95&7.80$\pm$0.48&5.97&9.19$\pm$0.31&7.99&9.95$\pm$0.46&-&-&-&-&9.07&9.70$\pm$0.27\\
\hline
&$\nu_{X2}$&$S_{X2}$&$\nu_{X3}$&$S_{X3}$&$\nu_{X4}$&$S_{X4}$&$\nu_{X5}$&$S_{X5}$&$\nu_{X6}$&$S_{X6}$&$\nu_{Ku1}$&$S_{Ku1}$\\
\hline
221812$-$01&8.75&9.26$\pm$0.17&9.83&8.86$\pm$0.26&11.05&7.56$\pm$0.23&-&-&-&-&14.30&5.88$\pm$0.12\\
010733$+$01&8.76&2.05$\pm$0.05&9.00&2.22$\pm$0.03&9.24&2.21$\pm$0.05&9.76&2.08$\pm$0.05&-&-&13.99&1.75$\pm$0.03\\
015411$-$01A&9.77&2.57$\pm$0.13&-&-&-&-&-&-&-&-&13.37&1.95$\pm$0.13\\
015411$-$01J&9.83&1.84$\pm$0.08&11.05&1.76$\pm$0.08&-&-&-&-&-&-&13.50&1.27$\pm$0.08\\
030533$+$00&10.73&4.24$\pm$0.08&-&-&-&-&-&-&-&-&13.24&1.58$\pm$0.06\\
030925$+$01F&9.77&11.53$\pm$0.80&-&-&11.05&10.70$\pm$0.11&-&-&-&-&14.30&9.00$\pm$0.10\\
030925$+$01J&9.83&9.27$\pm$0.11&-&-&-&-&-&-&-&-&13.50&6.95$\pm$0.13\\
031833$+$00&9.38&2.32$\pm$0.18&10.02&2.21$\pm$0.19&10.66&1.84$\pm$0.19&11.30&1.59$\pm$0.09&-&-&12.18&1.34$\pm$0.08\\
034526$+$00&10.02&9.50$\pm$0.48&10.67&9.30$\pm$0.47&11.30&9.10$\pm$0.48&-&-&-&-&12.18&9.00$\pm$0.49\\
\hline
\enddata
\end{deluxetable*}

\setcounter{table}{0}
\begin{deluxetable*}{ c c c c c c c c c c c c c }[hbt!]
\tablecolumns{13}
\tablewidth{0pt}
\tablecaption{VLA measurement points - continued.}
\tablehead{&$\nu_{Ku2}$&$S_{Ku2}$&$\nu_{Ku3}$&$S_{Ku3}$&$\nu_{Ku4}$&$S_{Ku4}$}
\startdata
\hline
221812$-$01&-&-&-&-&16.00&4.71$\pm$0.14\\
015411$-$01A&14.75&1.76$\pm$0.18&16.13&1.62$\pm$0.19&-&-\\
015411$-$01J&15.2&0.59$\pm$0.20&-&-&-&-\\
030533$+$00&13.50&3.70$\pm$0.09&15.20&3.50$\pm$0.11&16.25&3.40$\pm$0.06\\
030925$+$01F&16.00&8.38$\pm$0.12&-&-&-&-&\\
030925$+$01J&15.20&5.52$\pm$0.18&-&-&-&-\\
031833$+$00&14.43&0.95$\pm$0.09&16.13&0.80$\pm$0.06&-&-\\
034526$+$00&13.37&8.51$\pm$0.46&14.42&8.12$\pm$0.47&16.13&7.03$\pm$0.51\\
\hline
&$\nu_{L1}$&$S_{L1}$&$\nu_{L2}$&$S_{L2}$&$\nu_{S1}$&$S_{S1}$&$\nu_{S2}$&$S_{S2}$&$\nu_{S3}$&$S_{S3}$&$\nu_{C1}$&$S_{C1}$\\
\hline
221812$-$01&1.51&1.82$\pm$0.13&-&-&2.55&6.76$\pm$0.17&-&-&3.45&11.80$\pm$0.20&4.67&13.41$\pm$0.17\\
233001$-$00&1.26&2.40$\pm$0.18&1.90&4.25$\pm$0.24&2.37&7.10$\pm$0.19&3.00&7.51$\pm$0.11&3.44&7.27$\pm$0.09&4.54&7.80$\pm$0.07\\
030925$+$01A&1.52&2.39$\pm$0.20&-&-&2.37&4.59$\pm$0.40&3.00&7.37$\pm$0.45&3.63&10.45$\pm$0.55&5.65&12.59$\pm$0.45\\
030925$+$01J&1.52&4.38$\pm$0.10&-&-&2.55&13.41$\pm$0.30&3.50&15.97$\pm$0.42&-&-&4.67&16.21$\pm$0.18\\
\hline
& & & & & & {\bf2014} & & & & & &\\
\hline
&$\nu_{C2}$&$S_{C2}$&$\nu_{C3}$&$S_{C3}$&$\nu_{C4}$&$S_{C4}$&$\nu_{X1}$&$S_{X1}$&$\nu_{X2}$&$S_{X2}$&$\nu_{X3}$&$S_{X3}$\\
\hline
221812$-$01&6.16&11.27$\pm$0.13&7.53&9.88$\pm$0.17&-&-&8.55&9.20$\pm$0.14&9.83&7.71$\pm$0.12&11.05&6.50$\pm$0.12\\
233001$-$00&5.06&6.98$\pm$0.07&-&-&-&-&8.36&4.98$\pm$0.05&9.39&4.56$\pm$0.05&10.41&3.96$\pm$0.06\\
030925$+$01A&7.40&10.60$\pm$0.28&-&-&-&-&8.23&9.40$\pm$0.25&9.00&8.23$\pm$0.29&9.77&7.55$\pm$0.30\\
030925$+$01J&6.04&15.29$\pm$0.15&7.53&14.40$\pm$0.20&-&-&8.49&13.17$\pm$0.07&9.77&11.89$\pm$0.06&10.99&10.76$\pm$0.10\\
\hline
&$\nu_{X4}$&$S_{X4}$&$\nu_{Ku1}$&$S_{Ku1}$&$\nu_{Ku2}$&$S_{Ku2}$&$\nu_{Ku3}$&$S_{Ku3}$&$\nu_{Ku4}$&$S_{Ku4}$&$\nu_{Ku5}$&$S_{Ku5}$\\
\hline
221812$-$01&-&-&-&-&14.3&5.5$\pm$0.14&15.2&4.99$\pm$0.17&-&-&-&-\\
233001$-$00&11.43&3.47$\pm$0.08&13.24&3.30$\pm$0.07&13.76&3.51$\pm$0.07&15.74&2.94$\pm$0.08&16.26&2.95$\pm$0.08&-&-\\
030925$+$01A&10.92&6.56$\pm$0.33&13.37&4.92$\pm$0.32&14.88&4.17$\pm$0.46&16.13&3.66$\pm$0.39&21.85&3.70$\pm$0.45&24.37&3.30$\pm$0.43\\
030925$+$01J&-&-&14.30&9.29$\pm$0.11&15.19&8.62$\pm$0.14&-&-&-&-&-&-\\
\hline
& & & & & & {\bf2012} & & & & & &\\
\hline
&$\nu_{S1}$&$S_{S1}$&$\nu_{S2}$&$S_{S2}$&$\nu_{S3}$&$S_{S3}$&$\nu_{C1}$&$S_{C1}$&$\nu_{C2}$&$S_{C2}$&$\nu_{C3}$&$S_{C3}$\\
\hline
233001$-$00&2.40&6.85$\pm$0.95&3.20&9.29$\pm$0.71&3.80&9.64$\pm$1.08&4.50&10.52$\pm$0.73&5.10&10.56$\pm$0.71&7.10&9.61$\pm$0.70\\
\hline
&$\nu_{C4}$&$S_{C4}$&$\nu_{Ku1}$&$S_{Ku1}$&$\nu_{Ku2}$&$S_{Ku2}$&$\nu_{Ku3}$&$S_{Ku3}$&$\nu_{Ku4}$&$S_{Ku4}$\\
\hline
233001$-$00&7.70&9.38$\pm$0.73&13.20&7.27$\pm$1.02&13.80&6.84$\pm$0.97&14.20&6.49$\pm$1.03&14.80&6.22$\pm$0.98&-&-\\
\enddata
\label{VLA_measurement_points}
\tablecomments{In case of the sources that have been observed more than once a year, a letter indicates the month of observation. 
}
\end{deluxetable*}

\begin{deluxetable*}{ c c c c c c c }[h!]
\tablecaption{GMRT measurement points}
\tablecolumns{7}
\tablewidth{0pt}
\tablehead{
%& & & {\bf2018} & & &  &
%\hline
&$\nu_{Band4_1}$&$S_{Band4_1}$&$\nu_{Band4_2}$&$S_{Band4_2}$&$\nu_{Band4_3}$&$S_{Band4_3}$\\
Name&(MHz)&(mJy)&(MHz)&(mJy)&(MHz)&(mJy)\\
&(1)&(2)&(3)&(4)&(5)&(6)}
\startdata
221812$-$01 & 595 & 0.73$\pm$0.10 & 677 & 0.79$\pm$0.10 & - & -\\
223041$-$00&593&0.31$\pm$0.02&673&0.36$\pm$0.02&757&0.48$\pm$0.01\\
233001$-$00&593&0.97$\pm$0.10&673&1.11$\pm$0.10&757&1.21$\pm$0.10\\
010733$+$01&593&2.70$\pm$0.20&-&-&715&2.80$\pm$0.20\\
013815$+$00&593&0.51$\pm$0.05&633&0.53$\pm$0.03&757&0.73$\pm$0.04\\
015411$-$01&-&-&673&2.60$\pm$0.25&757&2.70$\pm$0.25\\
030925$+$01&-&-&670&0.50$\pm$0.10&757&0.60$\pm$0.10\\
031833$+$00&-&-&670&0.11$\pm$0.03&-&-\\
\enddata
\tablecomments{013815+00 quasar discussed in \citep{MKB2020} is included.}
\label{gmrt_points}
\end{deluxetable*}

\begin{deluxetable*}{ c c c c c c c c c c c c c c c c}[h!]
\tablecaption{Results of X-ray observations.}
\tablehead{
Name  & Instrument & Epoch &Counts &S$_{0.5-2 \rm{keV}}$ & S$_{2-10 \rm{keV}}$ & log L$_{2-10 \rm{keV}}$& HR &ObsID\\
    &            & & ${0.5-7\, \rm{keV}}$ &  \multicolumn {2}{c} ($\rm 10^{-15}~erg~s^{-1}~cm^{-2}$)                          &   ($\rm erg~s^{-1}$)          &             \\
(1) &  (2)  &   (3)    &   (4)            &    (5)                &   (6) & (7) & (8) & (9)    \\
}
\startdata
221650$+$00 & XMM-Newton &Nov 2016 &$<49.8$  & $<4.5$ & $<9.4$ &$<43.1$ & $-$&0783450101\\
221812$-$01 & Chandra    &Apr 2018& $<4.8$ & $<1.9$ & $<3.5$ & $-$& $-$&20421 \\
223041$-$00 & XMM-Newton & May 2016&$<58.2$ & $<17.4$ & $<36.5$ & $<44.1$ & $-$&0783450201\\
233001$-$00 & XMM-Newton & May 2012&$45.1^{+8.0}_{-7.7}$ & $12.9\pm2.2$ & $23.5\pm4.1$ & 44.6 & $-0.53^{+0.15}_{-0.14}$&0673002341\\
010733$+$01 & Chandra    & Aug 2018& $<3.6$ & $<1.4$ & $<2.5$ & $<41.0$ & $-$&20422 \\
015411$-$01 & Chandra    &Oct 2018& $10\pm3$ & 2.3 & 5.6 & 40.5 & $0.02^{+0.28}_{-0.29}$&20423 \\
030925$+$01 & Chandra    &Nov 2017& $9\pm3$ & 1.7 & 7.5 & 40.4 & $0.49^{+0.16}_{-0.45}$&20424\\
034526$+$00 & Chandra    &Dec 2017& $<3.6$& $<1.5$ & $<2.8$  & $<42.3$ & $-$&20425\\
\enddata

\vspace{0.1 in}
\tablecomments{Columns are listed as follows:(1) source name, (2) instrument used for observations, (3) epoch of observations; (4) number of photon counts; (5)\&(6) X-ray flux; (7) luminosity; (8) hardness ratio $\rm HR = {{H-S} \over {H+S}}$, where $\rm S = 0.5-2$\,keV and $\rm H = 2-7$\,keV; (9) observation ID. Errors in {\it Chandra} counts calculated as $\sqrt{counts}$. '$<$' marks 90\% upper limit.
$\Gamma$=1.7 was assumed for the flux estimation.} 
\label{chandra}
\end{deluxetable*}

\begin{deluxetable*}{ c c c c c c c c c c c c c c c c}[h!]
\tablecolumns{13}
\tablewidth{0pt}
\tablecaption{Emission lines measurements.}
\tablehead{
Name & Epoch&
\multicolumn{2}{c}{MgII} & 
\multicolumn{2}{c}{$\rm H\beta$}& 
\multicolumn{2}{c}{[O III]} & 
\multicolumn{2}{c}{$\rm H\alpha$}& 
[N II]& [S II]& [S II]\\
   & &
\multicolumn{2}{c}{$\lambda2800$}& 
\multicolumn{2}{c}{$\lambda4861$}& 
\multicolumn{2}{c}{$\lambda5007$}& 
\multicolumn{2}{c}{$\lambda6563$}& 
$\lambda6584$& $\lambda6717$& $\lambda6731$\\
 (1) & (2) & (3) & (4) & (5) & (6) & (7)& (8)& (9)&(10)& (11)&(12)& (13)\\
 }
\startdata
233001$-$00&Sep 2012 &54$\pm$2 &9294$\pm$251 &$-$ &$-$ &$-$ &$-$ &$-$ &$-$ &$-$ &$-$&$-$\\
010733$+$01  &Dec 2015 &$-$ &$-$ &8$\pm$1 & 182$\pm$28 &19$\pm$1 &230$\pm$11$*$ &25$\pm$1 &209$\pm$6$*$ & 9$\pm$1&$-$ &$-$\\
           &Jul 2019 &$-$ &$-$ &10$\pm$2& 297$\pm$41$*$& 17$\pm$2& 322$\pm$30$*$& 27$\pm$2& 292$\pm$13$*$& 10$\pm$1& $-$ &$-$\\
015411$-$01&Sep 2000 & $-$&$-$ &70$\pm$9 &747$\pm$13 &$-$ &$-$& 198$\pm$26& 719$\pm$72 & 370$\pm$13&$-$&$-$\\
020827$-$00&Sep 2001 &153$\pm$5 &3220$\pm$107& $-$ &$-$ &$-$ &$-$ &$-$ &$-$ &$-$ &$-$&$-$\\
030925$+$01  &Dec 2001 &$-$ &$-$ &62$\pm$4 &229$\pm$15$*$ &133$\pm$4 &189$\pm$8 &308$\pm$3 & 183$\pm$3&178$\pm$3 &51$\pm$3 &44$\pm$4\\
           &Aug 2019 &$-$ &$-$ &63$\pm$5 & 336$\pm$33 &169$\pm$5 &370$\pm$11$*$ &273$\pm$3 & 299$\pm$4$*$&128$\pm$3 &34$\pm$3 &34$\pm$3\\
\enddata
\vspace{0.1 in}
\tablecomments{Columns are listed as follows: (1) source name; (2) epoch of observations; (3,4) the flux and FWHM for MgII (5,6) the flux and FWHM for $\rm H_{\beta}$; (7,8) the flux and FWHM  for [O III]; (9,10) the flux and FWHM for $\rm H_{\alpha}$; (11, 12, 13) the flux for narrow lines [N II] and [S II]. The FWHM and flux are in units of $\rm km~s^{-1}$, and $\rm 10^{-17}~erg~s^{−1}~cm^{-2}$, respectively. $*$ marks upper limit, line is not corrected for instrumental resolution. We apply the instrumental correction for the line only if the observed FWHM exceeds the instrumental FWHM by at least 50\%.
}
\label{table_emission_lines}
\end{deluxetable*}

\begin{deluxetable*}{ c c c c c c c c c c c c c c c c}[hbt!]
\tablecolumns{9}
\tablewidth{0pt}
\tablecaption{Astrophysical properties of transient radio sources.}
\tablehead{
Name & Epoch& Activity&$\rm logL_{O III}$&
 $\rm logM_{BH}$& method& $\rm logL_{bol}$& method& $\rm log\lambda_{Edd}$& $\rm logP_{j}$& $\rm log P_j/L_{bol}$\\
  & & phase &($\rm erg~s{-1}$)&
($\rm M_{\odot}$)& & ($\rm erg~s{-1}$)&  & &($\rm erg~s{-1}$)& \\
 (1) & (2) & (3) & (4)&(5)&(6)&(7)&(8)&(9)&(10)&(11) \\
 }
\startdata
233001$-$00&Sep 2012 & RL &$-$&9.16 & MgII&45.49&$\rm L(3000\AA)$ &$-$1.78 &44.81&$-$0.68\\
010733$+$01  &Dec 2015 & RL & 39.84 &7.13$a$ &O[III]& 42.65& $\rm H\alpha/H\beta$  &$-$2.59$b$&42.46&$-$0.19 \\
015411$-$01&Sep 2000& RQ &$-$&8.90 &$\sigma_\star$  &42.85& $\rm H\alpha/H\beta$ &$-$4.16 &41.06&$-$1.78\\
020827$-$00&Sep 2001 & RQ &$-$&8.42 & MgII& 45.79&$\rm L(3000\AA)$ &$-$0.75 &43.89&$-$1.90\\
030925$+$01  &Dec 2001& RQ &39.69 &6.75 & O[III] &42.81& $\rm H\alpha/H\beta$ &$-$2.06 &40.80&$-$2.00 \\
  &Aug 2019& RL &39.80 &6.75$c$ & O[III] &42.81$c$& $\rm H\alpha/H\beta$ &$-$2.06$c$ &41.68&$-$1.13 \\
\tableline
013815$+$00& Oct 2001& RQ & $-$& 9.21& MgII& 45.55& $\rm L(3000\AA)$& $-$1.84& 43.33 & $-$2.22\\
        & Dec 2015& RL & 42.27& 9.36& MgII& 45.70& $\rm L(3000\AA)$&$-$1.69& 44.01&$-$1.70\\
        & Jan 2018& RL & 42.13& 9.26& MgII& 45.56& $\rm L(3000\AA)$&$-$1.83& 44.01&$-$1.56\\
\enddata
\vspace{0.1 in}
\tablecomments{Columns are listed as follows: (1) source name; (2) epoch of observations; (3) activity phase: RL (radio-loud) or RQ (radio-quiet);
(4) $\rm [O III]\lambda5700$ line luminosity; (5) black hole mass; (6) method used for black hole mass estimation; (7) bolometric luminosity; (8) method used to derive the bolometric disk luminosity as described in sec. \ref{physical_parameters} 
(9) Eddington ratio ($\rm L_{bol}/L_{Edd}$); (10) jet kinetic power; (11) jet production efficiency. Data for quasar 013815+00 has been taken from \citet{MKB2020}.
$a$ marks upper limit, $b$ marks lower limit, $c$ indicates adopted $\rm M_{BH}$ and calculations based on it.
}

\label{table_physical_parameters}
\end{deluxetable*}

\begin{deluxetable*}{ c c c c c }[hbt!]
\tablecaption{Infrared data.}
\tablehead{
   & W1 &W2 &W3 & W4\\
 Name  &(mag) &(mag) &(mag) &(mag)\\
  & 3.4$\mu m$&4.6$\mu m$ &12$\mu m$& 22$\mu m$
  }
\startdata
221650$+$00&15.82$\pm$0.05&15.73$\pm$0.14&$<$12.07&$<$8.98\\
221812$-$01&17.36$\pm$0.17&$<$16.59&11.91$\pm$0.36&$<$8.93\\
233001$-$00&16.20$\pm$0.07&14.99$\pm$0.08&12.37$\pm$0.46&$<$8.28\\
010733$+$01&15.56$\pm$0.05&14.41$\pm$0.06&11.06$\pm$0.17&$<$8.33\\
015411$-$01&12.45$\pm$0.02&12.43$\pm$0.03&10.91$\pm$0.11&$<$8.43\\
020827$-$00&15.20$\pm$0.03&14.03$\pm$0.04&11.00$\pm$0.12&$<$8.49\\
030533$+$00&15.98$\pm$0.05&15.84$\pm$0.14&$<$11.81&$<$8.69\\
030925$+$01&14.26$\pm$0.03&13.99$\pm$0.04&10.63$\pm$0.12&8.00$\pm$0.26\\
034526$+$00&15.42$\pm$0.04&15.27$\pm$0.10&11.50$\pm$0.19&8.91$\pm$0.48\\
\hline
013815$+$00&15.33$\pm$0.04&14.33$\pm$0.05&11.84$\pm$0.27& $<$8.84\\
\enddata
\tablecomments{WISE colors in Vega magnitudes \citep{Cutri}, as presented in Fig. \ref{WISE}. $<$ marks upper limit. 
%013815+00 quasar discussed in \citep{MKB2020} is included.
}
\label{table_wise}
\end{deluxetable*}
\newpage

%\numberwithin{figure}{section}
%\section{Figures} 
%Here we present the multi-frequency VLBA images (Figure \ref{vlba_images}) and optical spectra (Figure \ref{image_optical_spectra}) of our sources.

\begin{figure*}[ht!]
\begin{minipage}{\textwidth}
    \begin{center}
    \numberwithin{figure}{section}
    \section{Figures}
    Here we present the multi-frequency VLBA images (Figure \ref{vlba_images}) and optical spectra (Figure \ref{image_optical_spectra}) of our sources.
    \end{center}
\end{minipage}
\newline
\includegraphics[scale=0.30]{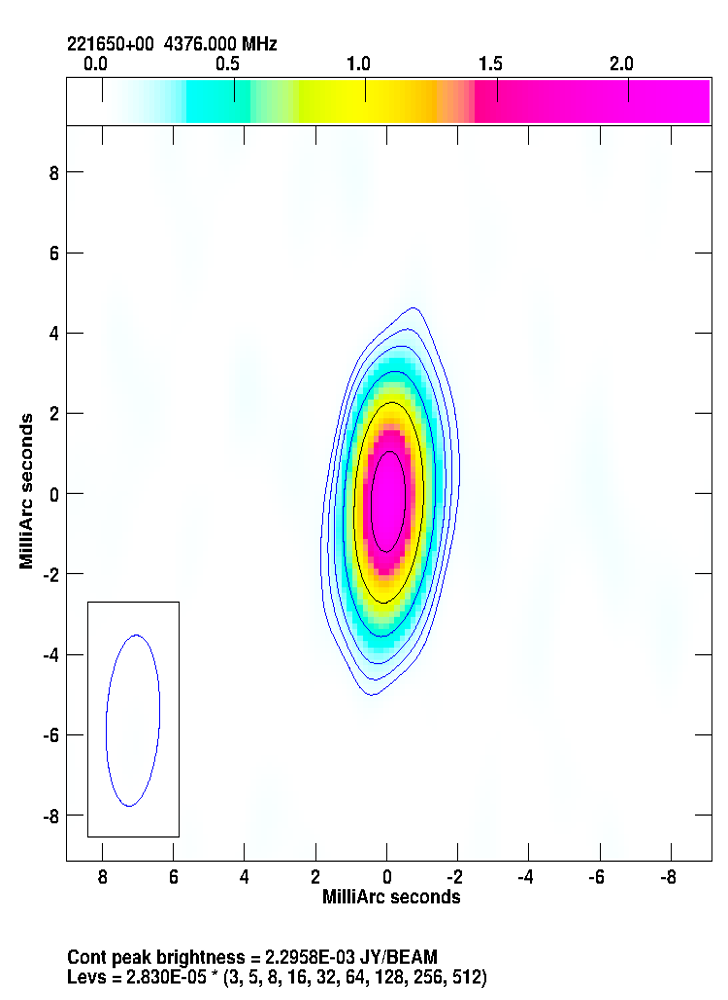}
\includegraphics[scale=0.30]{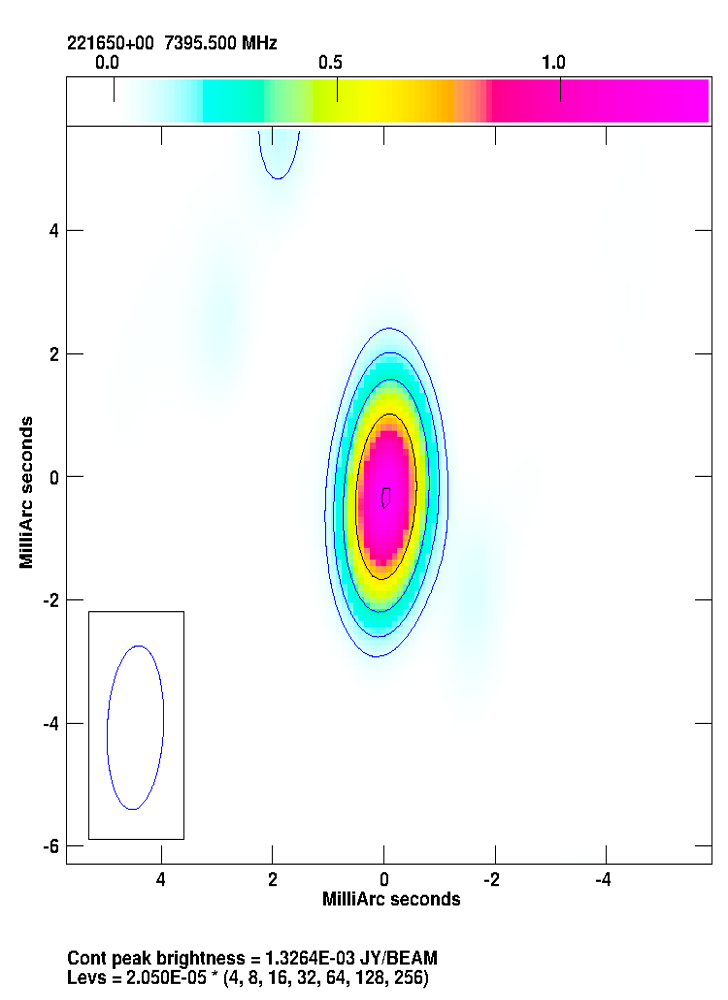}
\includegraphics[scale=0.30]{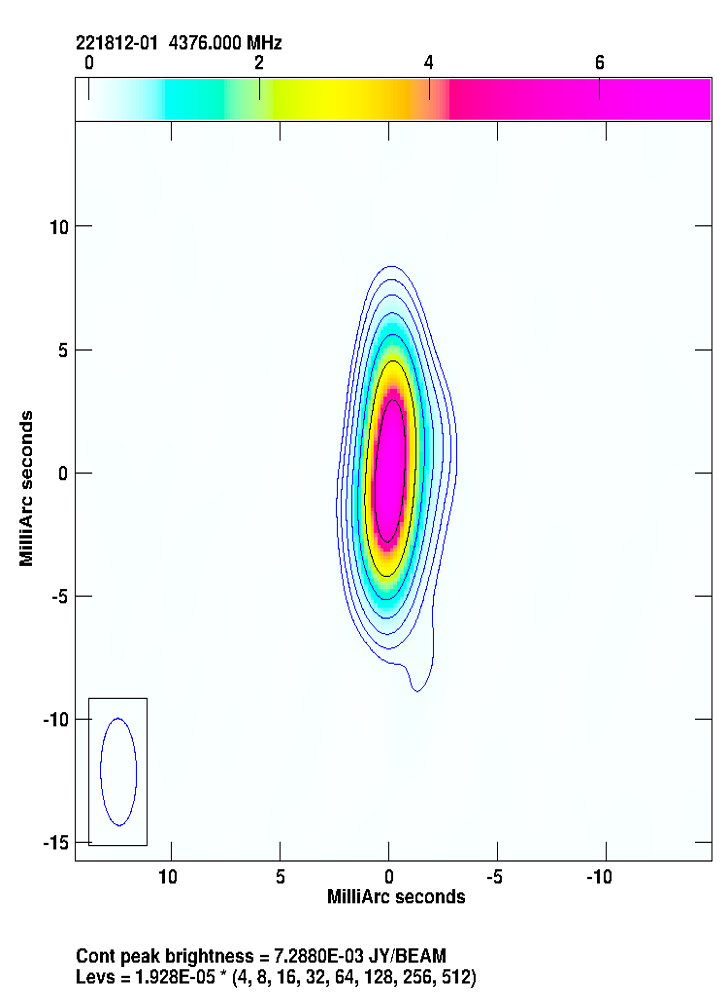}\\
\includegraphics[scale=0.30]{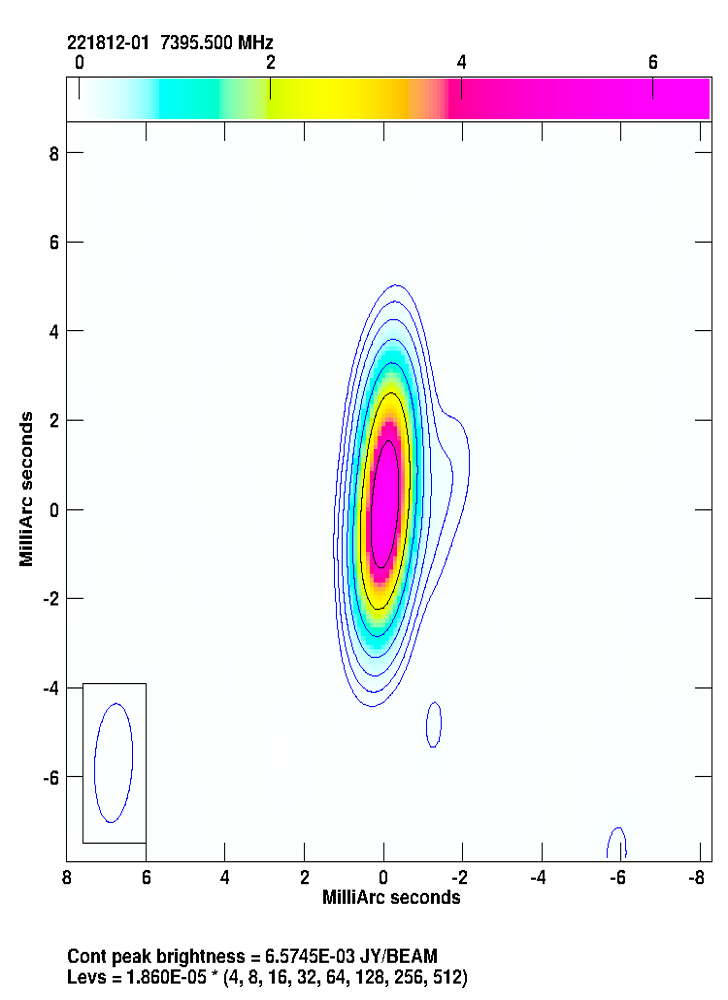}
\includegraphics[scale=0.30]{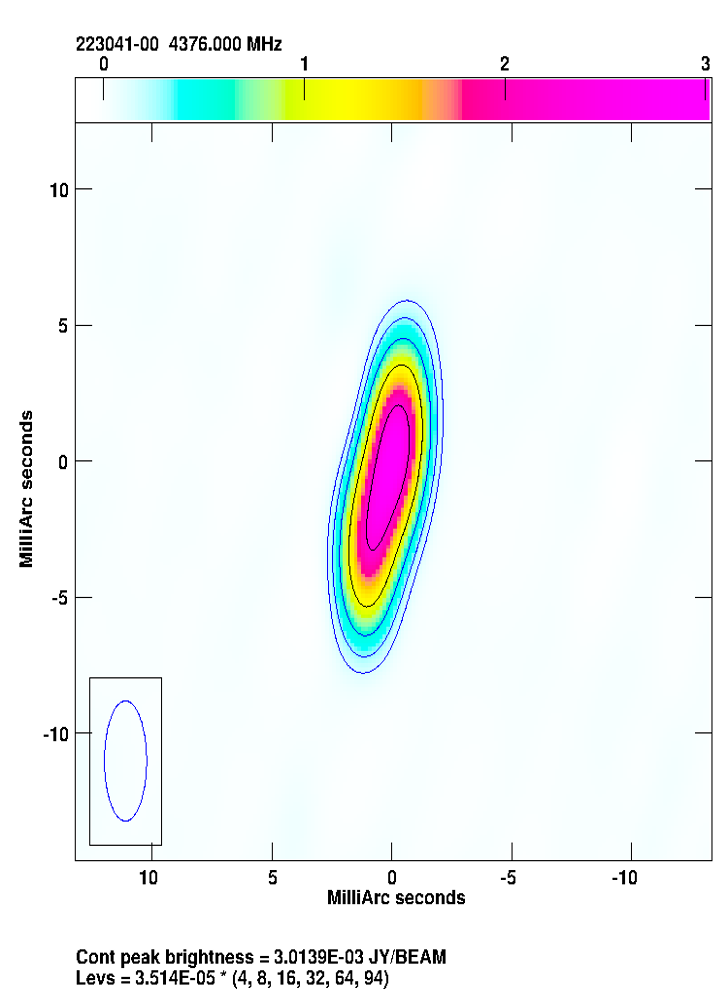}
\includegraphics[scale=0.30]{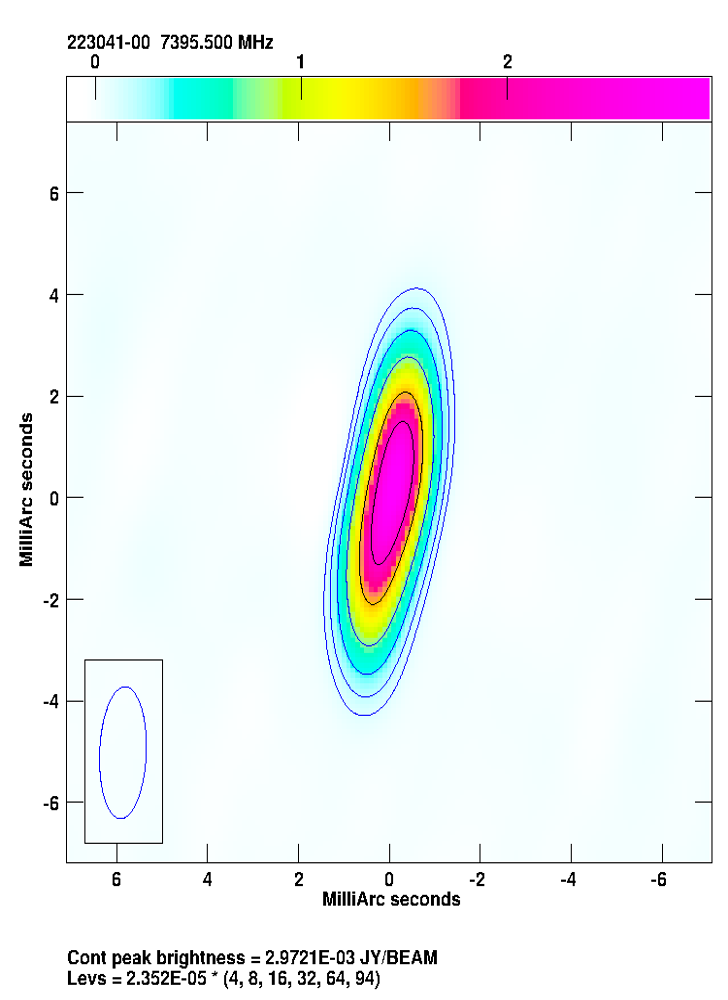}\\

\caption{VLBA 4.5 and 7.5 GHz images. If more than one component has been identified in the source, they are all marked with black dots. In the case of 030925+01 the VLBA images from two observational epochs, June 2015 and March 2016, are presented at the end. }
\label{vlba_images}
\end{figure*}
\setcounter{figure}{0}
\begin{figure*}[hbt!]
\includegraphics[scale=0.30]{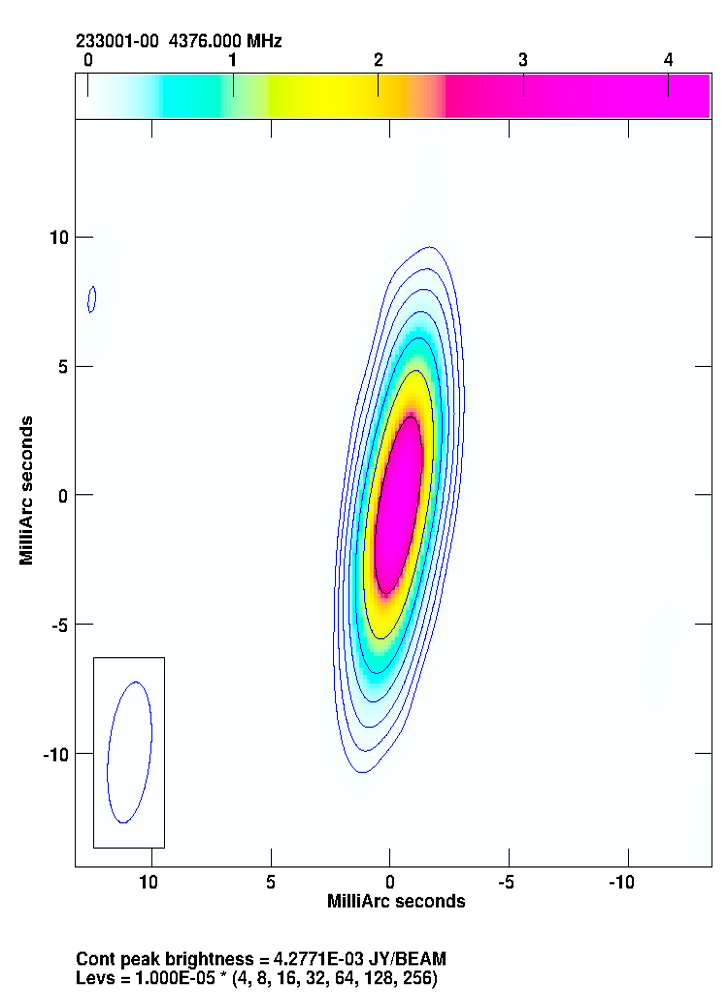}
\includegraphics[scale=0.30]{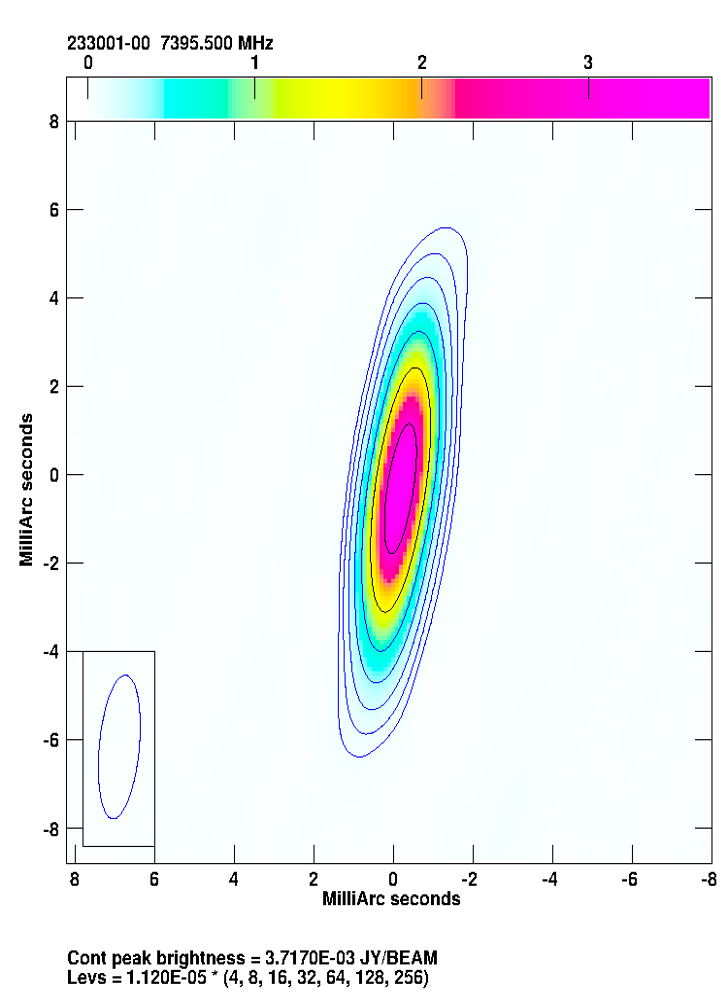}
\includegraphics[scale=0.30]{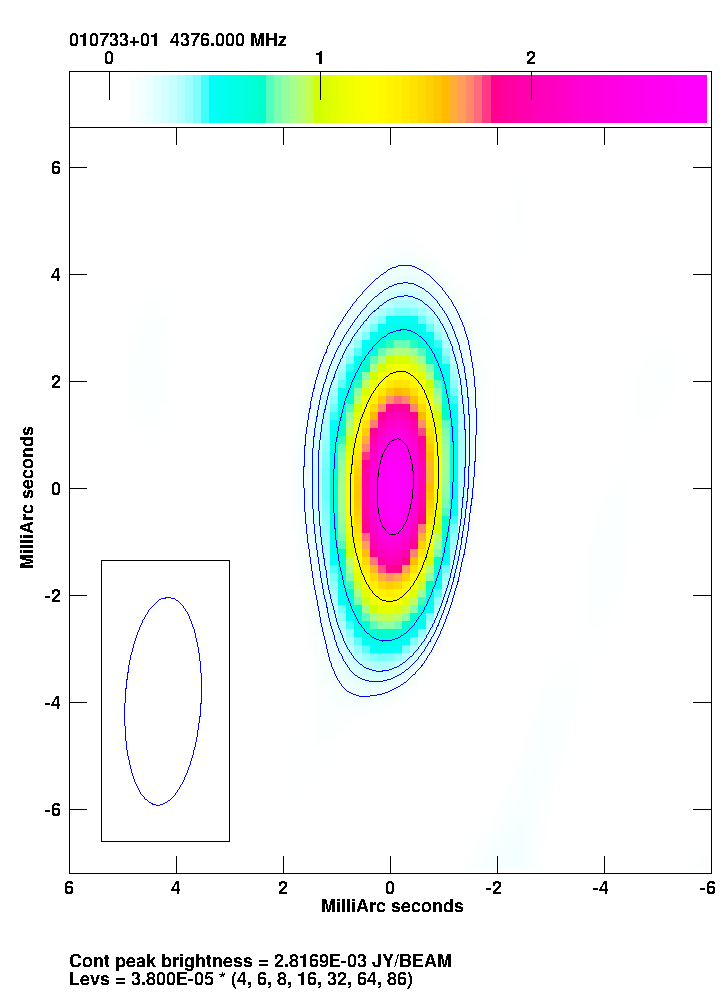}\\
\includegraphics[scale=0.30]{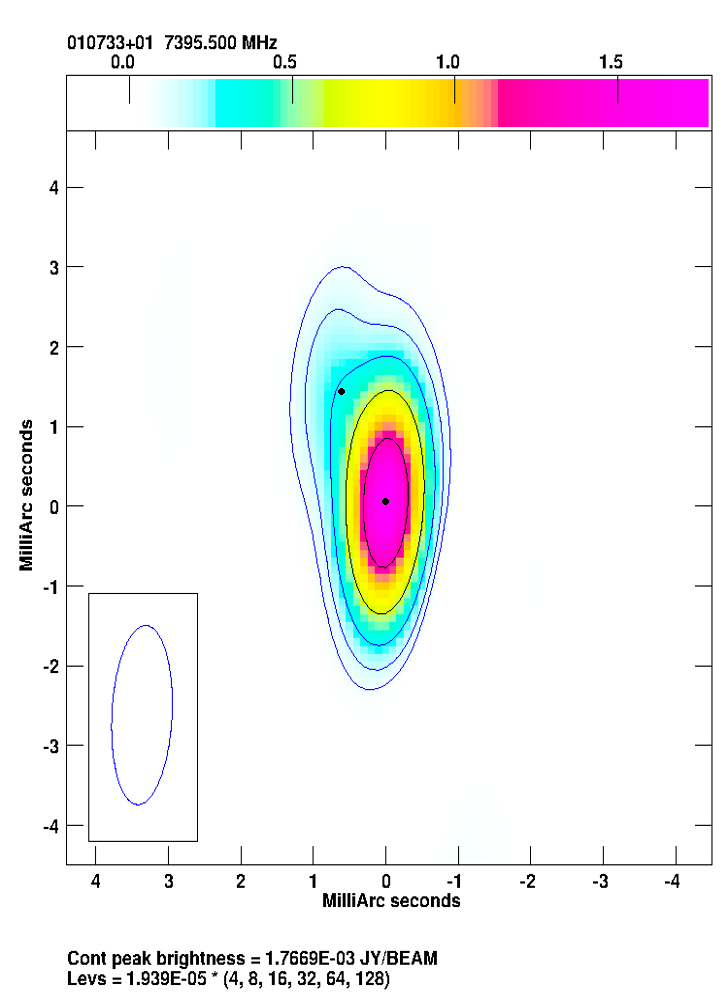}
\includegraphics[scale=0.30]{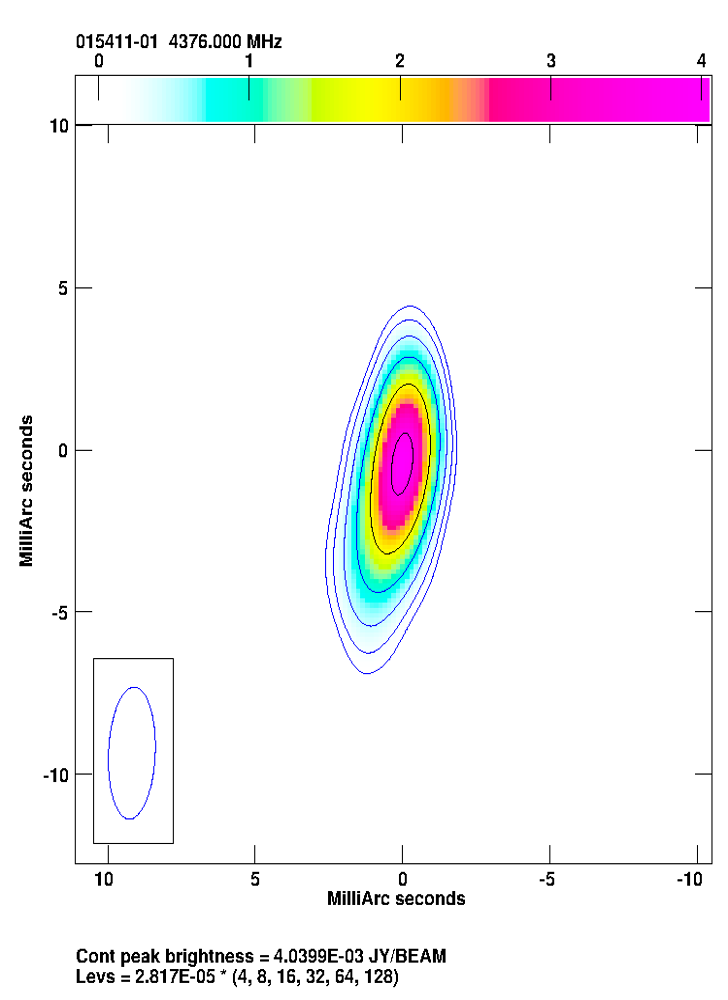}
\includegraphics[scale=0.30]{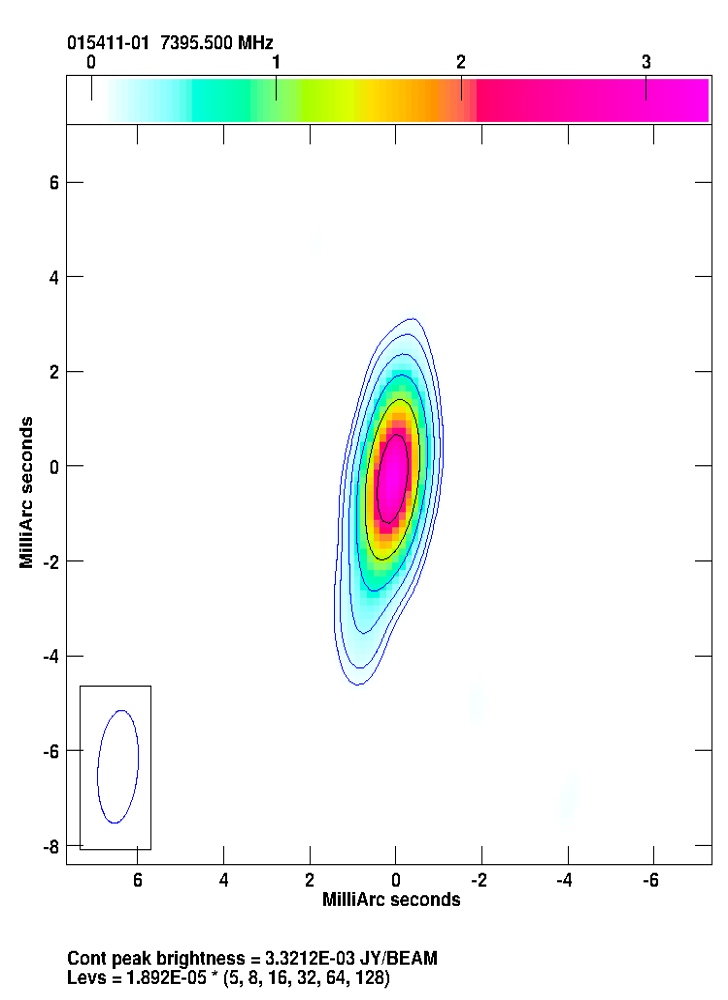}\\
\includegraphics[scale=0.30]{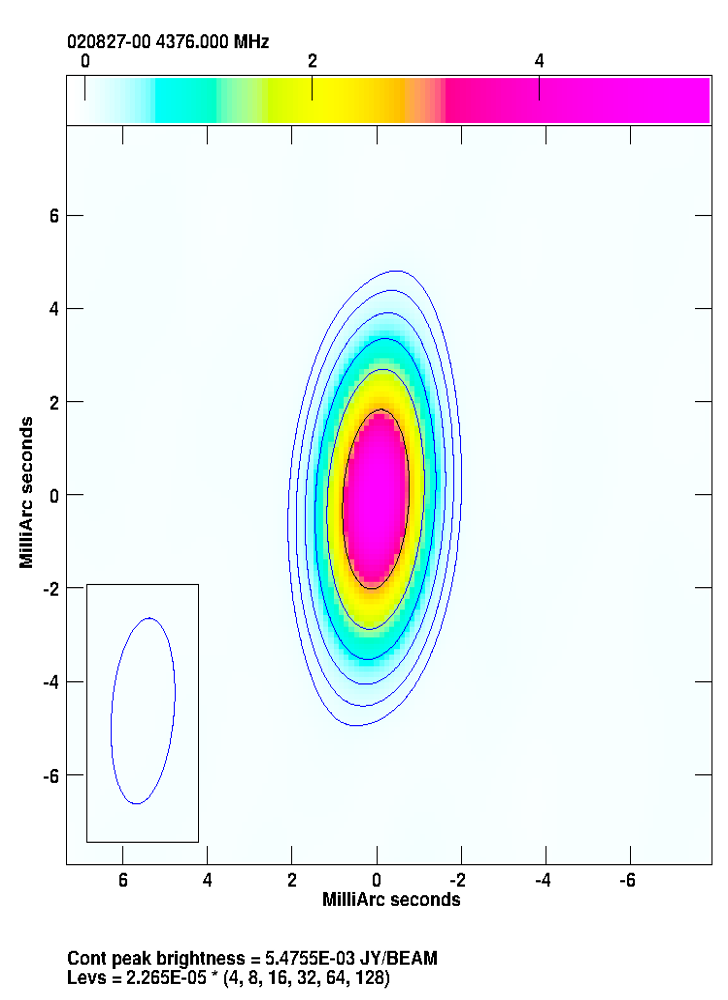}
\includegraphics[scale=0.30]{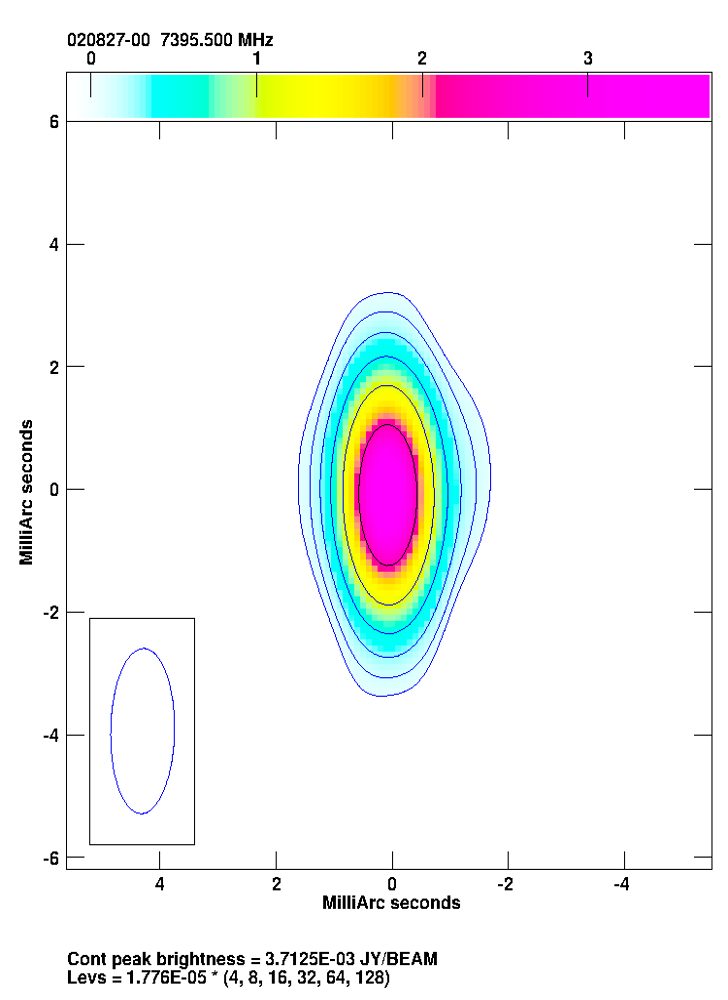}
\includegraphics[scale=0.30]{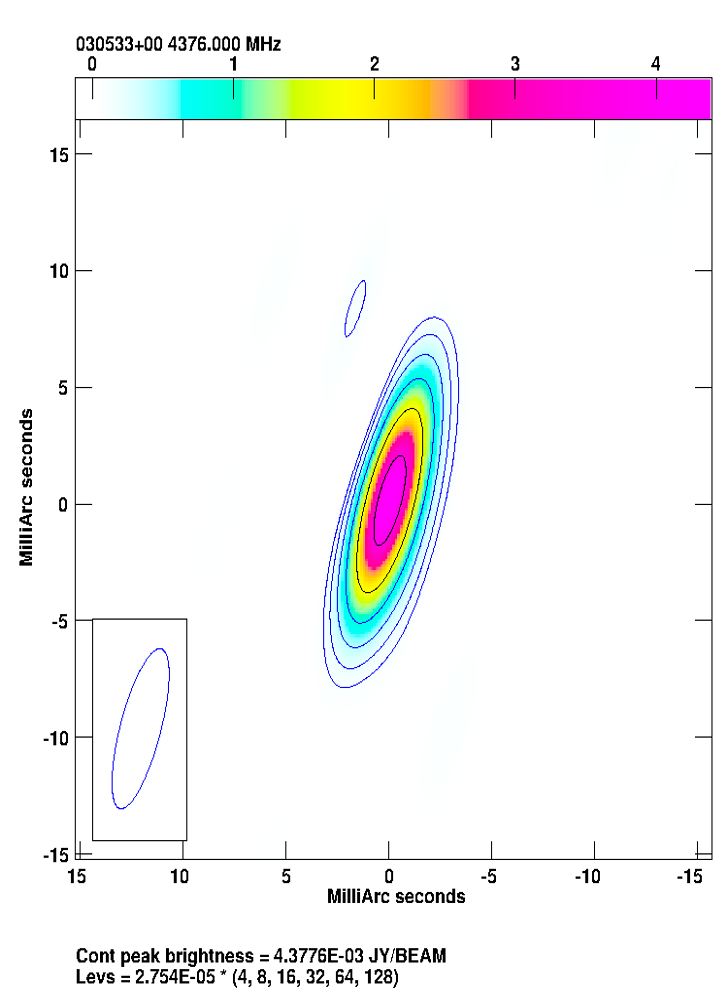}\\
\caption{VLBA 4.5 and 7.5 GHz images - continued.}
\end{figure*}

\setcounter{figure}{0}
\begin{figure*}[hbt!]
\includegraphics[scale=0.30]{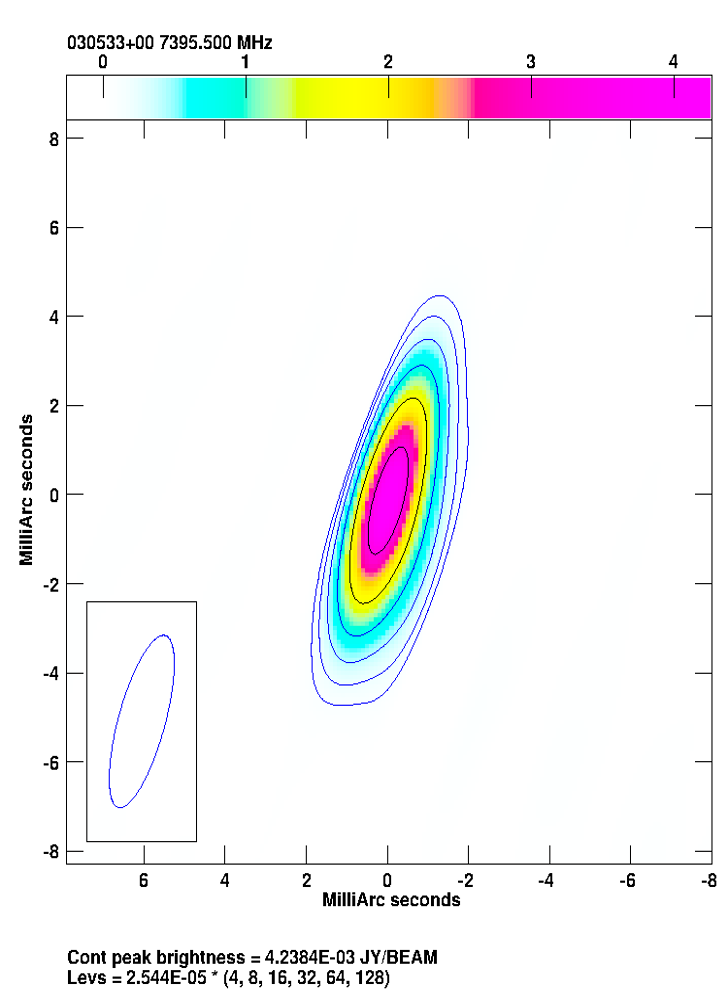}
\includegraphics[scale=0.30]{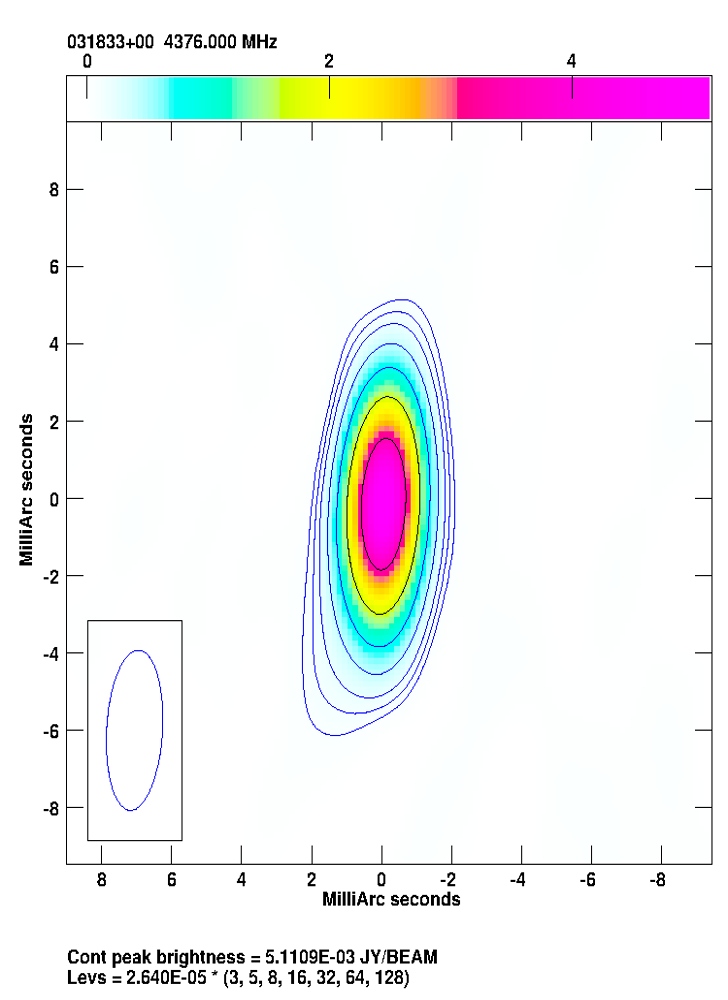}
\includegraphics[scale=0.30]{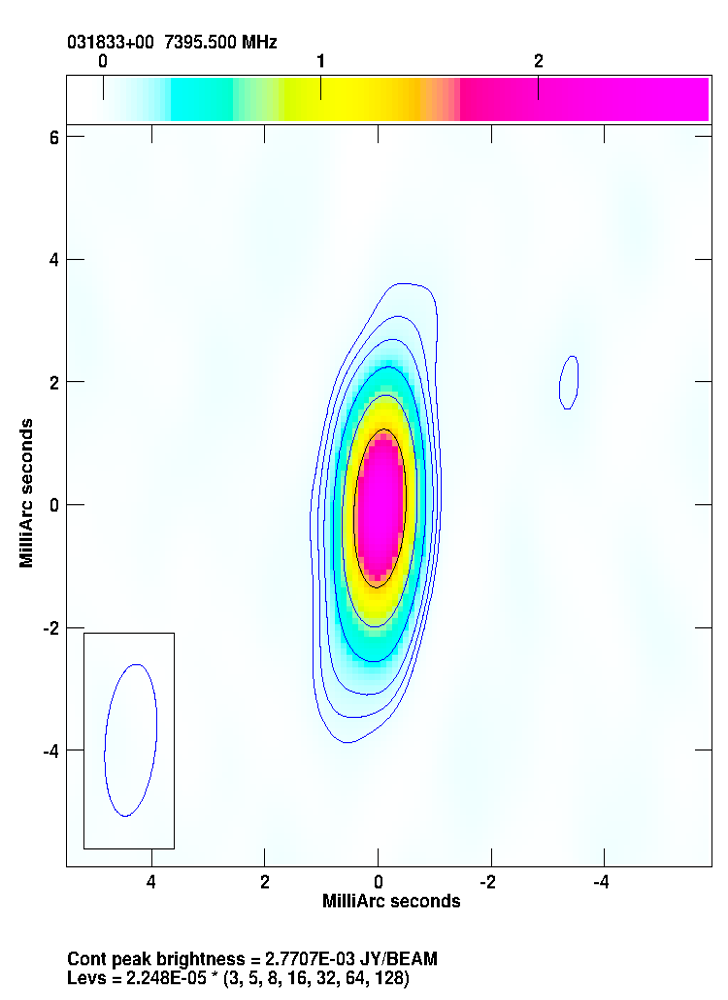}\\
\includegraphics[scale=0.30]{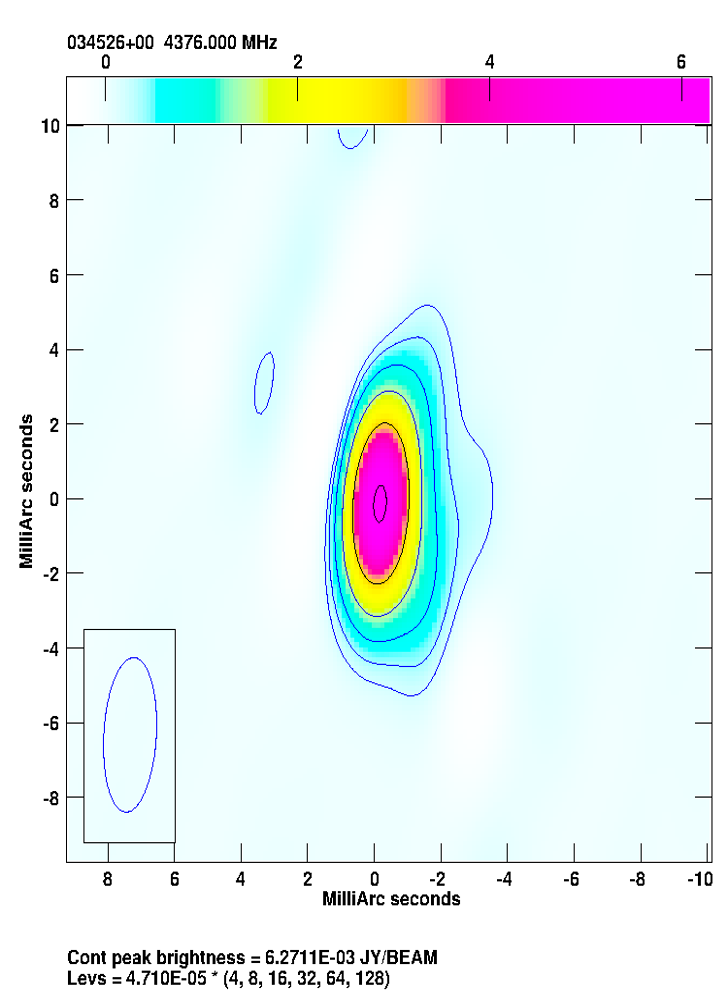}
\includegraphics[scale=0.30]{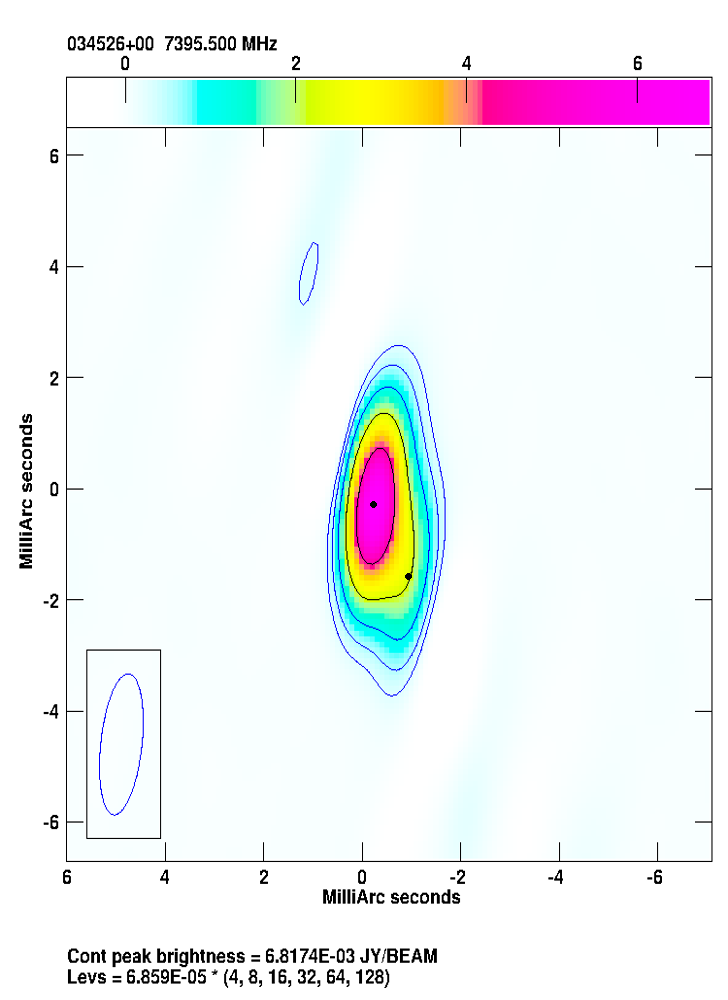}
\includegraphics[scale=0.3]{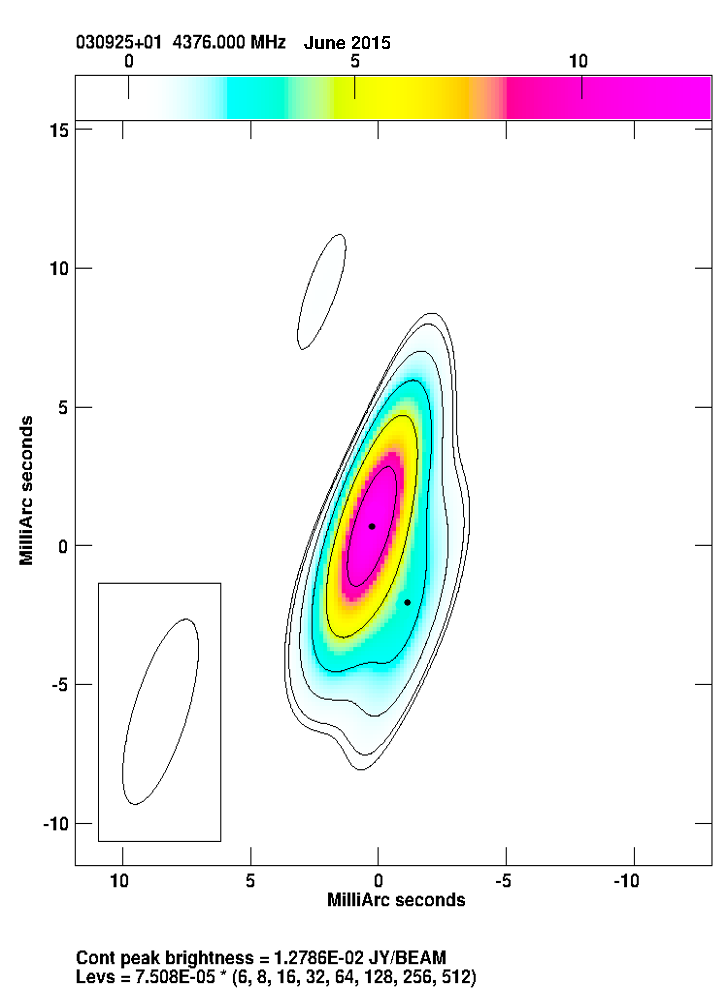}\\
\includegraphics[scale=0.3]{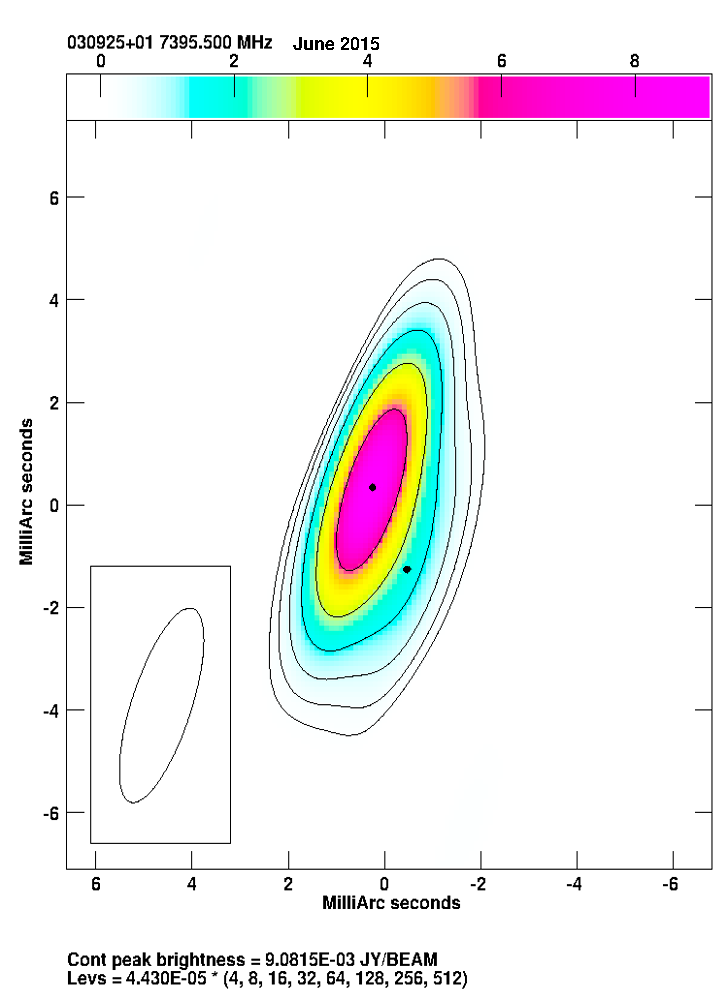}
\includegraphics[scale=0.3]{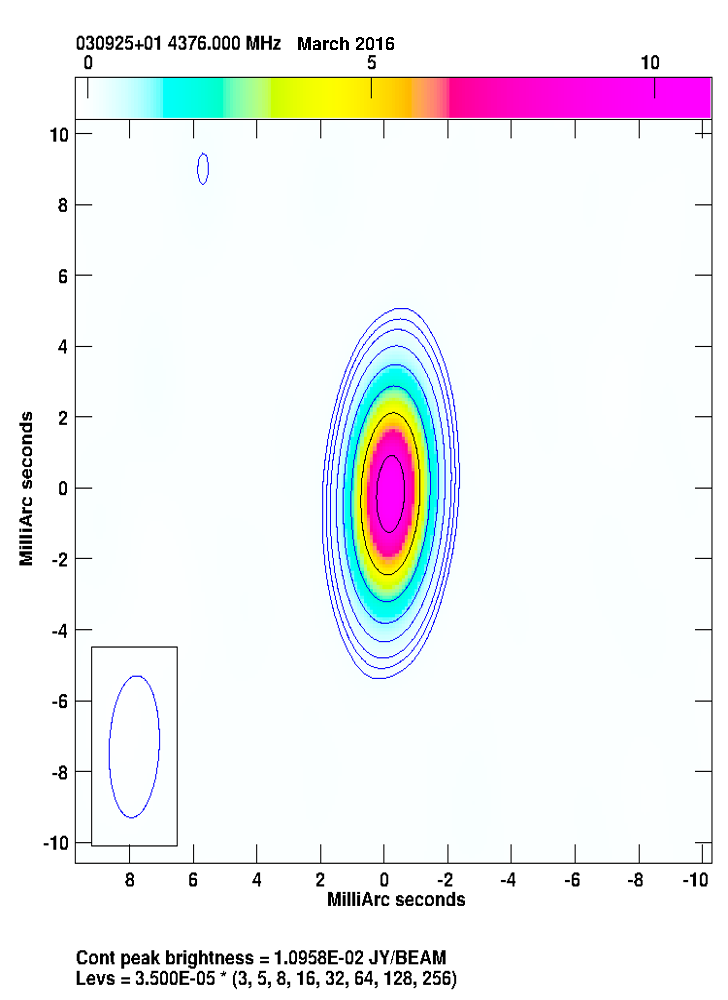}
\includegraphics[scale=0.3]{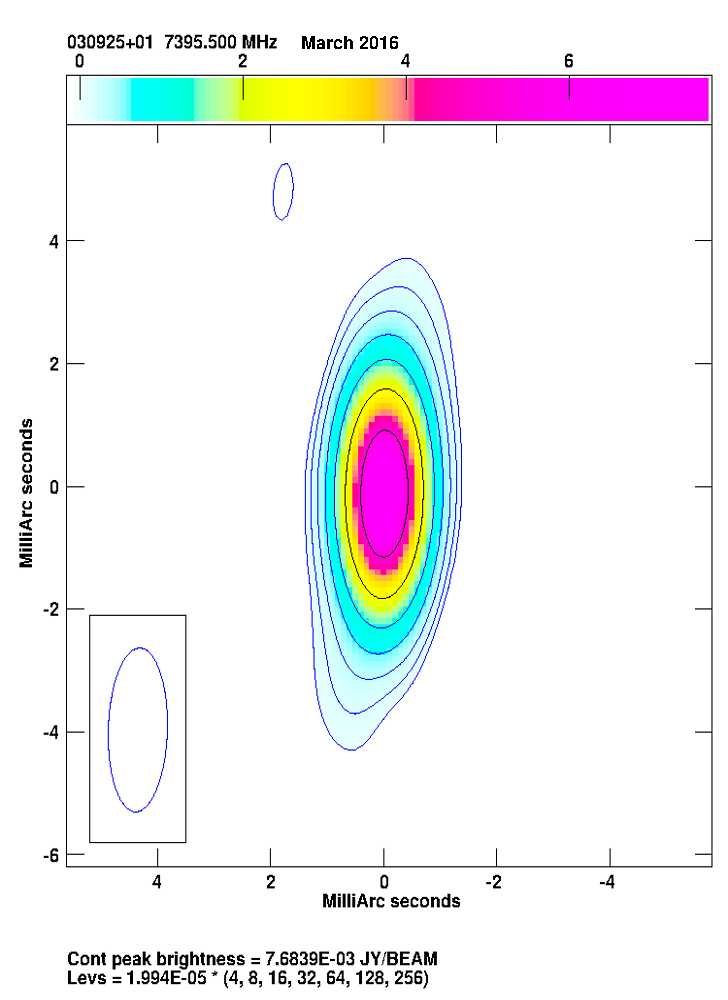}
\caption{VLBA 4.5 and 7.5 GHz images - continued.}
\end{figure*}

\begin{figure*}[hbt!]
\centering
\includegraphics[scale=1.45]{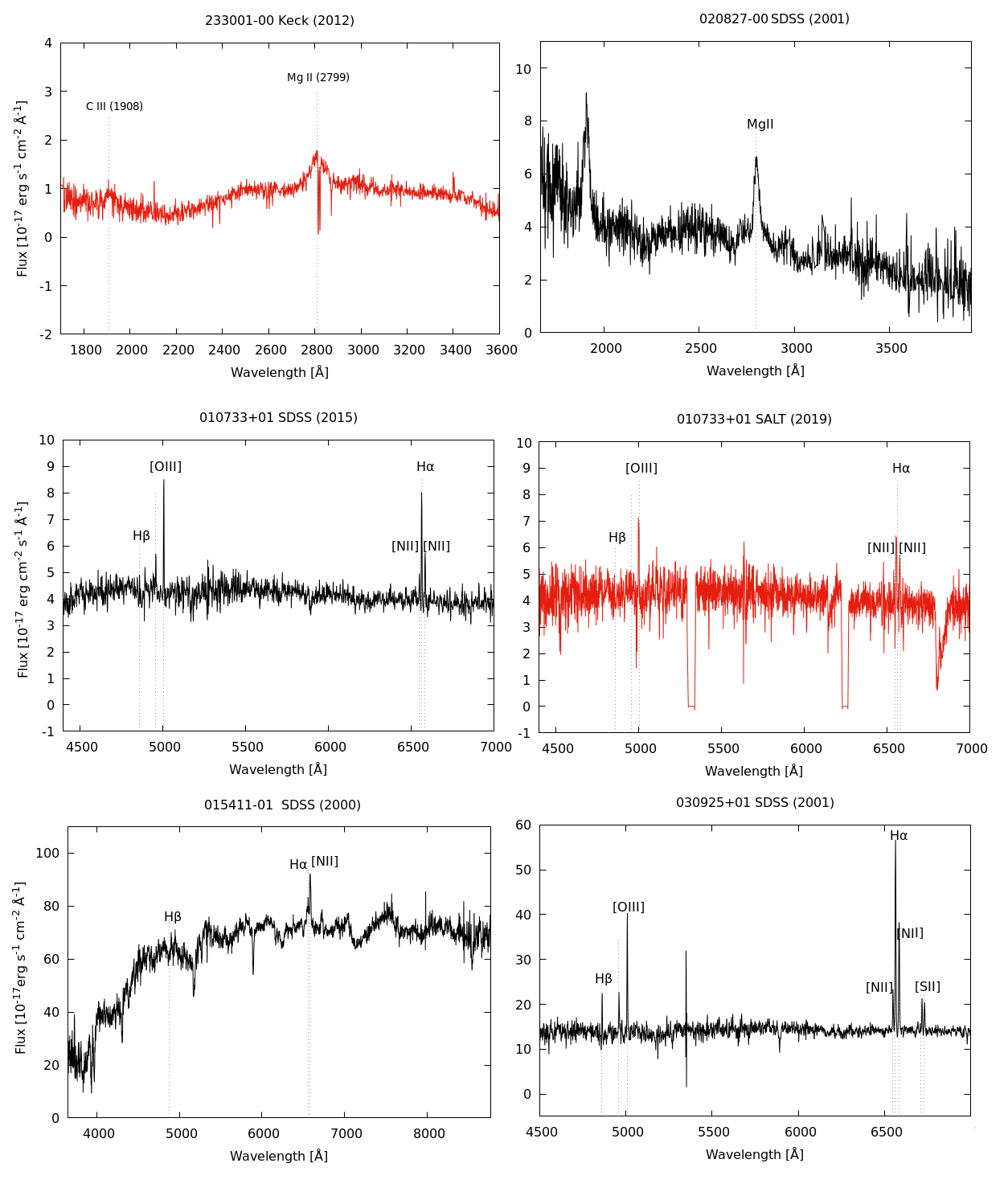}\\
\includegraphics[scale=0.35]{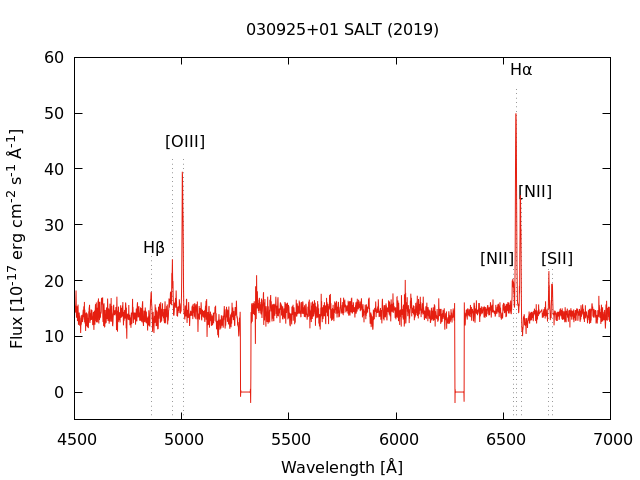}\\

\caption{Available spectra of transient sources. The measurements of marked emission lines are presented in Table \ref{table_emission_lines}. The SDSS spectra are marked in black. Gaps visible in SALT spectra are due to a physical separation between CCDs in acquisition camera.}
\label{image_optical_spectra}
\end{figure*}
\end{document}